\documentclass[aps,prb,twocolumn,preprintnumbers,amsmath,amssymb]{revtex4-1}
\usepackage{graphicx}
\usepackage{dcolumn}
\usepackage{bm}
\usepackage{color}

\begin{document}

\def\gsim{\mathop {\vtop {\ialign {##\crcr 
$\hfil \displaystyle {>}\hfil $\crcr \noalign {\kern1pt \nointerlineskip } 
$\,\sim$ \crcr \noalign {\kern1pt}}}}\limits}
\def\lsim{\mathop {\vtop {\ialign {##\crcr 
$\hfil \displaystyle {<}\hfil $\crcr \noalign {\kern1pt \nointerlineskip } 
$\,\,\sim$ \crcr \noalign {\kern1pt}}}}\limits}


\title{Gr\"{u}neisen Parameter and Thermal Expansion \\ near Magnetic Quantum Critical Points
in Itinerant Electron Systems}


\author{Shinji Watanabe$^1$ and Kazumasa Miyake$^2$}
\affiliation{$^1$Department of Basic Sciences, Kyushu Institute of Technology, Kitakyushu, Fukuoka 804-8550, Japan \\
$^2$Center for Advanced High Magnetic Field Science, Osaka University, Toyonaka, Osaka 560-0043, Japan}


\date{\today}

\begin{abstract}
Complete expressions of the thermal-expansion coefficient $\alpha$ and the Gr\"{u}neisen parameter $\Gamma$ are derived on the basis of the self-consistent renormalization (SCR) theory. 
By considering the zero-point as well as thermal spin fluctuation under the stationary condition, the specific heat for each class of the magnetic quantum critical point (QCP) specified by the dynamical exponent 
$z=3$ [feorromagnetism (FM)] and $z=2$ [antiferromagnetism (AFM)] 
and the spatial dimension $(d=3$ and $2)$ is shown to be expressed as $C_{V}=C_{\rm a}-C_{\rm b}$, where $C_{\rm a}$ is dominant at low temperatures, reproducing the past SCR criticality endorsed by the renormalization group theory. Starting from the explicit form of the entropy and using the Maxwell relation, $\alpha=\alpha_{\rm a}+\alpha_{\rm b}$ (with  $\alpha_{\rm a}$ and $\alpha_{\rm b}$ being related to $C_{\rm a}$ and $C_{\rm b}$, respectively) is derived, which is proven to be equivalent to $\alpha$ derived from the free energy. 
The temperature-dependent coefficient found to exist in $\alpha_{\rm b}$, which is dominant at low temperatures, 
contributes to the crossover from the quantum-critical regime to the Curie-Weiss regime. 
For sufficiently low temperatures, the thermal-expansion coefficient at the QCP behaves as $\alpha\approx\alpha_{\rm b}\sim T^{1/3}$ (3d FM), $T^{1/2}$ (3dAFM), $-\ln{T}$ (2dFM), and $-\ln(-\ln{T})/\ln\left(-\frac{T}{\ln{T}}\right)$ (2d AFM). 
Based on these correctly calculated $C_{V}$ and $\alpha$, Gr\"{u}neisen parameter $\Gamma=\Gamma_{\rm a}+\Gamma_{\rm b}$ is derived, where $\Gamma_{\rm a}$ and $\Gamma_{\rm b}$ contain $\alpha_{\rm a}$ and $\alpha_{\rm b}$, respectively.  
The inverse susceptibility (renormalized by the mode-mode coupling of spin fluctuations) coupled to the volume $V$ 
in $\Gamma_{\rm b}$ gives rise to the divergence of $\Gamma$ 
at the QCP for each class 
even though the characteristic energy scale of spin fluctuation $T_0$ is finite at the QCP, which gives a finite contribution in  
$\Gamma_{\rm a}=-\frac{V}{T_0}\left(\frac{\partial{T_0}}{\partial{V}}\right)_{T=0}$. 
For $T\ll T_{0}$, 
the Gr\"{u}neisen parameter at the QCP behaves as $\Gamma\approx\Gamma_{\rm b}\sim-{T^{-2/3}}/{\ln{T}}$ (3d FM), ${T^{-1/2}}/{({\rm const.}-T^{1/2})}$ (3d AFM), $-T^{-2/3}\ln{T}$ (2d FM), and ${\ln(-\ln{T})}/[{T\ln{T}}{\ln\left(-\frac{T}{\ln{T}}\right)}]$ (2d AFM). 
General properties of $\alpha$ and $\Gamma$ including their signs as well as the relation to $T_0$ and the Kondo temperature 
in temperature-pressure phase diagrams of Ce- and Yb-based heavy electron systems are discussed.  
\end{abstract}


\maketitle

\section{Introduction}

Quantum critical phenomena in itinerant electron systems have attracted much attention in condensed matter physics. 
When the transition temperature to the magnetically ordered phase is suppressed continuously to absolute zero by tuning control parameter of materials such as pressure, magnetic-field, and chemical substitution, the quantum critical point (QCP) is realized. 
Near the QCP, enhanced magnetic fluctuation causes a non-Fermi-liquid behavior in physical quantities, which is referred to as quantum critical phenomena. 
 
The self-consistent renormalization (SCR) theory of spin fluctuation has been developed by Moriya and Kawabata in 1973~\cite{MK1973,MK1973b}. The SCR theory succeeded in explaining not only the Curie-Weiss behavior but also quantum critical behavior at low temperatures in magnetic susceptibility, which  are caused by spin fluctuation in nearly ferromagnetic metals~\cite{Moriya,Lonzarich1985}. 
The spin fluctuation has been revealed to cause 
a non-Fermi liquid behavior in the specific heat~\cite{Makoshi1975} and the resistivity~\cite{UM1975} 
in nearly ferromagnetic metals and also in nearly antiferromagnetic metals~\cite{Moriya}. 

The quantum critical phenomena have been studied by the renormalization-group (RG) theory by Hertz in 1976~\cite{Hertz} 
and reexamined by Millis in 1993~\cite{Millis}, which has explained low-temperature properties of physical quantities in the vicinity of the QCP. 
The RG theory has been shown to yield the same critical exponents~\cite{ZM1995} as those found in the SCR theory~\cite{HM1995,MT,IM1996,IM1998,Ishigaki}. 

The magneto-volume effect in nearly ferromagnetic metals has been studied by Moriya and Usami in 1980~\cite{MU1980}. They discussed the effect of spin fluctuation on the thermal expansion and the effect was also studied in nearly antiferromagnetic metals~\cite{IM1998}. 
In 1997, 
Kambe {\it et al}. analyzed the thermal-expansion coefficient and the Gr\"{u}neisen parameter observed in Ce$_{1-x}$La$_x$Ru$_2$Si$_2$ by using the SCR theory and the RG theory and 
pointed out a possibility that the Gr\"{u}neisen parameter diverges at the QCP\cite{Kambe1997}. 
In 2003, by using the scaling hypothesis and the RG theory, Zhu {\it et al}. evaluated critical part of the thermal expansion coefficient. By taking the ratio to the critical part of the specific heat, they evaluated the critical part of the Gr\"{u}neisen parameter, which actually diverges at the QCP~\cite{Zhu2003,Garst2005}. 
Experimentally, in CeNi$_2$Ge$_2$, which is located closely to the 3d AFM QCP, the divergence of the Gr\"{u}neisen parameter has been observed~\cite{Kuchler2003}. 
Divergence of the Gr\"{u}neisen parameter has also been observed in CeIn$_{3-x}$Sn$_x$ $(x=0.65)$~\cite {Kuchler2006} and in CeRhIn$_{5-x}$Sn$_{x}$ $(x=0.48)$~\cite{Donath2009}, where the 3d AFM order is suppressed by the chemical doping. 

It is well known that if the system is dominated by a single energy scale 
$T^{*}$, the entropy is expressed as a scaled form $S=k_{\rm B}f(T/T^{*})$,  
where $k_{\rm B}$ is the Boltzmann constant and $T$ is temperature,  
so that  
the Gr\"{u}neisen parameter is given by 
$\Gamma=-\frac{V}{T^{*}}\left(\frac{\partial{T^*}}{\partial{V}}\right)_{S}$ with $V$ being the volume~\cite{Gruneisen1912,Takke1981,Thalmeier1986}. 
Normal metal with the Fermi temperature is known to be the case and lattice system with the Debye temperature where acoustic phonons give the dominant contribution is also the case. 
The expression of $\Gamma$ suggests that if $T^{*}$ becomes zero with non-vanishing $(\partial T^{*}/\partial V)_{S}$, the Gr\"{u}neisen parameter diverges. 

On the other hand, in the SCR theory, there exists the characteristic energy scale of spin fluctuation $T_0$, which is known to be finite even at the QCP in general~\cite{Moriya}. 
It is also known that the magnetic correlation length diverges at the QCP, the inverse of which gives the zero characteristic scale. Hence, it is interesting to clarify how these quantities affect $\Gamma$ at the FM QCP and also AFM QCP. 
This requires theoretical study to clarify how the Gr\"{u}neisen parameter behaves at the QCP in the SCR theory.

An advantageous point of the SCR theory is that it describes not only the quantum critical behavior in the vicinity of the QCP, but also the Curie-Weiss behavior at higher temperatures in the magnetic susceptibility in a unified way~\cite{Moriya}. The crossover from the quantum-critical regime to the high-temperature (Curie-Weiss) regime for the other physical quantities such as the specific heat and the resistivity can also be calculated~\cite{Moriya,MU2003,Moriya_text}. 

So far, critical parts of the thermal-expansion coefficient $\alpha$ and the Gr\"{u}neisen parameter $\Gamma$ were reported by the RG theory~\cite{Zhu2003,Garst2005}. 
It seems important to clarify their complete expressions with 
not only the critical part but also non-critical part including their 
coefficients of the temperature-dependent terms in the SCR theory. 
In many cases the critical part is observed in the very vicinity of the QCP, and 
in case experimentally accessible temperature does not reach the low-temperature regime, the crossover behavior is usually observed. Hence, it is useful to obtain the complete expressions of $\alpha$ and $\Gamma$ for comparison with experiments. 

In the original SCR theory,  
the specific heat was calculated with the zero-point spin fluctuation being neglected~\cite{HM1995,MT,IM1996}. 
Taking into account the zero-point spin fluctuation~\cite{Ishigaki,Takahashi1999} as well as the stationary condition of the free energy adequately~\cite{Takahashi1999}, the specific heat was calculated, which has shown that the dominant contribution to the quantum criticality comes from the thermal spin-fluctuation and the critical indices~\cite{HM1995,MT,IM1996} endorsed by the RG theory~\cite{Millis,ZM1995} do not change.  

However, in the calculation of the thermal-expansion coefficient and the Gr\"{u}neisen parameter in the SCR theory, the zero-point spin fluctuation as well as the stationary condition of the free energy should be taken into account correctly, which has not been addressed in Refs.~\cite{MU1980,Kambe1997}. 
Takahashi considered these effects in the 
extended SCR theory by introducing the conservation law of the total spin-fluctuation amplitude and  discussed the magneto-volume effect~\cite{Takahashi2006}.

In this paper, we derive the thermal expansion coefficient $\alpha$ and the Gr\"{u}neisen parameter $\Gamma$ in the {\it complete } framework of the original SCR theory. 
By taking into account zero-point spin fluctuation as well as the stationary condition of the free energy correctly, we reexamine the specific heat $C_{V}$ near the ferromagnetic (FM) QCP and the antiferromagnetic (AFM) QCP in three spatial dimension $(d=3)$ and two spatial dimension $(d=2)$. 
Then, we derive the thermal expansion coefficient $\alpha$ for each class starting from the entropy, which is proven to be equivalent to that obtained from the explicit form of the free energy with the use of the stationary condition in the SCR theory. 
On the basis of these correctly calculated $C_{V}$ and $\alpha$, we obtain $\Gamma$. 
By performing analytical and numerical calculations of $C_{V}$, $\alpha$, and $\Gamma$ near the magnetic QCP, their quantum-critical properties are clarified.

We find that the temperature dependent coefficient exists in the expression of $\alpha(T)$, which 
has not been reported in the past RG studies~\cite{Zhu2003,Garst2005}. 
Furthermore, the complete expressions of $\alpha(T)$ and $\Gamma(T)$ clarify the crossover from the quantum-critical regime at low temperatures to the Curie-Weiss regime at higher temperatures for each class of the QCP. 
Then, we give the answers to the following questions: 1) What is the relation to the divergence of $\Gamma$ at the QCP shown by the RG theory? 2) How to reconcile with finite $T_0$ at the QCP in the SCR theory? 
3) What is the relation to the Moriya-Usami theory?

The organization of this paper is as follows: 
In Sect.~2, 
the definitions of the thermal-expansion coefficient and 
the Gr\"{u}neisen parameter are explained by introducing thermodynamically equivalent expressions. 
In Sect.~3, the SCR theory is outlined and the properties of the specific heat near the QCP are  summarized. 
In Sect.~4, the Gr\"{u}neisen parameter is derived from the entropy in the SCR theory. 
In Sect.~5, the thermal-expansion coefficient near the QCP is derived from the entropy and the free energy, respectively, in the SCR theory and equivalence of both the results is proven. 
In Sect.~6, Sect.~7, and 
Sect.~8,
results of numerical calculations of the thermal expansion coefficient and the Gr\"{u}neisen parameter near the QCP for each class are analyzed, respectively. 
Sect.~9 
is devoted to discussions by comparing the present theory with other theories and experiments. 
In 
Sect.~10, 
the paper is summarized.
From Sect.~2 to 
Sect.~8, 
we concentrate on the electronic Gr\"{u}neisen parameter relevant for low temperatures where lattice degrees of freedom give minor contributions. 
In 
Sect.~9, 
the general case including phonons is discussed.  

\section{
Thermal-expansion coefficient and 
Gr\"{u}neisen parameter}
\label{sec:aG}

In this section, the definitions of the thermal-expansion coefficient $\alpha$ and and the Gr\"{u}neisen parameter $\Gamma$ are summarized. The equivalent expressions of $\alpha$ and $\Gamma$ are also derived for the use in discussions in the forthcoming sections.

\subsection{Thermal-expansion coefficient}

The thermal-expansion coefficient is defined as 
\begin{eqnarray}
\alpha=\frac{1}{V}\left(\frac{\partial{V}}{\partial{T}}\right)_{P}, 
\label{eq:a1}
\end{eqnarray}
where $P$ is the pressure. 
By using the relation 
\begin{eqnarray}
\left(\frac{\partial{V}}{\partial{T}}\right)_{P}=-\frac{\left(\frac{\partial{P}}{\partial{T}}\right)_{V}}{\left(\frac{\partial{P}}{\partial{V}}\right)_{T}}, 
\end{eqnarray}
Eq.~(\ref{eq:a1}) is expressed as
\begin{eqnarray}
\alpha
=\kappa_{T}\left(\frac{\partial{P}}{\partial{T}}\right)_{V},
\label{eq:a_PT_def}
\end{eqnarray}
where the isothermal compressibility $\kappa_{T}$ is defined as 
\begin{eqnarray}
\kappa_{T}=-\frac{1}{V}\left(\frac{\partial{V}}{\partial{P}}\right)_{T}.
\label{eq:comp}
\end{eqnarray}

On the other hand, by using the Maxwell relation $(\partial V/\partial T)_{P}=-(\partial S/\partial P)_{T}$ in Eq.~(\ref{eq:a1}), $\alpha$ can be expressed as 
\begin{eqnarray}
\alpha=-\frac{1}{V}\left(\frac{\partial{S}}{\partial{P}}\right)_{T}.  
\label{eq:a_SP_def}
\end{eqnarray}
%

\subsection{Gr\"{u}neisen parameter}
\label{sec:Grn}

The Gr\"{u}neisen parameter $\Gamma$ is defined by
\begin{eqnarray}
\Gamma=\frac{\alpha{V}}{C_{V}\kappa_{T}}, 
\label{eq:Grn}
\end{eqnarray}
where 
$C_{V}$ is the specific heat at a constant volume
\begin{eqnarray}
C_{V}=T\left(\frac{\partial{S}}{\partial{T}}\right)_{V}. 
\label{eq:Cv}
\end{eqnarray}
With the use of Eq.~(\ref{eq:a_SP_def}) for $\alpha$ in Eq.~(\ref{eq:Grn}), 
the Gr\"{u}neisen parameter $\Gamma$ 
is expressed as follows:
\begin{eqnarray}
\Gamma&=&
-\frac{\partial(S,T)}{\partial(P,T)}
\frac{1}{T\frac{\partial(S,V)}{\partial(T,V)}}
\frac{(-V)}{\frac{\partial(V,T)}{\partial(P,T)}}
\nonumber
\\
&=&-\frac{V}{T}\frac{\partial(S,T)}{\partial(S,V)}
\nonumber
\\
&=&-\frac{V}{T}\left(\frac{\partial{T}}{\partial{V}}\right)_{S}. 
\label{eq:Grn_TV}
\end{eqnarray}
%

If the entropy is expressed as $S=k_{\rm B}S(T/T^{*})$ with a single characteristic temperature  scale $T^{*}$ as in the Fermi-liquid region of metals,  
$(\partial{T}/\partial{V})_{S}$ is given in a form as
\begin{eqnarray}  
\left(\frac{\partial{T}}{\partial{V}}\right)_{S}=
\frac{T}{T^{*}}\left(\frac{\partial T^{*}}{\partial V}\right)_{S}. 
\label{eq:dTdV}
\end{eqnarray}
Then, the Gr\"{u}neisen parameter $\Gamma$ is expressed as a conventional form as~\cite{Takke1981,Flouquet2005}
\begin{eqnarray}
\Gamma=-\left(\frac{\partial{\ln}T^*}{\partial{{\ln}V}}\right)_{S}. 
\label{eq:Grn_TV2}
\end{eqnarray}
%


\section{SCR theory}
\label{sec:SCR}

In this section, 
the self-consistent renormalization (SCR) theory of spin fluctuation is outlined.  
By taking into account the zero-point as well as thermal spin fluctuation under consideration of  
the stationary condition of the SCR theory, the specific heat near the magnetic QCP is reexamined. 
Hereafter the energy units are taken as 
$\hbar=1$ and $k_{\rm B}=1$ unless otherwise noted. 

\subsection{Formulation of the SCR theory}

The action of the itinerant electrons with Coulomb interaction is expressed in the form of the Ginzburg-Landau-Wilson functional 
\begin{eqnarray}
\Phi[\varphi]&=&\frac{1}{2}\sum_{\bar{q}}\Omega_{2}(\bar{q})\varphi(\bar{q})\varphi(-\bar{q}) 
\nonumber
\\
&+&\sum_{\bar{q}_1,\bar{q}_2,\bar{q}_3,\bar{q}_4}
\Omega_{4}(\bar{q}_1, \bar{q}_2, \bar{q}_3, \bar{q}_4)
\nonumber
\\
& &
\times
\varphi(\bar{q}_1)\varphi(\bar{q}_2)\varphi(\bar{q}_3)\varphi(\bar{q}_4)
\delta\left(\sum_{i=1}^{4}\bar{q}_i\right), 
\nonumber
\\
& &
\label{eq:Action}
\end{eqnarray}
which can be derived from the Hamiltonian via the Stratonovich-Hubbard transformation applied to the onsite Coulomb interaction term~\cite{Hertz}. 
Hence, Eq. (11) describes the action for isotropic spin space~\cite{note_MW}. 
Here, $\bar{q}$ is abbreviation for $\bar{q}\equiv({\bf q},{\rm i}\omega_{l})$ where $\omega_l=2\pi lT$ with $l$ being integer. 
Since long wavelength $|{\bf q}|\ll q_{\rm c}$ around the magnetically ordered vector ${\bf Q}$ and low frequency $|\omega|\ll\omega_{\rm c}$ regions play the dominant role in the critical phenomena with $q_{\rm c}$ and $\omega_{\rm c}$ being the cutoffs for momentum and frequency, respectively,   
$\Omega_i$ for $i=2, 4$ are expanded for $q$ and $\omega$ around $({\bf Q}, 0)$: 
\begin{eqnarray}
\Omega_2({\bf q},{\rm i}\omega_l)\approx\frac{\eta_0+Aq^2+C_{q}|\omega_l|}{N_{\rm F}}, 
\end{eqnarray}
where $C_{q}$ is defined as 
$C_{q}\equiv C/q^{z-2}$ with $z$ being the dynamical exponent (e.g.,  $z=3$ for ferromagnetism and  $z=2$ for antiferromagnetism) 
and $N_{\rm F}$ is the density of states at the Fermi level, 
and 
$\Omega_{4}(\bar{q}_1, \bar{q}_2, \bar{q}_3, \bar{q}_4)\approx v_{4}/(\beta N)$ with $\beta\equiv1/T$.

To construct the effective action for the best Gaussian, taking account of the mode-mode coupling effects up to the 4th order in $\Phi[\varphi]$, we employ the Feynman's inequality~\cite{Feynman} on the free energy: 
\begin{eqnarray}
F\le F_{\rm eff}+T\langle\Phi-\Phi_{\rm eff}\rangle_{\rm eff}\equiv\tilde{F}(\eta). 
\label{eq:Free_SCR} 
\end{eqnarray}
Here, the effective action $\Phi_{\rm eff}$ is parametrized as 
\begin{eqnarray}
\Phi_{\rm eff}[\varphi]=\frac{1}{2}\sum_{l}\sum_{q}\frac{\eta+Aq^2+C_q|\omega_l|}{N_{\rm F}}\left|\varphi(q,{\rm i}\omega_{l})\right|^{2},     
\label{eq:Action_eff}
\end{eqnarray}
where $\eta$ expresses the effect of the mode-mode coupling of spin fluctuations and parameterizes the closeness to the quantum criticality. 
In Eq.~(\ref{eq:Free_SCR}),   
$\langle\cdots\rangle_{\rm eff}$ denotes the statistical average taken by the weight $\exp\left(-\Phi_{\rm eff}[\varphi]\right)$ and $F_{\rm eff}$ is given by  
\begin{eqnarray}
F_{\rm eff}=-T\ln\int{\cal D}\varphi\exp\left(-\Phi_{\rm eff}[\varphi]\right). 
\label{eq:F_eff}
\end{eqnarray}

By optimal condition $\frac{d\tilde{F}(\eta)}{d\eta}=0$, the self-consistent renormalization (SCR) equation for $\eta$ is given by
\begin{eqnarray}
\frac{\eta_{0}-\eta}{2N_{\rm F}}+\frac{6v_{4}}{N}\langle\varphi^2\rangle_{\rm eff}=0,   
\label{eq:SCReq}
\end{eqnarray}
where spin fluctuation $\langle\varphi^2\rangle_{\rm eff}$ is defined as
\begin{eqnarray}
\langle\varphi^2\rangle_{\rm eff}=T\sum_{q}\sum_{l}\frac{N_{\rm F}}{\eta+Aq^2+C_{q}|\omega_{l}|}.
\label{eq:phi2_def}  
\end{eqnarray}
Here, $\langle\varphi^2\rangle_{\rm eff}$ 
consists of the quantum (zero-point) fluctuation $\langle\varphi^2\rangle_{\rm zero}$ and thermal fluctuation $\langle\varphi^2\rangle_{\rm th}$ as 
\begin{eqnarray}
\langle\varphi^2\rangle_{\rm eff}=
\langle\varphi^2\rangle_{\rm zero}+\langle\varphi^2\rangle_{\rm th},  
\label{eq:psi2}
\end{eqnarray}
where $\langle\varphi^2\rangle_{\rm zero}$ and $\langle\varphi^2\rangle_{\rm th}$ are expressed as 
\begin{eqnarray}
\langle\varphi^2\rangle_{\rm zero}&=&
\frac{N_{\rm F}}{\pi}\sum_{q}\frac{1}{C_q}\int_{0}^{\omega_{\rm c}}d\omega
\frac{\omega}{\Gamma_{q}^2+\omega^2}, 
\\
\langle\varphi^2\rangle_{\rm th}&=&
\frac{N_{\rm F}}{\pi}\sum_{q}\frac{2}{C_q}\int_{0}^{\omega_{\rm c}}d\omega
\frac{1}{{\rm e}^{\beta\omega}-1}
\frac{\omega}{\Gamma_{q}^2+\omega^2}, 
\end{eqnarray}
respectively. 
Here, $\Gamma_{q}$ is defined by $\Gamma_{q}\equiv(\eta+Aq^2)/C_{q}$.

From Eq.~(\ref{eq:Action_eff}), 
the dynamical magnetic susceptibility is given by  
\begin{eqnarray}
\chi_{\bf Q}(q,\omega)=\frac{N_{\rm F}}{\eta+Aq^2-{\rm i}C_{q}\omega},
\label{eq:chi}
\end{eqnarray}
where $\bf Q$ is the wavenumber vector of the magnetically-ordered phase (e.q., ${\bf Q}={\bf 0}$ for ferromagnetism and ${\bf Q}\ne{\bf 0}$ for antiferromagnetism).

The free energy $\tilde{F}$ defined by Eq.~(\ref{eq:Free_SCR}) is expressed as
\begin{eqnarray}
\tilde{F}&=&\frac{1}{\pi}\sum_{q}\int_{0}^{\omega_{\rm c}}d\omega
\frac{\Gamma_{q}}{\omega^2+\Gamma_{q}^2}\left\{\frac{\omega}{2}+T{\ln}\left(1-{\rm e}^{-\frac{\omega}{T}}\right)\right\}
\nonumber
\\
&+&\frac{\eta_0-\eta}{2N_{\rm F}}\langle\varphi^2\rangle_{\rm eff}
+\frac{3v_{4}}{N}\langle\varphi^2\rangle_{\rm eff}^2
-\frac{1}{\pi}\sum_{q}\frac{\pi\omega_{\rm c}}{4}.    
\label{eq:freeE}
\end{eqnarray}

Here, let us define dimensionless parameters for $\eta$ 
\begin{eqnarray}
y\equiv\frac{\eta}{Aq_{\rm B}^2}
\label{eq:y}
\end{eqnarray}
and the wave number $x\equiv q/q_{\rm B}$ with $q_{\rm B}$ being the wave number of the Brillouin zone. Thus, 
$\Gamma_{q}$ is expressed as 
\begin{eqnarray}
\Gamma_{q}=2{\pi}T_{0}x^{z-2}(y+x^2), 
\label{eq:Gamma_def} 
\end{eqnarray}
where 
the characteristic temperature of spin fluctuation is defined as
\begin{eqnarray}
T_{0}\equiv\frac{Aq_{\rm B}^2}{2\pi{C_{q_{\rm B}}}}.  
\label{eq:T0}
\end{eqnarray}

Near the QCP, quantum spin fluctuation   
$\langle\varphi^2\rangle_{\rm zero}$ is calculated~\cite{IM1998} for the cases 
above and just at the upper critical dimension $4$, respectively, 
as
\begin{widetext}
\begin{eqnarray}
\langle\varphi^2\rangle_{\rm zero}=Nd\frac{T_0}{2T_A}
\left\{
\begin{array}{lr}
C_1-C_{2}y+\cdots, \quad \mbox{for $d+z>4$}  \\ 
C_1+y{\ln}y-C_{2}y+\cdots, \quad \mbox{for $d+z=4$}
\end{array}
\right.
\label{eq:psi2_zero}
\end{eqnarray}
\end{widetext}
where 
$T_{A}$ is defined as
\begin{eqnarray}
T_{A}\equiv\frac{Aq_{\rm B}^2}{2N_{\rm F}}.     
\label{eq:TA}
\end{eqnarray}
The constants $C_1$ and $C_2$ are given by
\begin{eqnarray}
C_1&=&
\int_{0}^{x_{\rm c}}dxx^{d+z-3}\ln\left|\frac{\omega_{{\rm c}T_0}^2+x^{2z}}{x^{2z}}\right|,
\label{eq:C1}
\\
C_2&=&
\left\{
\begin{array}{lr}
2\int_{0}^{x_{\rm c}}dxx^{d+z-5}\frac{\omega_{{\rm c}T_0}^2}{\omega_{{\rm c}T_0}^2+x^{2z}}, 
\quad \mbox{for $d+z>4$}  \\
1+\ln{x}_{\rm c}^2-\frac{1}{2}\ln\left|\frac{\omega_{{\rm c}T_0}^2+x_{\rm c}^4}{\omega_{{\rm c}T_0}^2}\right|, 
\quad \mbox{for $d+z=4$}
\end{array}
\right.
\label{eq:C2}
\end{eqnarray}
respectively. Here,    
the cut off of the wave number is set to be $q_{\rm c}$ in the $q$ integration, which 
is expressed as $x_{\rm c}\equiv\frac{q_{\rm c}}{q_{\rm B}}$ in the dimensionless scaled form and 
$\omega_{{\rm c}T}$ is defined as $\omega_{{\rm c}T}\equiv\frac{\omega_{\rm c}}{2\pi{T}}$.
The thermal spin fluctuation   
$\langle\varphi^2\rangle_{\rm th}$ is calculated as
\begin{eqnarray}
\langle\varphi^2\rangle_{\rm th}=Nd\frac{T_0}{T_A}
\int_{0}^{x_{\rm c}}dxx^{d+z-3}
\left\{
{\ln}u-\frac{1}{2u}-\psi(u)
\right\},  
\nonumber
\\
\label{eq:psi2_th}
\end{eqnarray}
where $\psi(u)$ is the digamma function 
with 
$u$ defined as 
\begin{eqnarray}
u\equiv\frac{\Gamma_{q}}{2\pi{T}}=\frac{x^{z-2}(y+x^2)}{t}. 
\label{eq:u}
\end{eqnarray}
Here, $t$ is defined as the dimensionless scaled temperature 
\begin{eqnarray}
t\equiv\frac{T}{T_{0}}. 
\label{eq:t}
\end{eqnarray}
By substituting Eq.~(\ref{eq:psi2_zero}) and Eq.~(\ref{eq:psi2_th}) into 
Eq.~(\ref{eq:psi2}),  
the SCR equation [Eq.~(\ref{eq:SCReq})] is written in the scaled form 
for $d+z>4$ as~\cite{MT,HM1995,IM1996}  
\begin{eqnarray}
y=y_0+\frac{d}{2}y_1\int_{0}^{x_{\rm c}}dxx^{d+z-3}
\left\{
{\ln}u-\frac{1}{2u}-\psi(u)
\right\}, 
\nonumber
\\
\label{eq:SCReq2}
\end{eqnarray}
and for $d+z=4$ as~\cite{IM1998} 
\begin{eqnarray}
y=y_0+\frac{y_1}{2}\left(
y{\ln}y
+d
\int_{0}^{x_{\rm c}}dxx
\left\{
{\ln}u-\frac{1}{2u}-\psi(u)
\right\}
\right), 
\nonumber
\\
\label{eq:SCReq3}
\end{eqnarray}
where $y_0$ and $y_1$ are given by 
\begin{eqnarray}
y_0&=&\frac{\frac{\eta_0}{Aq_{\rm B}^2}+3dv_4\frac{T_0}{T_A^2}{C_1}}
{1+3dv_4\frac{T_0}{T_A^2}{C_2}},
\label{eq:y0}
\\
y_1&=&\frac{12v_4\frac{T_0}{T_A^2}}{1+3dv_4\frac{T_0}{T_A^2}{C_2}},
\label{eq:y1}
\end{eqnarray}
respectively. 
Here, note that $y_{0}$ is different from that obtained by substituting $\eta_0$ for $\eta$ in the r.h.s. of Eq.~(\ref{eq:y}). 

\begin{figure*}
\includegraphics[width=15cm]{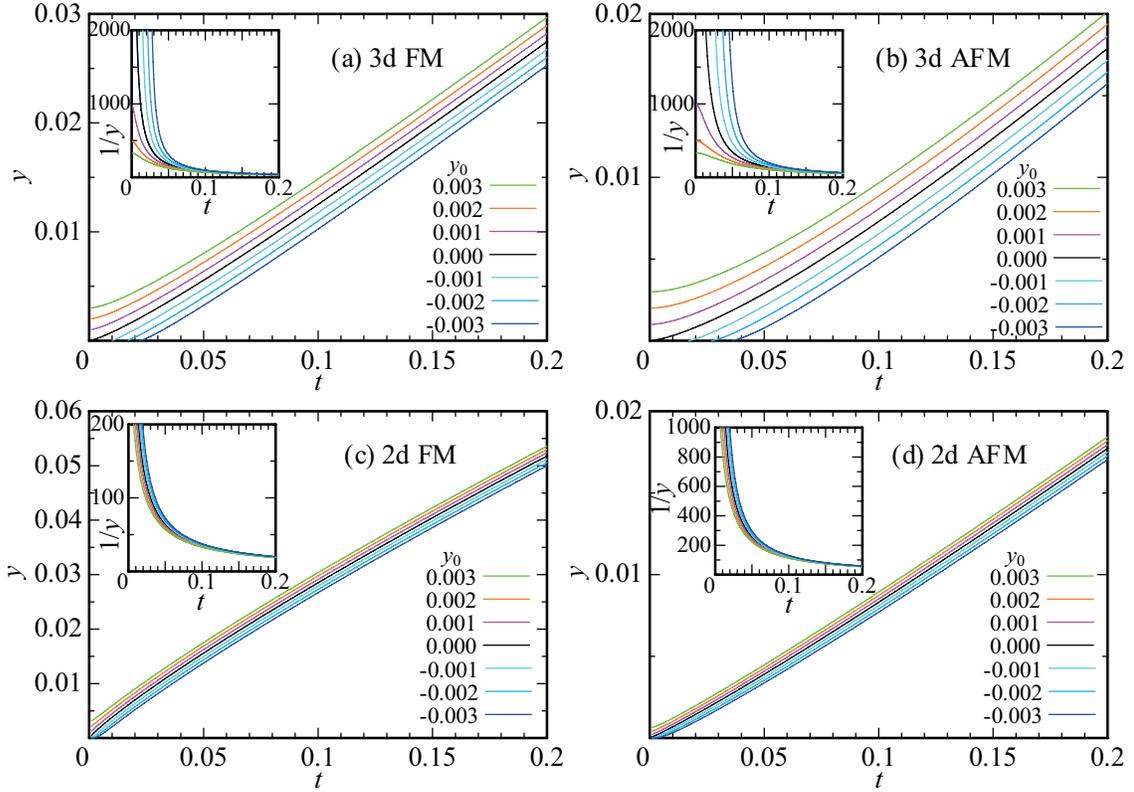}
\caption{
(Color online) 
Scaled temperature dependence of $y$ for (a) 3d FM, (b) 3d AFM, (c), 2d FM, and (d) 2d AFM. The inset shows the scaled temperature dependence of $1/y\propto\chi_{\bf Q}(0,0)$. 
}
\label{fig:y_t}
\end{figure*}

The solution of the SCR equation $y$ is proportional to the inverse susceptibility
\begin{eqnarray}
y=\frac{1}{2T_{\rm A}}\frac{1}{\chi_{\bf Q}(0,0)},
\end{eqnarray}
which is obtained by substituting Eq.~(\ref{eq:y}) into Eq.~(\ref{eq:chi}) with the use of Eq.~(\ref{eq:TA}). 
Numerical solutions of Eq.~(\ref{eq:SCReq2}) and Eq.~(\ref{eq:SCReq3}) are shown in Figs.~\ref{fig:y_t}(a)-(c) and Fig.~\ref{fig:y_t}(d), respectively~\cite{HM1995,MT,IM1996,IM1998}. The $t$ dependences of $y$ for the paramagnetic region $(y_0>0)$ and the region where the magnetic order occurs   $(y_0<0)$ and just at the QCP $(y_0=0)$ are shown. For $y_0<0$, $y=0$ is realized for $t>0$ in the case of $d=3$ (see Fig.~\ref{fig:y_t}(a) and Fig.~\ref{fig:y_t}(b)), where the magnetic phase transition takes place at finite temperature, while $y=0$ is realized only at $t=0$ for $d=2$ (see Fig.~\ref{fig:y_t}(c) and Fig.~\ref{fig:y_t}(d)), satisfying the Mermin-Wagner theorem~\cite{MerminWagner}. In each class, the Curie-Weiss behavior $\chi_{\bf Q}(0,0)\propto y^{-1}\sim t^{-1}$ appears in the high-$t$ regime (e.g., see $t\gsim 0.07$ in Fig.~\ref{fig:y_t}(a)). The quantum critical region appears in the low-$t$ regime at the QCP realized for $y_0=0$, whose property in each class is analyzed as follows.

In $d=3$, the $x$ integral in Eq.~(\ref{eq:SCReq2}) converges for $y\to 0$ and then 
the solution is obtained as 
\begin{eqnarray}
y\propto t^{1+\frac{1}{z}}
\label{eq:y_t}
\end{eqnarray}
at the QCP with $y_0=0.0$, 
 as shown in \ref{sec:Ld3}. 
This yields 
$y\sim t^{4/3}$ for the 3d FM QCP $(z=3)$ and 
$y\sim t^{3/2}$ for the 3d AF QCP $(z=2)$.

In $d=2$, the $x$ integral in Eq.~(\ref{eq:SCReq2}) shows logarithmic divergence for $y\to 0$. 
At the FM QCP for $z=3$, the solution of Eq.~(\ref{eq:SCReq2}) is obtained as 
$y=-\frac{y_1}{12}t\ln{t}$ (see Appendix~\ref{sec:d2z3}). 
At the AF QCP for $z=2$, the solution of Eq.~(\ref{eq:SCReq3}) is obtained as 
$y=-t\frac{\ln(-\ln{t})}{2\ln{t}}$ (see Appendix~\ref{sec:d2z2}).  
The criticality $y$ $(\propto \eta)$ for each class is summarized in Table~\ref{tb:mag_QCP}.

\begin{table*}[b]
\begin{center}
\begin{tabular}{l|cccccc} \hline
{class} &{$\eta$} & {$\rho$} & {$C/T$} & {$\chi$} & {$(T_{1}T)^{-1}$}  & Refs.
\\ \hline
3d FM &{$T^{4/3}$} & {$T^{5/3}$} & {$-\ln{T}$} & {$T^{-4/3} \to {\rm C. W.}$}  & {$T^{-4/3}$}  
& \cite{UM1975,IM1996}
\\ 
3d AFM  &{$T^{3/2}$} & {$T^{3/2}$} & {const.$-T^{1/2}$}   & {const.${-T^{1/4}}$ $\to$ C. W.} & {$T^{-3/4}$}  
& \cite{MT,HMN}
\\
2d FM &{$-T\ln{T}$} & {$T^{4/3}$} & {$T^{-1/3}$} & {$-1/(T\ln{T}) \to {\rm C. W.}$} & {$-1/(T\ln{T})^{3/2}$}    
& \cite{HM1995}
\\  
2d AFM &{$\frac{-T\ln(-\ln{T})}{\ln{T}}$} & {$T$} & {$-\ln{T}$} &  - & {$-\ln{T}/T$}   
& \cite{IM1998,note_d2_z2}
\\ \hline
\end{tabular}
\end{center}
\caption{Quantum criticality at the magnetic QCP for each class specified by $z=3$ (FM) and $z=2$ (AFM) in $d=3$ and $2$~\cite{Moriya}. 
Electrical resistivity $\rho(T)$, specific-heat coefficient $C/T$, 
uniform susceptibility $\chi(T)$, and 
NMR relaxation rate $(T_1T)^{-1}$.
For $\chi$, $\to$ C.W. denotes the crossover to the Curie-Weiss behavior.
Note that $\eta\propto y$ holds [see Eq.~(\ref{eq:y})]. 
}
\label{tb:mag_QCP}
\end{table*}

\subsection{Entropy and specific heat}

The entropy $S=-\left(\frac{\partial{\tilde{F}}}{\partial{T}}\right)_{V}$ is obtained by differentiating the free energy $\tilde{F}$ in Eq.~(\ref{eq:freeE}) with respect to 
the
 temperature. 
Noting 
that the terms with $\left(\frac{\partial\eta}{\partial{T}}\right)_{V}$ 
and also $\left(\frac{\partial\langle\varphi^2\rangle_{\rm eff}}{\partial{T}}\right)_{V}$ vanish with the use of the SCR equation [Eq.~(\ref{eq:SCReq})], the entropy is derived as~\cite{Takahashi1999}
\begin{eqnarray}
S&=&-Nd\int_{0}^{x_{\rm c}}dxx^{d-1}
\left\{
{\ln}\sqrt{2\pi}-u+\left(u-\frac{1}{2}\right){\ln}u-{\ln}\Gamma(u)
\right\}
\nonumber
\\
&+&Nd\int_{0}^{x_{\rm c}}dxx^{d-1}u
\left\{
{\ln}u-\frac{1}{2u}-\psi(u)
\right\},
\label{eq:S}
\end{eqnarray}
where $\Gamma(u)$ is the Gamma function.

The specific heat under a constant volume is obtained by differentiating the entropy $S$ in Eq.~(\ref{eq:S}) with respect to 
the
 temperature~\cite{Takahashi1999,Ishigaki} as 
\begin{eqnarray}
C_V&=&T\left(\frac{\partial{S}}{\partial{T}}\right)_{V}, 
\nonumber
\\
&=&
C_{\rm a}
-C_{\rm b},
\label{eq:CvT}
\end{eqnarray}
where 
$C_{\rm a}$ and $C_{\rm b}$ are given by 
\begin{eqnarray}
C_{\rm a}&=&-Nd\int_{0}^{x_{\rm c}}dxx^{d-1}u^2\left\{
\frac{1}{u}+\frac{1}{2u^2}-\psi'(u)
\right\}, 
\label{eq:Ca}
\\
C_{\rm b}&=&
\tilde{C}_{\rm b}
\left(\frac{\partial{y}}{\partial{t}}\right)_{V},
\label{eq:Cb}
\end{eqnarray}
respectively. Here, $\psi'(u)$ is the trigamma function and $\tilde{C}_{\rm b}$ is given by 
\begin{eqnarray}
\tilde{C}_{\rm b}&=&-Nd\int_{0}^{x_{\rm c}}dxx^{d+z-3}u\left\{
\frac{1}{u}+\frac{1}{2u^2}-\psi'(u)
\right\}. 
\label{eq:tildaCb}
\end{eqnarray}

As for $C_{\rm a}$, the $x$ integral in Eq.~(\ref{eq:Ca}) shows no divergence from $x=0$ even for $y\to 0$ 
irrespective of spatial dimensions~\cite{Takahashi1999}. Hence, $C_{\rm a}$ for $t\ll 1$ at the QCP has no explicit $y$ dependence.

As for $\tilde{C}_{\rm b}$, in $d=3$, the $x$ integral in Eq.~(\ref{eq:tildaCb}) shows no divergence from $x=0$ for $y\to 0$ and is evaluated as~\cite{Takahashi1999} 
\begin{eqnarray}
\frac{\tilde{C}_{\rm b}}{N}\sim t^{1+\frac{1}{z}}, 
\end{eqnarray}
for $t\to 0$ at the QCP. Namely, $\tilde{C}_{\rm b}$ shows the same temperature dependence as $y$, i.e., $\tilde{C}_{\rm b}\sim y$, as seen in Eq.~(\ref{eq:y_t}).  
In $d=2$, the $x$ integral in Eq.~(\ref{eq:tildaCb}) shows logarithmic divergence arising from $x=0$ for $y\to 0$. Hence, $\tilde{C}_{\rm b}$ has the ${\ln}y$ dependence at the QCP. 

\begin{figure*}
\includegraphics[width=15cm]{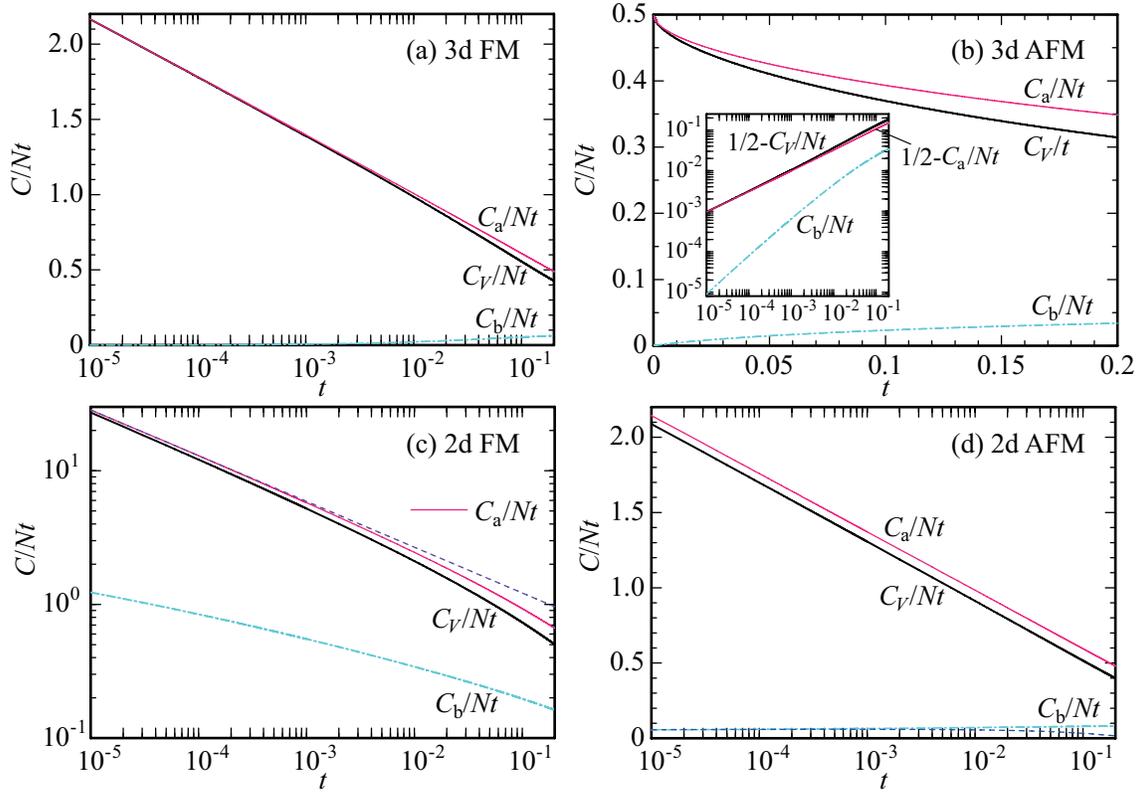}
\caption{(Color online) Specific-heat coefficient vs. scaled temperature just at the QCP. $C_{V}/Nt$ (bold solid line), $C_{\rm a}/Nt$ (thin line), and $C_{\rm b}/Nt$ (dash dotted line) are calculated numerically in Eqs.~(\ref{eq:CvT}), (\ref{eq:Ca}), and (\ref{eq:Cb}), respectively for $y_0=0.0$ and $y_1=1.0$. 
(a) 3d 
FM
 QCP.
(b) 3d 
AFM
 QCP. The inset shows log-log plot of $1/2-C_{V}/Nt$ (thick solid line), $1/2-C_{\rm a}/Nt$ (thin solid line), and $C_{\rm b}/Nt$ (dash-dotted line).  
(c) 2d 
FM
 QCP. The dashed line represents the least-square fit of $C_{\rm a}/Nt$ for $10^{-5}\le t\le 10^{-4}$ with $at^{-1/3}$.
(d) 2d 
AFM
 QCP. The dashed line represents the least-square fit of $C_{\rm b}/t$ for $10^{-5}\le t\le 10^{-4}$ with $-a\frac{\{{\ln}(-{\ln}t)\}^2}{-{\ln}t}$.
}
\label{fig:Ct_t2dFM}
\end{figure*}

The temperature dependence of the specific heat at the magnetic QCP 
for each class is summarized in the following subsections.
The numerical calculation of Eq.~(\ref{eq:CvT}) is also performed. 
To calculate $\left(\frac{\partial{y}}{\partial{t}}\right)_{V}$, 
by differentiating the SCR equation [Eq.~(\ref{eq:SCReq2}) and Eq.~(\ref{eq:SCReq3})] with respect to the scaled temperature $t$, we have 
\begin{eqnarray}
\left(\frac{\partial{y}}{\partial{t}}\right)_{V}
=
\left\{
\begin{array}{lr}
\frac{\frac{y_1}{2t}\tilde{C}_{\rm b}\frac{1}{N}}{1-\frac{dy_1}{2t}M} 
\quad \mbox{for $d+z>4$},  \\
\frac{\frac{y_1}{2t}\tilde{C}_{\rm b}\frac{1}{N}}{1-\frac{y_1}{2}(\ln{y}+1)-\frac{dy_1}{2t}M}  
\quad \mbox{for $d+z=4$},
\end{array}
\right.
\label{eq:dydt}
\end{eqnarray}
where $M$ is given by
\begin{eqnarray}
M=
\int_{0}^{x_{\rm c}}dxx^{d+2z-5}
\left\{
\frac{1}{u}+\frac{1}{2u^2}-\psi'(u)
\right\}. 
\label{eq:M}
\end{eqnarray}
To calculate the $t$ dependence of the specific heat just at the QCP, we set $y_0=0.0$ in the SCR equation [Eq.~(\ref{eq:SCReq2}) and Eq.~(\ref{eq:SCReq3})] with setting as    
$y_1=1.0$ and $x_{\rm c}=1.0$. 
By solving the SCR equation, we obtain the solution $y(t)$. 
Then, by inputting $y(t)$ to Eq.~(\ref{eq:dydt}), we obtain $\left(\frac{\partial{y}}{\partial{t}}\right)_{V}$. 
Finally, by substituting $y(t)$ and $\left(\frac{\partial{y}}{\partial{t}}\right)_{V}$ 
into
 Eq.~(\ref{eq:CvT}) for each class, we obtain $C_{V}(t)$, 
which is shown in Figs.~\ref{fig:Ct_t2dFM}(a)-\ref{fig:Ct_t2dFM}(d), respectively. 

\subsubsection{3d Ferromagnetic case}

For $t\ll1$, $C_{\rm a}$ is evaluated as~\cite{Makoshi1975,IM1996,Takahashi1999} 
\begin{eqnarray}
\frac{C_{\rm a}}{N}&\approx& -\frac{t}{6}{\ln}t.  
\end{eqnarray}
For $t\ll1$, $\tilde{C}_{\rm b}$ is evaluated as~\cite{Takahashi1999} 
\begin{eqnarray}
\frac{\tilde{C}_{\rm b}}{N}&\sim&t^{\frac{4}{3}}.  
\end{eqnarray}
Since $C_{\rm b}$ behaves as $C_{\rm b}\sim t^{5/3}$ in Eq.~(\ref{eq:Cb}) with $\partial y/\partial t\sim t^{1/3}$ for $t\ll 1$,  
the specific heat in Eq.~(\ref{eq:CvT}) behaves as
\begin{eqnarray}
\frac{C_{V}}{N}\approx\frac{C_{\rm a}}{N}
\sim
-t {\ln}t 
\label{eq:Cv_d3z3} 
\end{eqnarray}
for $t\ll 1$, where the dominant contribution comes from $C_{\rm a}$, 
as seen in Fig.~\ref{fig:Ct_t2dFM}(a). 
This reproduces the criticality shown by the past SCR theory, which is summarized in Table~\ref{tb:mag_QCP} (see $C/T$ for 3d FM).
It is noted that the same temperature dependence as Eq.~(\ref{eq:Cv_d3z3}) was also derived from the RG  theory~\cite{ZM1995}.

\subsubsection{3d Antiferromagnetic case}

For $t\ll1$, $C_{\rm a}$ is evaluated as~\cite{MT,Takahashi1999} 
\begin{eqnarray}
\frac{C_{\rm a}}{N}&\approx&\frac{1}{2}t\left(x_{\rm c}-
\frac{15}{2}a_{\frac{5}{2}}^{*}
t^{\frac{1}{2}}\right),   
\end{eqnarray}
where $a_{\frac{5}{2}}^{*}$ is a constant given by  $a_{\frac{5}{2}}^{*}=\int_{0}^{\infty}duu^{3/2}\left\{\frac{1}{2u^2}-\ln{u}+\frac{1}{2u}+\psi(u)\right\}$.
For $t\ll1$, $\tilde{C}_{\rm b}$ is evaluated as~\cite{Takahashi1999} 
\begin{eqnarray}
\frac{\tilde{C}_{\rm b}}{N}&\sim&t^{\frac{3}{2}}. 
\end{eqnarray}
Since $C_{\rm b}$ behaves as $C_{\rm b}\sim t^2$ in Eq.~(\ref{eq:Cb}) with $\partial y/\partial t\sim t^{1/2}$ for $t\ll 1$, 
the specific heat in Eq.~(\ref{eq:CvT}) behaves as 
\begin{eqnarray}
\frac{C_{V}}{N}\approx\frac{C_{\rm a}}{N}
\sim
\frac{1}{2}t\left(x_{\rm c}-\frac{15}{2}a_{\frac{5}{2}}^{*}t^{\frac{1}{2}}\right) 
\label{eq:Cv_d3z2}
\end{eqnarray}
for $t\ll 1$, where the dominant contribution comes from $C_{\rm a}$, 
as seen in Fig.~\ref{fig:Ct_t2dFM}(b).
This reproduces the criticality shown by the past SCR theory (see $C/T$ for 3d AFM in Table~\ref{tb:mag_QCP}).
It is noted that the same temperature dependence as Eq.~(\ref{eq:Cv_d3z2}) was also derived from the RG  theory~\cite{ZM1995}. 

\subsubsection{2d Ferromagnetic case}

For $t\ll1$, $C_{\rm a}$ is evaluated as~\cite{HM1995,Takahashi1999} 
\begin{eqnarray}
\frac{C_{\rm a}}{N}\approx
\frac{10}{9}a_{\frac{5}{3}}
t^{\frac{2}{3}},
\label{eq:Ca_2dFM} 
\end{eqnarray}
where $a_{\nu}$ is a constant given by $a_{\nu}\equiv{\pi\zeta(\nu)\Gamma(\nu)}/{[(2\pi)^{\nu}\sin(\nu\pi/2)]}$ and $a_{5/3}=0.5629\cdots$.
For $y\ll t^{\frac{2}{3}}\ll 1$, we have 
\begin{eqnarray}
\frac{\tilde{C}_{\rm b}}{N}\approx\frac{t}{2}
{\ln}\left[\frac{1}{y}\left(\frac{t}{6}\right)^{\frac{2}{3}}\right].  
\end{eqnarray}
Since $y\approx -\frac{y_1}{12}t{\ln}t$ at the QCP is obtained from the SCR equation [Eq.~(\ref{eq:SCReq2})]
for $d=2$ and $z=3$ by setting $y_0=0.0$~\cite{HM1995}, we have
\begin{eqnarray}
\frac{\tilde{C}_{\rm b}}{N}\approx -\frac{t}{6}{\ln}t. 
\end{eqnarray}
Since the coefficient $\frac{10}{9}a_{\frac{5}{3}}$ in Eq.~(\ref{eq:Ca_2dFM}) is the quantity of $O(1)$, which is much larger than that in $\tilde{C}_{\rm b}\left(\frac{\partial{y}}{\partial{t}}\right)_{V}\sim\frac{y_1}{72}t{\ln}t({\ln}t+1)$, 
the specific heat in Eq.~(\ref{eq:CvT}) is dominated by $C_{\rm a}$ as 
\begin{eqnarray}
\frac{C_{V}}{N}\approx\frac{C_{\rm a}}{N}
\sim
t^{\frac{2}{3}} 
\label{eq:Cv_d2z3} 
\end{eqnarray}
for $t\ll 1$.  
This can be confirmed by 
numerical calculation of Eq.~(\ref{eq:CvT}), as shown 
in Fig.~\ref{fig:Ct_t2dFM}(c). 
This reproduces the criticality shown by the past SCR theory (see $C/T$ for 2d FM in Table~\ref{tb:mag_QCP}).
It is noted that the same temperature dependence as Eq.~(\ref{eq:Cv_d2z3}) was also derived from the RG  theory~\cite{Zhu2003}.

\subsubsection{2d Antiferromagnetic case}

For $t\ll1$, $C_{\rm a}$ is evaluated as~\cite{Takahashi1999} 
\begin{eqnarray}
\frac{C_{\rm a}}{N}\approx-\frac{t}{6}{\ln}t.  
\end{eqnarray}
For $y\ll t\ll 1$, $\tilde{C}_{\rm b}$ is evaluated as 
\begin{eqnarray}
\frac{\tilde{C}_{\rm b}}{N}\approx
-\frac{t}{2}{\ln}\left(6\frac{y}{t}\right). 
\end{eqnarray}
Since $y\approx -t\frac{{\ln}(-{\ln}t)}{2{\ln}t}$ at the QCP is obtained from the SCR equation [Eq.~(\ref{eq:SCReq3})]
for $d=2$ and $z=2$ by setting $y_0=0.0$~\cite{note_d2_z2}, 
we have
\begin{eqnarray}
\frac{\tilde{C}_{\rm b}}{N}\sim
t{\ln}\left(-{\ln}t\right). 
\label{eq:Cb_tld_d2z2}
\end{eqnarray}
Since $C_{\rm b}$ behaves as $C_{\rm b}\sim-t\frac{\left\{\ln(-\ln{t})\right\}^2}{\ln{t}}$ 
in Eq.~(\ref{eq:Cb}) with $\partial y/\partial t\sim-\ln(-\ln{t})/\ln{t}$ for $0<t\ll 1$ [see the dashed line in Fig.~\ref{fig:Ct_t2dFM}(c)],  
the specific heat in Eq.~(\ref{eq:CvT}) behaves as
\begin{eqnarray}
\frac{C_{V}}{N}\approx\frac{C_{\rm a}}{N}
\sim
-t{\ln}t 
\label{eq:Cv_d2z2}  
\end{eqnarray}
for $t\ll 1$, where the dominant contribution comes from $C_{\rm a}$, 
as seen in Fig.~\ref{fig:Ct_t2dFM}(d). 
This reproduces the criticality shown by the past SCR theory (see $C/T$ for 2d AFM in Table~\ref{tb:mag_QCP}).
It is noted that the same temperature dependence as Eq.~(\ref{eq:Cv_d2z2}) was also derived from the RG  theory~\cite{Zhu2003}.

\section{Gr\"{u}neisen parameter near the magnetic QCP}
\label{sec:Grn_deriv}

The Gr\"{u}neisen parameter $\Gamma$ near the magnetic QCP is derived on the basis of Eq.~(\ref{eq:Grn_TV}) in the SCR theory.  
The calculation starts from the entropy $S$ in Eq.~(\ref{eq:S}). 
By differentiating both sides of Eq.~(\ref{eq:S}) with respect to the volume $V$ under a constant entropy $S$, we have
\begin{eqnarray}
0=Nd\int_{0}^{x_{\rm c}}dxx^{d-1}
\left(\frac{\partial{u}}{\partial{V}}\right)_{S}
u
\left\{
\frac{1}{u}+\frac{1}{2u^2}-\psi'(u)
\right\},
\nonumber
\\
\label{eq:G1}
\end{eqnarray}
where $\left({\partial{u}}/{\partial{V}}\right)_{S}$ is given by
\begin{eqnarray}
\left(\frac{\partial{u}}{\partial{V}}\right)_{S}
=\frac{1}{t}\left\{
x^{z-2}\left(\frac{\partial{y}}{\partial{V}}\right)_{S}
-u\left(\frac{\partial{t}}{\partial{V}}\right)_{S}
\right\}.
\label{eq:G2}
\end{eqnarray}
Then, substituting Eq.~(\ref{eq:G2}) into Eq.~(\ref{eq:G1}), we have
\begin{eqnarray}
\left(\frac{\partial{t}}{\partial{V}}\right)_{S}
=\frac{\tilde{C}_{\rm b}\left(\frac{\partial{y}}{\partial{V}}\right)_{S}}{C_{\rm a}}, 
\label{eq:dtdV}
\end{eqnarray}
where $C_{\rm a}$ and $\tilde{C}_{\rm b}$ are defined by Eqs.~(\ref{eq:Ca}) and (\ref{eq:tildaCb}), respectively. 
By differentiating Eq.~(\ref{eq:t}) with respect to the volume $V$ under a constant entropy $S$, 
we have
\begin{eqnarray}
\left(\frac{\partial{t}}{\partial{V}}\right)_{S}
=
\frac{1}{T_0}
\left(\frac{\partial{T}}{\partial{V}}\right)_{S}
-\frac{T}{T_0^2}
\left(
\frac{\partial{T_0}}{\partial{V}}
\right)_{S}. 
\label{eq:dtdV2}
\end{eqnarray}
By substituting Eq.~(\ref{eq:dtdV2}) into Eq.~(\ref{eq:dtdV}), 
the Gr\"{u}neisen parameter [Eq.~(\ref{eq:Grn_TV})] is expressed as follows:
\begin{eqnarray}
\Gamma=-\frac{\tilde{C}_{\rm b}}{C_{\rm a}}\frac{V}{t}\left(\frac{\partial{y}}{\partial{V}}\right)_{S}
-\frac{V}{T_0}
\left(\frac{\partial{T_0}}{\partial{V}}\right)_{S}. 
\label{eq:Grn_QCP1}
\end{eqnarray}
This is one of the central results of the present paper, whose property will be discussed in detail in Sect.~\ref{sec:Grn}.
The second term expresses the volume derivative of 
the characteristic temperature of spin fluctuation.  
The first term is proportional to $\tilde{C}_{\rm b}$, which gives a minor contribution to $C_{V}$ as shown in Fig.~\ref{fig:Ct_t2dFM}. However, this term gives the dominant contribution to $\Gamma$ at low temperatures, which will be shown in Sect.~\ref{sec:Grn}. 

The Gr\"{u}neisen parameter $\Gamma$ can also be derived from  
Eq.~(\ref{eq:Grn}) with 
the specific heat $C_{V}$ in Eq.~(\ref{eq:CvT}) and 
the thermal-expansion coefficient $\alpha$ defined by Eq.~(\ref{eq:a_PT_def}) 
or by Eq.~(\ref{eq:a_SP_def}).  
Each derivation will be shown in the following Sect.~\ref{sec:a_PT_QCP} and Sect.~\ref{sec:a_SP_QCP}, respectively.

\section{Thermal-expansion coefficient near 
the magnetic QCP}
\label{sec:alpha}
 
So far, in the theory of spin fluctuations, the thermal-expansion coefficient $\alpha$ in itinerant magnets has been discussed on the basis of Eq.~(\ref{eq:a_PT_def})~\cite {Takahashi1999}. In Sect.~\ref{sec:a_SP_QCP}, we will show that $\alpha$ can be derived from Eq.~(\ref{eq:a_SP_def}) in a much simpler form in the SCR theory, which enables us to capture the physical picture. Next, we will derive $\alpha$ by the standard way from Eq.~(\ref{eq:a_PT_def}) in Sects.~\ref{sec:P} and \ref{sec:a_PT_QCP}. It will be shown that the result is lengthy, which is hard to see immediate correspondence to the result obtained in Sect.~\ref{sec:a_SP_QCP}, although both should be equivalent from the viewpoint of the thermodynamic relation as shown in Sect.~\ref{sec:aG}. To show the equivalence in the SCR theory explicitly, the proof will be given in Sect.~\ref{sec:proof}.

\subsection{Derivation from $\alpha=-\frac{1}{V}\left(\frac{\partial{S}}{\partial{P}}\right)_{T}$}
\label{sec:a_SP_QCP}

First, let us derive the thermal-expansion coefficient $\alpha$ defined by Eq.~(\ref{eq:a_SP_def}).
Then, calculation starts from the entropy $S$ in Eq.~(\ref{eq:S}). 
By differentiating the entropy $S$ with respect to the pressure $P$ under a constant temperature, we obtain 
\begin{eqnarray} 
\left(\frac{\partial{S}}{\partial{P}}\right)_{T}
=-\frac{\tilde{C}_{\rm b}}{t}\left(\frac{\partial{y}}{\partial{P}}\right)_{T}
-\frac{C_{\rm a}}{T_0}\left(\frac{\partial{T_0}}{\partial{P}}\right)_{T}, 
\end{eqnarray}
where $\tilde{C}_{\rm b}$ and $C_{\rm a}$ appeared in the formula of the specific heat, 
which are given by Eqs.~(\ref{eq:tildaCb}) and (\ref{eq:Ca}), respectively.
Then, the thermal-expansion coefficient is obtained as
\begin{eqnarray}
\alpha
&=&-\frac{1}{V}\left(\frac{\partial{S}}{\partial{P}}\right)_{T}
\label{eq:a_SP_def2}
\\
&=&\alpha_{\rm a}+\alpha_{\rm b}, 
\label{eq:a_SP}
\end{eqnarray}
where $\alpha_{\rm a}$ and $\alpha_{\rm b}$ are defined by 
\begin{eqnarray}
\alpha_{\rm a}
&\equiv&
\frac{1}{V}
\frac{C_{\rm a}}{T_0}\left(\frac{\partial{T_0}}{\partial{P}}\right)_{T},
\label{eq:a_a}
\\
\alpha_{\rm b}
&\equiv&\frac{1}{V}
\frac{\tilde{C}_{\rm b}}{t}\left(\frac{\partial{y}}{\partial{P}}\right)_{T}, 
\label{eq:a_b}
\end{eqnarray}
respectively. 
This is one of the central results of the present paper, whose property will be discussed in detail in Sect.~\ref{sec:a}. 

\subsection{Pressure near the magnetic QCP}
\label{sec:P}

Next, let us derive the pressure $P=-\left(\frac{\partial\tilde{F}}{\partial{V}}\right)_{T}$ starting from the free energy in Eq.~(\ref{eq:freeE}).
By differentiating $\Gamma_{q}$ [Eq.~(\ref{eq:Gamma_def})] with respect to the volume $V$ under a constant temperature, we have
\begin{eqnarray}
\left(\frac{\partial\Gamma_{q}}{\partial{V}}\right)_{T}
=2\pi\left(\frac{\partial{T_{0}}}{\partial{V}}\right)_{T}
x^{z-2}(y+x^2)+2\pi{T_{0}}x^{z-2}\left(\frac{\partial{y}}{\partial{V}}\right)_{T}.
\nonumber
\\
\label{eq:dGdV}
\end{eqnarray}
In the calculation of $\left(\frac{\partial\tilde{F}}{\partial{V}}\right)_{T}$,  
the terms with $\left(\frac{\partial{y}}{\partial{V}}\right)_{T}$ and also $\left(\frac{\partial\langle\varphi^2\rangle_{\rm eff}}{\partial{V}}\right)_{T}$ vanish because of the SCR equation~[Eq.~(\ref{eq:SCReq})] or optimization condition $d{\tilde F}(y)/dy=0$~\cite{Takahashi2006}. 
The details are given in \ref{sec:Pressure}.
Then, only the first term with $\left(\frac{\partial{T_0}}{\partial{V}}\right)_{T}$ in Eq.~(\ref{eq:dGdV}) remains and we have  
\begin{widetext}
\begin{eqnarray}
P&=&
-\left(\frac{\partial{\tilde{F}}}{\partial{V}}\right)_{T}, 
\nonumber
\\
&=&-\frac{1}{T_0}\left(\frac{\partial{T_0}}{\partial{V}}\right)_{T}I
-
\left[\frac{\partial}{\partial{V}}\left(\frac{\eta_0}{Aq_{\rm B}^2}\right)\right]_{T}
T_{A} 
\langle\varphi^2\rangle_{\rm eff},
\nonumber
\\
&-&
\left(\frac{\eta_{0}}{Aq_{\rm B}^2}-y\right)
\left(\frac{\partial{T_{A}}}{\partial{V}}\right)_{T}
\langle\varphi^2\rangle_{\rm eff}
-\frac{3}{N}\left(\frac{\partial{v_4}}{\partial{V}}\right)_{T}\langle\varphi^2\rangle_{\rm eff}^2, 
\label{eq:dFdV}
\end{eqnarray}
\end{widetext}
where $I$ is given by 
\begin{eqnarray}
I=
\frac{1}{\pi}
\sum_{q}\int_{0}^{\omega_{\rm c}}d\omega
\Gamma_{q}\frac{\partial}{\partial\Gamma_{q}}\left(\frac{\Gamma_{q}}{\omega^2+\Gamma_{q}^2}\right)
\left[\frac{\omega}{2}+T{\ln}\left(1-{\rm e}^{-\frac{\omega}{T}}\right)\right].  
\nonumber
\\
\label{eq:I_integral}
\end{eqnarray}
Here, by using the relation~\cite{Takahashi1999}
\begin{eqnarray}
\frac{\partial}{\partial\Gamma_{q}}\left(\frac{\Gamma_{q}}{\omega^2+\Gamma_{q}^2}\right)
=-\frac{\partial}{\partial\omega}\left(\frac{\omega}{\omega^2+\Gamma_{q}^2}\right),
\end{eqnarray}
the partial integration with respect to $\omega$ can be performed. 
Then, we have
\begin{eqnarray}
I&=&\frac{1}{\pi}\sum_{q}\Gamma_{q}\left\{\left.
-\frac{\omega}{\omega^2+\Gamma_{q}^2}
\left[\frac{\omega}{2}+T{\ln}\left(1-{\rm e}^{-\frac{\omega}{T}}\right)\right]
\right|_{0}^{\rm \omega_{\rm c}}
\right.
\nonumber
\\
& & 
\left.
+\int_{0}^{\omega_{\rm c}}d\omega\frac{\omega}{\omega^2+\Gamma_{q}^2}
\left(\frac{1}{2}+\frac{1}{{\rm e}^{\frac{\omega}{T}}-1}\right)
\right\}.
\label{eq:Ipi}
\end{eqnarray}
The first line in Eq.~(\ref{eq:Ipi}) is neglected since the spectrum of the spin fluctuation is considered to decrease faster than the Lorentzian in the high-frequency regime~\cite{Takahashi1999}. Hence, the first and second terms in the last line in Eq.~(\ref{eq:Ipi}) are expressed as 
\begin{eqnarray}
I=I_{\rm zero}+I_{\rm th},
\end{eqnarray}
respectively, 
where $I_{\rm zero}$ is given by 
\begin{eqnarray}
I_{\rm zero}&=&\frac{NdT_{0}t}{2}\int_{0}^{x_{\rm c}}dx{x}^{d-1} u {\ln}\left|\frac{\omega_{{\rm c}T}^2+u^2}{u^2}\right|, 
\label{eq:I_zero}
\end{eqnarray}
and $I_{\rm th}$ is given by 
\begin{eqnarray}
I_{\rm th}&=&NdT_{0}t\int_{0}^{x_{\rm c}}dx{x}^{d-1}u\left\{{\ln} u-\frac{1}{2u}-\psi(u)\right\}.
\label{eq:I_th}
\end{eqnarray}
%

\subsection{Derivation from $\alpha=\kappa_{T}\left(\frac{\partial{P}}{\partial{T}}\right)_{V}$}
\label{sec:a_PT_QCP}

Let us derive the thermal-expansion coefficient $\alpha$ defined by   $\alpha\equiv\kappa_{T}\left(\frac{\partial{P}}{\partial{T}}\right)_{V}$ in Eq.~(\ref{eq:a_PT_def}), 
where the isothermal compressibility is given by Eq.~(\ref{eq:comp}). 
By differentiating the pressure in Eq.~(\ref{eq:dFdV}) with respect to the temperature $T$ 
under a constant volume $V$, we have  
\begin{widetext}
\begin{eqnarray}
\frac{\alpha}{\kappa_{T}}&=&\left(\frac{\partial{P}}{\partial{T}}\right)_{V}, 
\nonumber
\\
&=&
-
\frac{1}{T_0}\left(\frac{\partial{T_0}}{\partial{V}}\right)_{T}
\left[
\left(\frac{\partial{I_{\rm zero}}}{\partial{T}}\right)_{V}
+\left(\frac{\partial{I_{\rm th}}}{\partial{T}}\right)_{V}
\right]
\nonumber
\\
&-&\frac{\partial}{\partial{T}}
\left.
\left\{
\left[\frac{\partial}{\partial{V}}\left(\frac{\eta_0}{Aq_{\rm B}^2}\right)\right]_{T}
T_{A}\langle\varphi^2\rangle_{\rm eff}
+
\left(\frac{\eta_{0}}{Aq_{\rm B}^2}-y\right)\left(\frac{\partial{T_A}}{\partial{V}}\right)_{T}
\langle\varphi^2\rangle_{\rm eff}
\right.
\right.
\nonumber
\\
& &
\ \ \ \ \ \ 
\left.
\left.
+\frac{3}{N}\left(\frac{\partial{v_4}}{\partial{V}}\right)_{T}\langle\varphi^2\rangle_{\rm eff}^2
\right\}\right|_{V},
\label{eq:ak}
\end{eqnarray}
\end{widetext}
where
\begin{eqnarray}
\left(\frac{\partial{I_{\rm zero}}}{\partial{T}}\right)_{V}&=&
Nd
\left(\frac{\partial{y}}{\partial{t}}\right)_{V}
\left\{
\frac{1}{2}\int_{0}^{x_{\rm c}}dxx^{d+z-3}{\ln}\left|\frac{\omega_{{\rm c}T}^2+u^2}{u^2}\right|
\right.
\nonumber
\\
& &
\left.
\ \ \ \ \ \ \ \ \ \ \ \ \ \ \ \ \ \ 
-
\int_{0}^{x_{\rm c}}dxx^{d+z-3}\frac{\omega_{{\rm c}T}^2}{\omega_{{\rm c}T}^2+u^2}
\right\},
\nonumber
\\
& &
\label{eq:dIdT_zero}
\end{eqnarray}
\begin{eqnarray}
\left(\frac{\partial{I_{\rm th}}}{\partial{T}}\right)_{V}&=&
\left(\frac{\partial{y}}{\partial{t}}\right)_{V}
\left(
Nd
L
-
\tilde{C}_{\rm b}
\right)
+
C_{\rm a}.   
\label{eq:dIdT_th}
\end{eqnarray}
Here, $L$ is given by
\begin{eqnarray}
L=
\int_{0}^{x_{\rm c}}dxx^{d+z-3}
\left\{
{\ln}u-\frac{1}{2u}-\psi(u)
\right\}.
\label{eq:L_def}
\end{eqnarray}
By substituting Eq.~(\ref{eq:dIdT_zero}) and Eq.~(\ref{eq:dIdT_th}) into Eq.~(\ref{eq:ak}), we obtain 
\begin{widetext}
\begin{eqnarray}
\frac{\alpha}{\kappa_T}=
\frac{1}{T_0}\left(\frac{\partial{T_0}}{\partial{V}}\right)_{T}
\left[
\left(\frac{\partial{y}}{\partial{t}}\right)_{V}
\left\{
-\frac{Nd}{2}\int_{0}^{x_{\rm c}}dxx^{d+z-3}{\ln}\left|\frac{\omega_{{\rm c}T}^2+u^2}{u^2}\right|
\right.
\right.
\nonumber
\\ 
\left.
\left.
+Nd\int_{0}^{x_{\rm c}}dxx^{d+z-3}\frac{\omega_{{\rm c}T}^2}{\omega_{{\rm c}T}^2+u^2}
-Nd
L
+\tilde{C}_{\rm b}
\right\}
-C_{\rm a}
\right]
\nonumber
\\ 
-\left(\frac{\partial{y}}{\partial{t}}\right)_{V}
\left\{
\frac{N}{6v_4}\frac{T_{A}^2}{T_0}
\left[\frac{\partial}{\partial{V}}\left(\frac{\eta_0}{Aq_{\rm B}^2}\right)\right]_{T}
\right.
\nonumber
\\ 
\left.
-
\frac{2}{T_0}\langle\varphi^2\rangle_{\rm eff}
\left(\frac{\partial{T_A}}{\partial{V}}\right)_{T}
+
\frac{T_A}{T_0}\langle\varphi^2\rangle_{\rm eff}
\frac{1}{v_4}\left(\frac{\partial{v_4}}{\partial{V}}\right)_{T}
\right\}, 
\label{eq:ak2}
\end{eqnarray}
\end{widetext}
where the last three terms have been obtained by using the SCR equation [Eq.~(\ref{eq:SCReq})]. 
Details are given in Appendix~\ref{sec:last3}.

\subsection{Equivalence of the expressions of thermal-expansion coefficients}
\label{sec:proof}

In Sect.~\ref{sec:a_SP_QCP} and Sect.~\ref{sec:a_PT_QCP}, each expression of the thermal-expansion coefficient $\alpha$ has been derived from Eq.~(\ref{eq:a_SP_def}) and Eq.~(\ref{eq:a_PT_def}), respectively. 
At first glance, it seems unclear whether Eq.~(\ref{eq:a_SP}) and Eq.~(\ref{eq:ak2}) are equivalent. 
However, with the use of the stationary condition of the SCR theory, it can be shown that both expressions are equivalent, 
which will be proven in this subsection.

Multiplying $\kappa_{T}$ on both sides of Eq.~(\ref{eq:ak2}) and using Eq.~(\ref{eq:comp})  
with the relation 
$\left(\frac{\partial{V}}{\partial{P}}\right)_{T}\left(\frac{\partial{Y}}{\partial{V}}\right)_{T}=\left(\frac{\partial{Y}}{\partial{P}}\right)_{T}$ for $Y=T_0, \eta_0, N_{\rm F}$ and $v_4$, we obtain 
\begin{eqnarray}
\alpha&=&\frac{1}{V}
\frac{1}{T_0}\left(\frac{\partial{T_0}}{\partial{P}}\right)_{T}
\left[
C_{\rm a}
\right.
\nonumber
\\
& &
\left.
+
\left(\frac{\partial{y}}{\partial{t}}\right)_{V}
\left\{
Nd
L
-\tilde{C}_{\rm b}
\right.
\right.
\nonumber
\\
& &
\left.
\left.
+\frac{Nd}{2}
\int_{0}^{x_{\rm c}}dxx^{d+z-3}{\ln}\left|\frac{\omega_{{\rm c}T}^2+u^2}{u^2}\right|
\right.
\right.
\nonumber
\\
& &
\left.
\left.
-Nd
\int_{0}^{x_{\rm c}}dxx^{d+z-3}\frac{\omega_{{\rm c}T}^2}{\omega_{{\rm c}T}^2+u^2}
\right\}
\right]
\nonumber
\\
& &
+\frac{1}{V}
\left(\frac{\partial{y}}{\partial{t}}\right)_{V}
\left\{
\frac{N}{6v_4}\frac{T_{A}^2}{T_0}
\left[\frac{\partial}{\partial{P}}\left(\frac{\eta_0}{Aq_{\rm B}^2}\right)\right]_{T}
\right.
\nonumber
\\
& &
\left.
-
\frac{2}{T_0}\langle\varphi^2\rangle_{\rm eff}
\left(\frac{\partial{T_A}}{\partial{P}}\right)_{T}
+
\frac{T_A}{T_0}\langle\varphi^2\rangle_{\rm eff}
\frac{1}{v_4}\left(\frac{\partial{v_4}}{\partial{P}}\right)_{T}
\right\}. 
\nonumber
\\
& &
\label{eq:a_P}
\end{eqnarray}

Near the QCP, the $x$ integration on the third line of Eq.~(\ref{eq:a_P}) 
is expanded around $y=0$ as
\begin{eqnarray}
\int_{0}^{x_{\rm c}}dxx^{d+z-3}\ln\left|\frac{\omega_{{\rm c}T}^2+u^2}{u^2}\right|
\nonumber
\\
=
\left\{
\begin{array}{lr}
C_1-C_{2}y+\cdots, \quad \mbox{$(d+z>4)$}  \\ 
C_1+y{\ln}y-C_{2}y+\cdots, \quad \mbox{$(d+z=4)$}
\end{array}
\right. 
\label{eq:Izero_log_res}
\end{eqnarray}
where $C_1$ and $C_2$ are given by Eqs.~(\ref{eq:C1}) and (\ref{eq:C2}), respectively. 
On the fourth line of Eq.~(\ref{eq:a_P}), the $x$ integration is also expanded around $y=0$  and we obtain 
\begin{eqnarray}
-\frac{1}{T_0}\left(\frac{\partial{T_0}}{\partial{P}}\right)_{T}
\int_{0}^{x_{\rm c}}dxx^{d+z-3}\frac{\omega_{{\rm c}T}^2}{\omega_{{\rm c}T}^2+u^2}
\nonumber
\\
=\frac{1}{2}
\left\{
\left(\frac{\partial{C_1}}{\partial{P}}\right)_{T}
-\left(\frac{\partial{C_2}}{\partial{P}}\right)_{T}y
\right\}+\cdots,
\label{eq:Izero_1_res}
\end{eqnarray}
whose derivation is given in \ref{sec:Izero_log}.

By substituting the SCR equation [Eq.~(\ref{eq:SCReq2})] into $y$ which appears in the r.h.s. of 
Eqs.~(\ref{eq:Izero_log_res}) and (\ref{eq:Izero_1_res}), and $\langle\varphi^2\rangle_{\rm eff}$ in the last line of Eq.~(\ref{eq:a_P}), Eq.~(\ref{eq:a_P}) is expressed as
\begin{eqnarray}
\alpha&=&\frac{1}{V}
\frac{1}{T_0}\left(\frac{\partial{T_0}}{\partial{P}}\right)_{T}
C_{\rm a}
+\frac{1}{V}\left(\frac{\partial{y}}{\partial{t}}\right)_{V}
\left[
-\frac{1}{T_0}\left(\frac{\partial{T_0}}{\partial{P}}\right)_{T}
\tilde{C}_{\rm b}
\right.
\nonumber
\\
& &
\left.
+N\frac{2}{y_1}
\left\{
\left(\frac{\partial{y_0}}{\partial{P}}\right)_{T}
+\left(\frac{\partial{y_1}}{\partial{P}}\right)_{T}
\frac{d}{2}
L
\right\}
\right]  
\label{eq:a_P2} 
\end{eqnarray}
for $d+z>4$. 
The details of the derivation are given in \ref{sec:deriv_a}. 
Here, the following equations, which are obtained by differentiating Eqs.~(\ref{eq:y0}) and (\ref{eq:y1}) with respect to 
the
pressure, respectively, are used to derive the second line of Eq.~(\ref{eq:a_P2}):
\begin{widetext}
\begin{eqnarray}
\left(\frac{\partial{y_0}}{\partial{P}}\right)_{T}
&=&
\left\{
\frac{1}{T_0}\left(\frac{\partial{T_0}}{\partial{P}}\right)_{T}
-\frac{2}{T_A}\left(\frac{\partial{T_A}}{\partial{P}}\right)_{T}
+\frac{1}{v_4}\left(\frac{\partial{v_4}}{\partial{P}}\right)_{T}
\right\}
\nonumber
\\
& &
\times
\frac{d}{4}y_1\left(C_1-C_2y_0\right)
+\frac{d}{4}y_1
\left\{
\left(\frac{\partial{C_1}}{\partial{P}}\right)_{T}
-
\left(\frac{\partial{C_2}}{\partial{P}}\right)_{T}
y_0
\right\}
\nonumber
\\
& &
+
\left[\frac{\partial}{\partial{P}}\left(\frac{\eta_0}{Aq_{\rm B}^2}\right)\right]_{T}
\frac{1}{Aq_{\rm B}^2}\frac{y_1}{12v_4\frac{T_0}{T_A^2}},
\label{eq:dy0dP}
\\
\left(\frac{\partial{y_1}}{\partial{P}}\right)_{T}
&=&
\left\{
\frac{1}{T_0}\left(\frac{\partial{T_0}}{\partial{P}}\right)_{T}
-\frac{2}{T_A}\left(\frac{\partial{T_A}}{\partial{P}}\right)_{T}
+\frac{1}{v_4}\left(\frac{\partial{v_4}}{\partial{P}}\right)_{T}
\right\}
\nonumber
\\
& &
\times
y_1\left(1-\frac{d}{4}C_2y_1\right)
-\frac{d}{4}y_1^2
\left(\frac{\partial{C_2}}{\partial{P}}\right)_{T}.
\label{eq:dy1dP}
\end{eqnarray}
\end{widetext}
For $d+z=4$, 
by substituting the SCR equation [Eq.~(\ref{eq:SCReq3})] into Eq.~(\ref{eq:Izero_1_res}) and 
with the use of Eqs.~(\ref{eq:dy0dP}) and (\ref{eq:dy1dP}), Eq.~(\ref{eq:a_P}) is shown to lead to 
\begin{eqnarray}
\alpha&=&\frac{1}{V}
\frac{1}{T_0}\left(\frac{\partial{T_0}}{\partial{P}}\right)_{T}
C_{\rm a}
+\frac{1}{V}\left(\frac{\partial{y}}{\partial{t}}\right)_{V}
\left[
-\frac{1}{T_0}\left(\frac{\partial{T_0}}{\partial{P}}\right)_{T}
\tilde{C}_{\rm b}
\right.
\nonumber
\\
& &
\left.
+N\frac{2}{y_1}
\left\{
\left(\frac{\partial{y_0}}{\partial{P}}\right)_{T}
+\left(\frac{\partial{y_1}}{\partial{P}}\right)_{T}
\left(
L
+\frac{1}{2}y\ln{y}
\right)
\right\}
\right],  
\nonumber
\\
& & 
\label{eq:a_P2_d2} 
\end{eqnarray}
which has a form with a logarithmic term in the last term in Eq.~(\ref{eq:a_P2}). 

The results of Eqs.~(\ref{eq:a_P2}) and (\ref{eq:a_P2_d2}) are consequences of the fact that the quantities 
$T_0$, 
$T_{A}$,
 $v_4$, and 
$\eta_0/(Aq_{\rm B}^2)$,
 which are included in $y_0$ and/or $y_1$ 
defined by Eqs.~(\ref{eq:y0}) and (\ref{eq:y1}), respectively, 
have
the
 pressure dependence. Since $T_0$ is included in the constants $C_1$ and $C_2$, as seen in Eqs.~(\ref{eq:C1}) and (\ref{eq:C2}), respectively, the pressure dependences appear via $T_0$.  

By substituting Eq.~(\ref{eq:dydt}) into Eqs.~(\ref{eq:a_P2}) and (\ref{eq:a_P2_d2}), the thermal-expansion coefficient is expressed as
%
\begin{eqnarray}
\alpha&=&\alpha_{\rm a}+\alpha_{\rm b},
\label{eq:a_P3}
\end{eqnarray}
where $\alpha_{\rm a}$ is given by
\begin{eqnarray}
\alpha_{\rm a}=
\frac{1}{V}
\frac{1}{T_0}\left(\frac{\partial{T_0}}{\partial{P}}\right)_{T}
C_{\rm a}, 
\label{eq:a_a2}
\end{eqnarray}
and $\alpha_{\rm b}$ is given by
\begin{eqnarray}
\alpha_{\rm b}
=\frac{1}{V}
\frac{
\frac{\tilde{C}_{\rm b}}{t}
\left[
\left(\frac{\partial{y_0}}{\partial{P}}\right)_{T}
+\left(\frac{\partial{y_1}}{\partial{P}}\right)_{T}
\frac{d}{2}
L
-\frac{1}{T_0}\left(\frac{\partial{T_0}}{\partial{P}}\right)_{T}
\tilde{C}_{\rm b}
\frac{y_1}{2}\frac{1}{N}
\right]
}
{
1-\frac{dy_1}{2t}M
}
\nonumber
\\
\label{eq:a_b2}
\end{eqnarray}
for $d+z>4$, and 
\begin{widetext}
\begin{eqnarray}
\alpha_{\rm b}
=\frac{1}{V}
\frac{
\frac{\tilde{C}_{\rm b}}{t}
\left[
\left(\frac{\partial{y_0}}{\partial{P}}\right)_{T}
+\left(\frac{\partial{y_1}}{\partial{P}}\right)_{T}
\left(
\frac{d}{2}
L
+\frac{1}{2}y\ln{y}
\right)
-\frac{1}{T_0}\left(\frac{\partial{T_0}}{\partial{P}}\right)_{T}
\tilde{C}_{\rm b}
\frac{y_1}{2}\frac{1}{N}
\right]
}
{
1
-\frac{y_1}{2}(\ln{y}+1)
-\frac{dy_1}{2t}M
}
\label{eq:a_b2_d2}
\end{eqnarray}
\end{widetext}
for $d+z=4$, respectively.

On the other hand, let us turn to Eq.~(\ref{eq:a_SP}) to be compared with Eq.~(\ref{eq:a_P3}). 
We see that Eq.~(\ref{eq:a_a}) is exactly the same as Eq.~(\ref{eq:a_a2}).
As for $\alpha_{\rm b}$, 
to calculate $\left(\frac{\partial{y}}{\partial{P}}\right)_{T}$, 
by differentiating the SCR equation [Eq.~(\ref{eq:SCReq2}) and Eq.~(\ref{eq:SCReq3})] with respect to 
the
pressure $P$, we obtain
\begin{widetext}
\begin{eqnarray}
\left(\frac{\partial{y}}{\partial{P}}\right)_{T}
=
\left\{
\begin{array}{lr}
\frac{
\left(\frac{\partial{y_0}}{\partial{P}}\right)_{T}
+\left(\frac{\partial{y_1}}{\partial{P}}\right)_{T}
\frac{d}{2}
L
-\frac{1}{T_0}\left(\frac{\partial{T_0}}{\partial{P}}\right)_{T}
\tilde{C}_{\rm b}
\frac{y_1}{2}\frac{1}{N}
}
{
1-\frac{dy_1}{2t}M
}
\quad \ \mbox{$(d+z>4)$},  \\
\frac{
\left(\frac{\partial{y_0}}{\partial{P}}\right)_{T}
+\left(\frac{\partial{y_1}}{\partial{P}}\right)_{T}
\left(
\frac{d}{2}
L
+\frac{1}{2}y\ln{y}
\right)
-\frac{1}{T_0}\left(\frac{\partial{T_0}}{\partial{P}}\right)_{T}
\tilde{C}_{\rm b}
\frac{y_1}{2}\frac{1}{N}
}
{
1-\frac{y_1}{2}(\ln{y}+1)
-\frac{dy_1}{2t}M
}
\quad \ \mbox{$(d+z=4)$}, 
\end{array}
\right. 
\label{eq:dydP}
\end{eqnarray}
\end{widetext}
respectively. 
By substituting Eq.~(\ref{eq:dydP}) into Eq.~(\ref{eq:a_b}), $\alpha_{\rm b}$ is obtained as 
\begin{eqnarray}
\alpha_{\rm b}=
\frac{1}{V}
\frac{
\frac{\tilde{C}_{\rm b}}{t}
\left[
\left(\frac{\partial{y_0}}{\partial{P}}\right)_{T}
+\left(\frac{\partial{y_1}}{\partial{P}}\right)_{T}
\frac{d}{2}
L
-\frac{1}{T_0}\left(\frac{\partial{T_0}}{\partial{P}}\right)_{T}
\tilde{C}_{\rm b}
\frac{y_1}{2}\frac{1}{N}
\right]
}
{
1-\frac{dy_1}{2t}M
}
\nonumber
\\
\label{eq:a_b2_2}
\end{eqnarray}
for $d+z>4$, and 
\begin{widetext}
\begin{eqnarray}
\alpha_{\rm b}
=\frac{1}{V}
\frac{
\frac{\tilde{C}_{\rm b}}{t}
\left[
\left(\frac{\partial{y_0}}{\partial{P}}\right)_{T}
+\left(\frac{\partial{y_1}}{\partial{P}}\right)_{T}
\left(
\frac{d}{2}
L
+\frac{1}{2}y\ln{y}
\right)
-\frac{1}{T_0}\left(\frac{\partial{T_0}}{\partial{P}}\right)_{T}
\tilde{C}_{\rm b}
\frac{y_1}{2}\frac{1}{N}
\right]
}
{
1
-\frac{y_1}{2}(\ln{y}+1)
-\frac{dy_1}{2t}M
}
\label{eq:a_b2_d2_2}
\end{eqnarray}
\end{widetext}
for $d+z=4$, respectively. 
Now we see that Eq.~(\ref{eq:a_b2_2}) is exactly the same as Eq.~(\ref{eq:a_b2}) 
for $d+z>4$. 
We also see that Eq.~(\ref{eq:a_b2_d2_2}) is exactly the same as Eq.~(\ref{eq:a_b2_d2}) 
for $d+z=4$. 
Hence, it is proven that both expressions on $\alpha$ of Eq.~(\ref{eq:a_SP}) and Eq.~(\ref{eq:ak2}) are equivalent.

\section{Numerical results on thermal-expansion coefficient $\alpha$ at
the
 magnetic QCP}
\label{sec:a}

Since equivalence of Eq.~(\ref{eq:a_SP}) and Eq.~(\ref{eq:ak2}) has been proven, 
let us analyze the temperature dependence of the thermal-expansion coefficient $\alpha$ on the basis of Eq.~(\ref{eq:a_SP}), which has a simpler expression. 
In Eq.~(\ref{eq:a_SP}), $\alpha_{\rm b}$ [Eq.~(\ref{eq:a_b})] includes $\left(\frac{\partial{y}}{\partial{P}}\right)_{T}$, which can be obtained by calculating the r.h.s. of Eq.~(\ref{eq:dydP}).
Namely, when the pressure derivatives of $T_0$, $T_A$, $v_4$, and $\eta_0/(Aq_{\rm B}^2)$ are given, 
one can obtain $\left(\frac{\partial{y_0}}{\partial{P}}\right)_{T}$ by Eq.~(\ref{eq:dy0dP}) and $\left(\frac{\partial{y_1}}{\partial{P}}\right)_{T}$ by Eq.~(\ref{eq:dy1dP}). 
Then, substituting them into Eq.~(\ref{eq:dydP}), one obtains the temperature dependence of   $\left(\frac{\partial{y}}{\partial{P}}\right)_{T}$. 
At the QCP tuned to the critical pressure $P=P_{\rm c}$, $y_0=0$ is realized. 
However, $\left(\frac{\partial{y_0}}{\partial{P}}\right)_{T=0}$ 
can have
 a non-zero value at the QCP in general, which will be shown for $d+z>4$ below.  
Since $y_0$ defined by Eq.~(\ref{eq:y0}) is the quantity for $T=0$, 
the first terms in the numerator and denominator of the r.h.s. of Eq.~(\ref{eq:dydP}) are constants. 
Note that $y_1$ defined by Eq.~(\ref{eq:y1}) is the quantity for $T=0$ and hence $\left(\frac{\partial{y_1}}{\partial{P}}\right)_{T}$ has no temperature dependence. 
Since $T_0$ is defined by Eq.~(\ref{eq:T0}), $\left(\frac{\partial{T_0}}{\partial{P}}\right)_{T}$ also has no temperature dependence. 

To see the temperature dependence of $\alpha$ at the QCP,  
numerical calculation of Eq.~(\ref{eq:a_SP}) is performed.
First, we solve the SCR equation [Eq.~(\ref{eq:SCReq2}) or Eq.~(\ref{eq:SCReq3})] 
by setting $y_0=0.0$ and $y_1=1.0$. 
With the use of the solution $y$ and $\left(\frac{\partial{y}}{\partial{T}}\right)_{V}$ obtained by Eq.~(\ref{eq:dydt}), we calculate $C_{\rm a}(t)$ in Eq.~(\ref{eq:Ca}) and 
$\tilde{C}_{\rm b}(t)$ in Eq.~(\ref{eq:tildaCb}). 
Then, we calculate $\left(\frac{\partial{y}}{\partial{P}}\right)_{T}$ in Eq.~(\ref{eq:dydP}) by setting 
$\left(\frac{\partial{y_0}}{\partial{P}}\right)_{T}=1.0$,  $\left(\frac{\partial{y_1}}{\partial{P}}\right)_{T}=1.0$, and 
$\frac{1}{T_0}\left(\frac{\partial{T_0}}{\partial{P}}\right)_{T}=1.0$ 
as representative values (the reason for this parametrization is explained below). 
Finally, by substituting $C_{\rm a}(t)$, $\tilde{C}_{\rm b}(t)$, and $\left(\frac{\partial{y}}{\partial{P}}\right)_{T}$ into Eq.~(\ref{eq:a_SP}), 
we obtain the temperature dependence of $\alpha(t)$ at the QCP. 
In the plot of $\alpha(t)$, the lattice constant is set as unity. 
The results for each
universality
 class are shown in Figs.~\ref{fig:a_t_d3z3}, \ref{fig:a_t_d3z2}, \ref{fig:a_t_d2z3}, and \ref{fig:a_t_d2z2}, respectively, whose properties are analyzed in the following subsections. 

Before going to detailed analysis of $\alpha(t)$, here we comment on the unit and parametrization. In Eq.~(\ref{eq:a_SP}), the volume $V$ is regarded as the molar volume. Then, 
by multiplying the value of $\frac{1}{T_0}\left(\frac{\partial{T_0}}{\partial{P}}\right)_T$ in the unit of GPa$^{-1}$ to the restored Boltzmann constant over the unit-cell volume $k_{\rm B}/V_{\rm unit}$,  where $V_{\rm unit}$ is given by $V_{\rm unit}=a~[\AA]\times b~[\AA]\times c~[\AA]$, we obtain $\alpha_{\rm a}$ in the unit of K$^{-1}$, as follows: 
\begin{eqnarray}
\alpha_{\rm a}=\underline{
\frac{1.38}{abc}\times 10^{-2}\times \frac{1}{T_0}\left(\frac{\partial{T_0}}{\partial{P}}\right)_{T}
}
\times \frac{C_{\rm a}}{N} \ \ \ [{\rm K}^{-1}]. 
\label{eq:a_a_unit}
\end{eqnarray}
As for $\alpha_{\rm b}$, 
by multiplying the value of $\left(\frac{\partial{y}}{\partial{P}}\right)_T$ in the unit of GPa$^{-1}$ to $k_{\rm B}/V_{\rm unit}$, we obtain $\alpha_{\rm b}$ in the unit of K$^{-1}$, as follows:
\begin{eqnarray}
\alpha_{\rm b}=\underline{
\frac{1.38}{abc}\times 10^{-2}\times \left(\frac{\partial{y}}{\partial{P}}\right)_{T}
}
\times \frac{\tilde{C}_{\rm b}}{Nt} \ \ \ [{\rm K}^{-1}]. 
\label{eq:a_b_unit}
\end{eqnarray}
Hence, multiplying the numerical value of the underlined part of Eq.~(\ref{eq:a_a_unit}) and (\ref{eq:a_b_unit}) for each material to $\alpha_{\rm a}$ and $\alpha_{\rm b}$, respectively, in the following Figs.~\ref{fig:a_t_d3z3}, \ref{fig:a_t_d3z2}, \ref{fig:a_t_d2z3}, and \ref{fig:a_t_d2z2}, direct comparison with experiments can be made. 
More detailed discussion about experiments will be given in Sect.~\ref {sec:P_Ce}.

\subsection{3d Ferromagnetic case}
\label{sec:a_d3z3}

\begin{figure*}
\includegraphics[width=15cm]{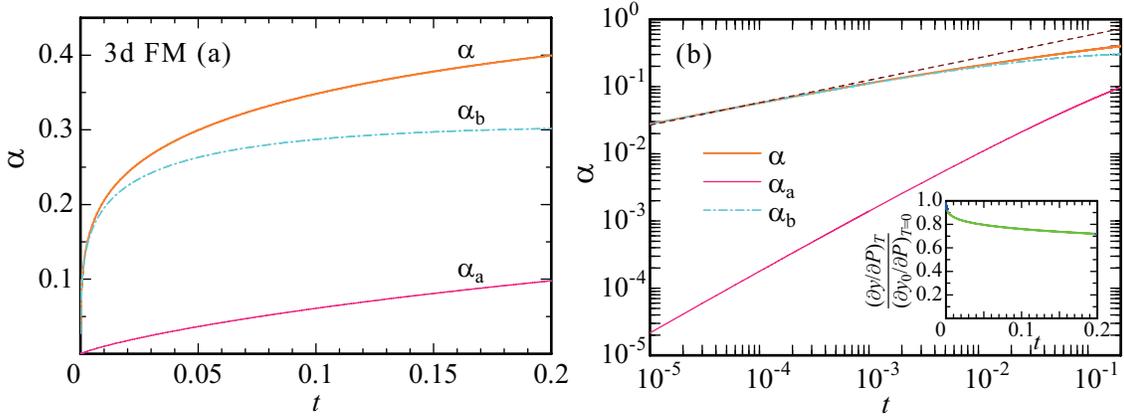}
\caption{(Color online) (a) Temperature dependence of the thermal-expansion coefficient $\alpha$ (thick solid line), $\alpha_{\rm a}$ (thin solid line), and $\alpha_{\rm b}$ (dash-dotted line) at the 3d FM QCP. 
(b) Log-log plot of (a). The dashed line represents the least-square fit of $\alpha$ with $at^{1/3}$ for $10^{-5}\le t\le 10^{-4}$. The inset shows the $t$ dependence of $\left(\frac{\partial{y}}{\partial{P}}\right)_{T}/\left(\frac{\partial{y_0}}{\partial{P}}\right)_{T=0}$. 
}
\label{fig:a_t_d3z3}
\end{figure*}

Figure~\ref{fig:a_t_d3z3}(a) shows 
the
 temperature dependence of 
the thermal-expansion coefficient 
$\alpha$ at the FM QCP $(z=3)$ in $d=3$.
As $t$ decreases, $\alpha_{\rm b}$ in Eq.~(\ref{eq:a_SP}) contributes to $\alpha$ dominantly, 
$\alpha\approx\frac{1}{V}\frac{\tilde{C}_{\rm b}}{t}\left(\frac{\partial{y}}{\partial{P}}\right)_{T}$, while   
contribution from $\alpha_{\rm a}$ becomes not negligible as $t$ increases. 

For $t\ll 1$, we estimate  
$L\sim t^{4/3}$, $\tilde{C}_{\rm b}\sim t^{4/3}$, and 
$M\sim -t^{4/3}$ in Eq.~(\ref{eq:dydP}). 
Hence, we have
$\left(\frac{\partial{y}}{\partial{P}}\right)_{T}\approx \left(\frac{\partial{y_0}}{\partial{P}}\right)_{T=0}-b_{1}t^{1/3}$ with $b_1$ being a positive constant. 
This can be seen in the inset of Fig.~\ref{fig:a_t_d3z3}(b) where 
the
 $t$ dependence of 
$\left(\frac{\partial{y}}{\partial{P}}\right)_{T}/\left(\frac{\partial{y_0}}{\partial{P}}\right)_{T=0}$ is plotted. 
At sufficiently low temperatures where $\left(\frac{\partial{y}}{\partial{P}}\right)_{T}$ can be regarded as a constant, 
$\left(\frac{\partial{y}}{\partial{P}}\right)_{T}\approx \left(\frac{\partial{y_0}}{\partial{P}}\right)_{T=0}$, 
 $\alpha$ behaves as 
\begin{eqnarray}
\alpha\propto\frac{\tilde{C}_{\rm b}}{t}
\sim t^{\frac{1}{3}}, 
\label{eq:a_3dFM}
\end{eqnarray}
where $\frac{\tilde{C}_{\rm b}}{t}\sim t^{1/3}$ dominates over $C_{\rm a}\sim -t\ln{t}$ [Eq.~(\ref{eq:Cv_d3z3})] in Eq.~(\ref{eq:a_SP}). 
This coincides with the temperature dependence of the critical part shown by the RG theory~\cite{Zhu2003}. 
However, it should be noted that $\alpha\sim t^{1/3}$ appears at sufficiently low temperatures 
for $t\lsim 10^{-3}$, as shown in Fig.~\ref{fig:a_t_d3z3}(b).
This is because the temperature dependent $\left(\frac{\partial{y}}{\partial{P}}\right)_{T}$ exists in $\alpha_{\rm b}$ in Eq.~(\ref{eq:a_b}), as noted above. 

\subsection{3d Antiferromagnetic case}
\label{sec:a_d3z2}

\begin{figure*}
\includegraphics[width=15cm]{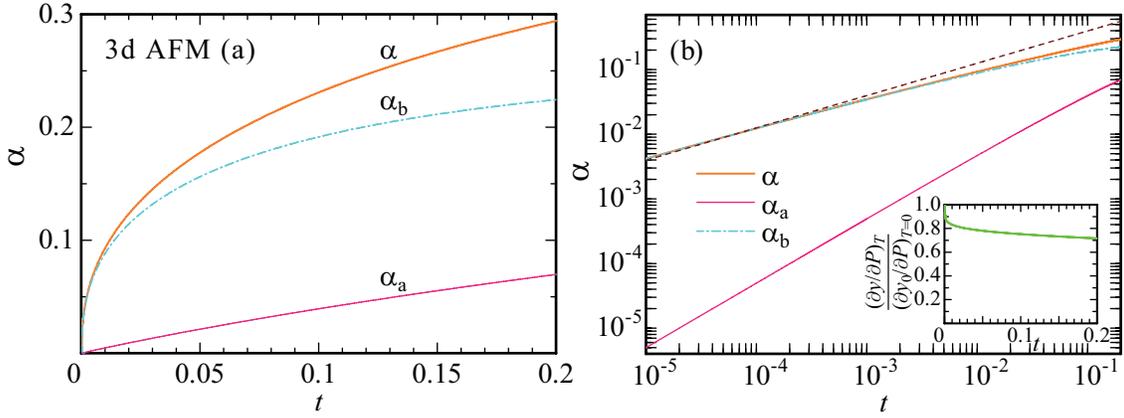}
\caption{(Color online) (a) Temperature dependence of the thermal-expansion coefficient $\alpha$ (thick solid line), $\alpha_{\rm a}$ (thin solid line), and $\alpha_{\rm b}$ (dash-dotted line) at the 3d AFM QCP. 
(b) Log-log plot of (a). The dashed line represents the least-square fit of $\alpha$ with $at^{1/2}$ for $10^{-5}\le t\le 10^{-4}$. The inset shows the $t$ dependence of $\left(\frac{\partial{y}}{\partial{P}}\right)_{T}/\left(\frac{\partial{y_0}}{\partial{P}}\right)_{T=0}$. 
}
\label{fig:a_t_d3z2}
\end{figure*}

Figure~\ref{fig:a_t_d3z2}(a) shows the temperature dependence of 
the thermal-expansion coefficient 
$\alpha$ at the AFM QCP $(z=2)$ in $d=3$.
As $t$ decreases, $\alpha_{\rm b}$ in Eq.~(\ref{eq:a_SP}) contributes to $\alpha$ dominantly, 
$\alpha\approx\frac{1}{V}\frac{\tilde{C}_{\rm b}}{t}\left(\frac{\partial{y}}{\partial{P}}\right)_{T}$, while   
contribution from $\alpha_{\rm a}$ becomes not negligible as $t$ increases. 

For $t\ll 1$, we estimate  
$L\sim t^{3/2}$, $\tilde{C}_{\rm b}\sim t^{3/2}$, and 
$M\sim -t^{5/4}$ in Eq.~(\ref{eq:dydP}). 
Hence, we have
$\left(\frac{\partial{y}}{\partial{P}}\right)_{T}\approx \left(\frac{\partial{y_0}}{\partial{P}}\right)_{T=0}-b_{2}t^{1/4}$ with $b_2$ being a positive constant. 
This can be seen in the inset of Fig.~\ref{fig:a_t_d3z2}(b) where the $t$ dependence of 
$\left(\frac{\partial{y}}{\partial{P}}\right)_{T}/\left(\frac{\partial{y_0}}{\partial{P}}\right)_{T=0}$ is plotted. 
At sufficiently low temperatures where $\left(\frac{\partial{y}}{\partial{P}}\right)_{T}$ can be regarded as a constant, 
$\left(\frac{\partial{y}}{\partial{P}}\right)_{T}\approx \left(\frac{\partial{y_0}}{\partial{P}}\right)_{T=0}$, 
 $\alpha$ behaves as 
\begin{eqnarray}
\alpha\propto\frac{\tilde{C}_{\rm b}}{t}
\sim t^{\frac{1}{2}}, 
\label{eq:a_3dAF}
\end{eqnarray}
where $\frac{\tilde{C}_{\rm b}}{t}\sim t^{1/2}$ dominates over $C_{\rm a}\sim t({\rm const.}-t^{1/2})$ [Eq.~(\ref{eq:Cv_d3z2})] in Eq.~(\ref{eq:a_SP}). 
This coincides with the temperature dependence of the critical part shown by the RG theory~\cite{Zhu2003}. 
It should be noted however that $\alpha\sim t^{1/2}$ appears at sufficiently low temperatures 
for $t\lsim 10^{-3}$, as shown in Fig.~\ref{fig:a_t_d3z2}(b).
This is due to the temperature dependence of $\left(\frac{\partial{y}}{\partial{P}}\right)_{T}$ in $\alpha_{\rm b}$ in Eq.~(\ref{eq:a_b}), as noted above. 

\subsection{2d Ferromagnetic case}
\label{sec:a_d2z3}

\begin{figure*}
\includegraphics[width=15cm]{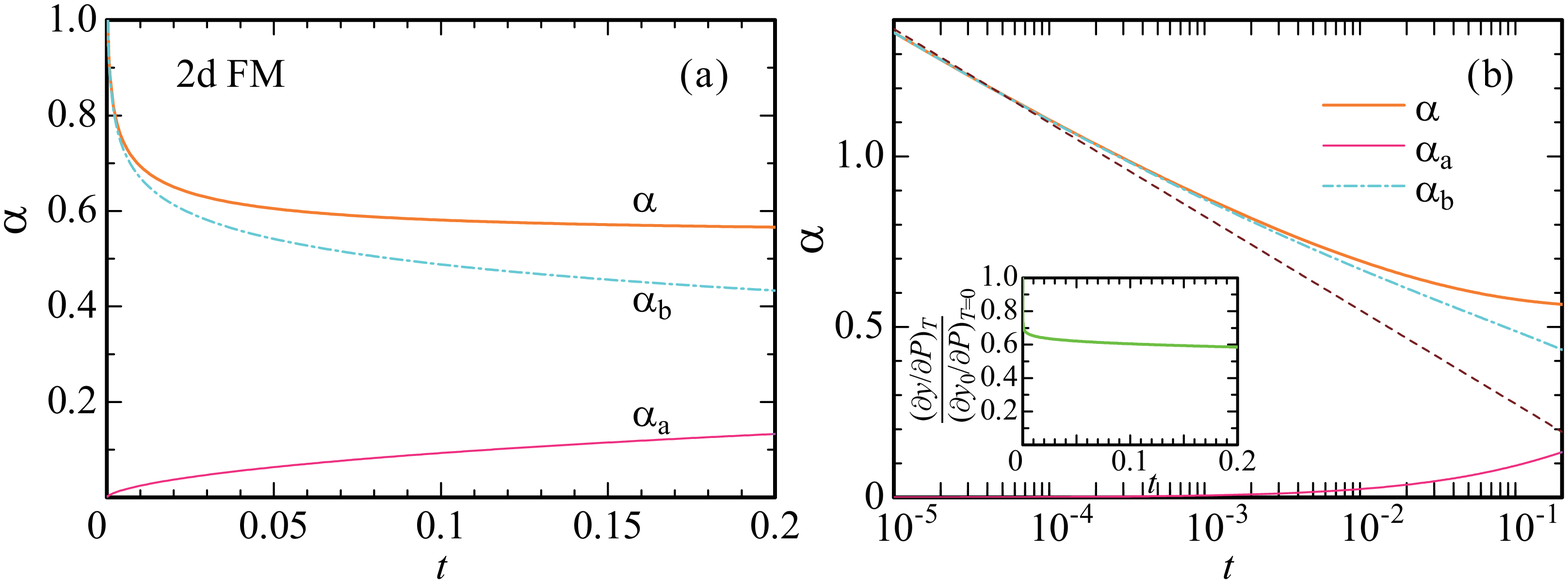}
\caption{(Color online) (a) Temperature dependence of the thermal-expansion coefficient $\alpha$ (thick solid line), $\alpha_{\rm a}$ (thin solid line), and $\alpha_{\rm b}$ (dash-dotted line) at the 2d FM QCP. 
(b) Semi-log plot of (a). The dashed line represents the least-square fit of $\alpha$ with $a\ln{t}$ for $10^{-5}\le t\le 10^{-4}$. The inset shows the $t$ dependence of $\left(\frac{\partial{y}}{\partial{P}}\right)_{T}/\left(\frac{\partial{y_0}}{\partial{P}}\right)_{T=0}$. 
}
\label{fig:a_t_d2z3}
\end{figure*}

Figure~\ref{fig:a_t_d2z3}(a) shows the temperature dependence of 
the thermal-expansion coefficient 
$\alpha$ at the FM QCP $(z=3)$ in $d=2$.
As $t$ decreases, $\alpha_{\rm b}$ in Eq.~(\ref{eq:a_SP}) contributes to $\alpha$ dominantly, 
$\alpha\approx\frac{1}{V}\frac{\tilde{C}_{\rm b}}{t}\left(\frac{\partial{y}}{\partial{P}}\right)_{T}$, while   
contribution from $\alpha_{\rm a}$ becomes not negligible as $t$ increases. 

For $t\ll 1$, we estimate  
$L\sim -t\ln{t}$, $\tilde{C}_{\rm b}\sim -t\ln{t}$, and 
$M\sim t/\ln{t}$ in Eq.~(\ref{eq:dydP}). 
Hence, we have
$\left(\frac{\partial{y}}{\partial{P}}\right)_{T}\approx \left(\frac{\partial{y_0}}{\partial{P}}\right)_{T=0}+b_{3{\rm a}}/\ln{t}+b_{3{\rm b}}t^{1/3}$ with $b_{3{\rm a}}$ and $b_{3{\rm b}}$ being positive constants. 
This can be seen in the inset of Fig.~\ref{fig:a_t_d2z3}(b), where the $t$ dependence of 
$\left(\frac{\partial{y}}{\partial{P}}\right)_{T}/\left(\frac{\partial{y_0}}{\partial{P}}\right)_{T=0}$ is plotted. 
At sufficiently low temperatures where $\left(\frac{\partial{y}}{\partial{P}}\right)_{T}$ can be regarded as a constant, 
$\left(\frac{\partial{y}}{\partial{P}}\right)_{T}\approx \left(\frac{\partial{y_0}}{\partial{P}}\right)_{T=0}$, 
 $\alpha$ behaves as 
\begin{eqnarray}
\alpha\propto\frac{\tilde{C}_{\rm b}}{t}
\sim -\ln{t}, 
\label{eq:a_2dFM}
\end{eqnarray}
where $\frac{\tilde{C}_{\rm b}}{t}\sim -\ln{t}$ dominates over $C_{\rm a}\sim t^{2/3}$ [Eq.~(\ref{eq:Cv_d2z3})] in Eq.~(\ref{eq:a_SP}). 
This coincides with the temperature dependence of the critical part shown by the RG theory~\cite{Zhu2003}. 
It should be noted however that $\alpha\sim -\ln{t}$ appears at sufficiently low temperatures 
for $t\lsim 10^{-4}$, as shown in Fig.~\ref{fig:a_t_d2z3}(b).
This is due to the temperature dependence of $\left(\frac{\partial{y}}{\partial{P}}\right)_{T}$ in $\alpha_{\rm b}$ in Eq.~(\ref{eq:a_b}), as noted above. 
As shown in Fig.~\ref{fig:a_t_d2z3}(a), $\alpha(t)$ shows a slight increase down to $t\sim 10^{-2}$, which is seen as almost flat-$t$ behavior, and divergent-$t$ behavior becomes visible for lower temperatures $t\lsim 10^{-2}$. 

Note that although $\alpha$ diverges for $t\to 0$, the entropy becomes zero, i.e., $S\to 0$, for $t\to 0$, which is confirmed with the use of Eq.~(\ref{eq:S}), satisfying the third law of the thermodynamics.
This can also be seen by integrating $\alpha{V}$ with respect to $P$ [see Eq.~(\ref{eq:a_SP_def})] as\cite{Zhu2003} 
\begin{eqnarray}
S(P,T)=S(P_{\rm c},T)-\int_{P_{\rm c}}^{P^{*}}\alpha{V}dP'
-\int_{P^{*}}^{P}\alpha{V}dP', 
\nonumber
\\
\label{eq:S_P}
\end{eqnarray}
where $P_{\rm c}$ denotes the QCP and $P^{*}$ characterizes the crossover from the quantum-critical to Fermi-liquid regimes. 
Let us consider the case for $P>P^{*}$ that $S(P,T)$ and the last term of the r.h.s. in Eq.~(\ref{eq:S_P}) are in the Fermi-liquid regime, both of which vanish for $t\to 0$. 
Since $y(t=0)$ has a finite slope at $P=P_{\rm c}$, i.e., $\left(\frac{\partial{y}}{\partial{P}}\right)_{T=0}\ne 0$, the crossover line in the $P$-$t$ phase diagram behaves as $P^{*}-P_{\rm c}\sim \left(\frac{\partial{P}}{\partial{y}}\right)_{T=0}t^{2/3}$. 
Here, $t^{2/3}$ is the crossover temperature between the quantum-critical region and the Fermi-liquid region in the $P$-$t$ phase diagram, where $y/t^{2/z}$ much smaller (larger) than 1 for $t\to 0$ gives the quantum-critical (Fermi-liquid) region 
[see Eq.~(\ref{eq:yt2z}) in Appendix~\ref{sec:Ld3}]. 
Then, the second term in the r.h.s. of Eq.~(\ref{eq:S_P}) becomes zero for $t\to 0$ since the integration region vanishes as $P^{*}-P_{\rm c}\sim t^{2/3}\to 0$ over which $\alpha$ is divergent as Eq.~(\ref{eq:a_2dFM}). 
This yields $S(P_{\rm c},T\to 0)=0$. 

\subsection{2d Antiferromagnetic case}
\label{sec:a_d2z2}

\begin{figure*}
\includegraphics[width=15cm]{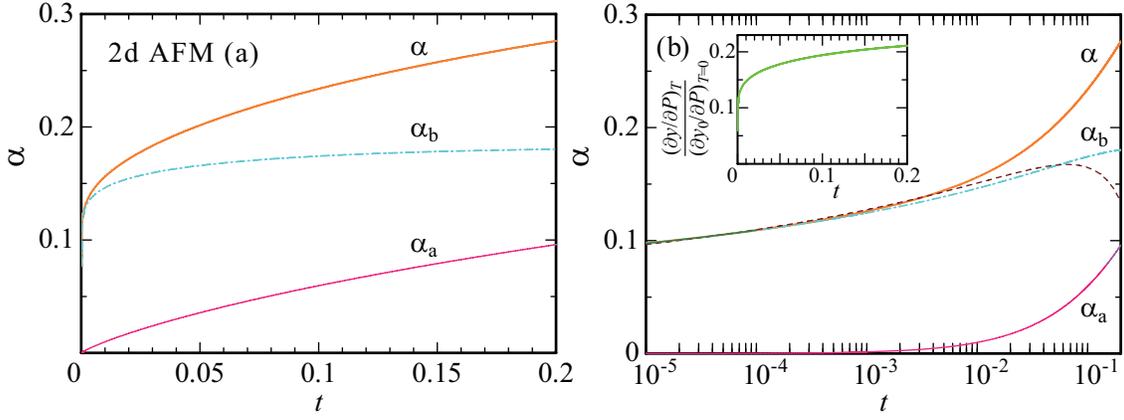}
\caption{(Color online) (a) Temperature dependence of the thermal-expansion coefficient $\alpha$ (thick solid line), $\alpha_{\rm a}$ (thin solid line), and $\alpha_{\rm b}$ (dash-dotted line) at the 2d AFM QCP. 
(b) Semi-log plot of (a). The dashed line represents the least-square fit of $\alpha$ with   $-a\frac{\ln(-\ln{t})}{\ln{t}}$ for $10^{-7}\le t\le 10^{-4}$. The inset shows the $t$ dependence of $\left(\frac{\partial{y}}{\partial{P}}\right)_{T}/\left(\frac{\partial{y_0}}{\partial{P}}\right)_{T=0}$. 
}
\label{fig:a_t_d2z2}
\end{figure*}

Figure~\ref{fig:a_t_d2z2}(a) shows the temperature dependence of 
the thermal-expansion coefficient 
$\alpha$ at the AFM QCP $(z=2)$ in $d=2$.
As $t$ decreases, $\alpha_{\rm b}$ in Eq.~(\ref{eq:a_SP}) contributes to $\alpha$ dominantly, 
$\alpha\approx\frac{1}{V}\frac{\tilde{C}_{\rm b}}{t}\left(\frac{\partial{y}}{\partial{P}}\right)_{T}$, while   
contribution from $\alpha_{\rm a}$ becomes not negligible as $t$ increases. 

In Eq.~(\ref{eq:dydP}) for $d+z=4$, 
we estimate  
$\frac{y_1}{2}(y\ln{y}+2L)=y\approx -\frac{t}{2}\frac{\ln(-\ln{t})}{\ln{t}}$ [see Eq.~(\ref{eq:SCReq3}) and Appendix~D], $\tilde{C}_{\rm b}\sim t\ln(-\ln{t})$ [Eq.~(\ref{eq:Cb_tld_d2z2})], and 
$M\sim t\frac{\ln{t}}{\ln(-\ln{t})}$ for $t\ll 1$. 
Hence, we have
$\left(\frac{\partial{y}}{\partial{P}}\right)_{T}/ \left(\frac{\partial{y_0}}{\partial{P}}\right)_{T=0}\approx
-{b_4}/\ln\left(-\frac{t}{\ln{t}}\right)$ with $b_4$ being a positive constant. 
Note that 
divergence of the denominator of Eq.~(\ref{eq:dydP}) occurs 
because the $\ln{y}$ and $M$ terms diverge for $t\to 0$ 
and then we have  $\lim_{T\to0}\left(\frac{\partial{y}}{\partial{P}}\right)_{T}=0$ irrespective of 
the input values of $\left(\frac{\partial{y_0}}{\partial{P}}\right)_{T=0}$ in Eq.~(\ref{eq:dydP}). 
This can be confirmed by the numerical calculation of Eq.~(\ref{eq:dydP}), which is 
shown in the inset of Fig.~\ref{fig:a_t_d2z2}(b). 

For $t\ll 1$, $\alpha$ behaves as
\begin{eqnarray}
\alpha
\propto\frac{\tilde{C}_{\rm b}}{t}\left(\frac{\partial{y}}{\partial{P}}\right)_{T}
\sim-\frac{\ln(-\ln{t})}{\ln\left(-\frac{t}{\ln{t}}\right)},
\label{eq:a_t_d2z2}
\end{eqnarray}
where $\frac{\tilde{C}_{\rm b}}{t}\left(\frac{\partial{y}}{\partial{P}}\right)_{T}\sim-\frac{\ln(-\ln{t})}{\ln\left(-\frac{t}{\ln{t}}\right)}$ 
dominates over $C_{\rm a}\sim-t\ln{t}$ [Eq.~(\ref{eq:Cv_d2z2})]
in Eq.~(\ref{eq:a_SP}). 
Although Eq.~(\ref{eq:a_t_d2z2}) gives the accurate expression of $\alpha$ for $t\ll 1$, here taking the leading term of the denominator, we plot $\alpha\sim-\frac{\ln(-\ln{t})}{\ln{t}}$ in Fig.~\ref{fig:a_t_d2z2}(b) as a dashed line, which well reproduces $\alpha_{\rm b}$ and also $\alpha$ for $t<10^{-2}$. 
It is noted that 
the low-temperature behavior $\ln(-\ln{t})$ in 
Eq.~(\ref{eq:a_t_d2z2}) coincides with the temperature dependence of the critical part shown by the RG theory~\cite{Zhu2003} except for 
the prefactor $\left(\partial{y}/\partial{P}\right)_{T}$, which gives the $-\ln\left(-\frac{t}{\ln{t}}\right)$ contribution to the denominator of Eq.~(\ref{eq:a_t_d2z2})
~\cite{GarstDrT}
.

The temperature dependence of $\alpha$ for $t\ll 1$ at the QCP for each class 
is summarized in Table~\ref{tb:alpha_QCP}.

\begin{table}
\begin{center}
\begin{tabular}{l|cc} \hline
{class} &{$\alpha$} & {$\Gamma$}  
\\ \hline
3d FM &{$T^{1/3}$} & {$-\frac{T^{-2/3}}{\ln{T}}$} 
\\ 
3d AFM &{$T^{1/2}$} & {$\frac{T^{-1/2}}{{\rm const.}-T^{1/2}}$}  
\\
2d FM &{$-\ln{T}$} & {$-T^{-2/3}\ln{T}$} 
\\  
2d AFM &
{$-\frac{\ln{(-\ln{T})}}{\ln\left(-\frac{T}{\ln{T}}\right)}$}
& $\frac{1}{T\ln{T}}\frac{\ln(-\ln{T})}{\ln\left(-\frac{T}{\ln{T}}\right)}$ 
\\ \hline
\end{tabular}
\end{center}
\caption{Temperature dependence of the thermal-expansion coefficient $\alpha$ and the Gr\"{u}neisen parameter $\Gamma$ just at the QCP for each class specified by $z=3$ (FM) and $z=2$ (AFM) in $d=3$ and $2$.  
}
\label{tb:alpha_QCP}
\end{table}

It is noted that Takahashi derived $\alpha(t)$ from the volume derivative of the free energy in the extended SCR theory by introducing the conservation law of the total spin fluctuation amplitude and discussed the 3d FM case with finite transition temperatures~\cite{Takahashi2006}. The present study has shown that $\alpha(t)$ derived from the volume derivative of the SCR free energy (Sect.~\ref {sec:a_PT_QCP}) is equivalent to $\alpha(t)$ derived from the pressure derivative of the SCR entropy (Sect.~\ref {sec:a_SP_QCP}), On the basis of the latter result, which has a much simpler expression, the critical properties of $\alpha(t)$ at the QCP for each class $(z=3, 2$ in $d=3, 2)$ 
have
 been clarified. 

\section{Numerical results on Gr\"{u}neisen parameter $\Gamma$ at  
the magnetic QCP}
\label{sec:Grn}

The Gr\"{u}neisen parameter $\Gamma$ is defined by Eq.~(\ref{eq:Grn}). On the other hand, $\Gamma$ in the adiabatic process has been derived in Sect.~\ref{sec:Grn_deriv}, whose explicit form is given by Eq.~(\ref{eq:Grn_QCP1}). It can be shown that for $t\to 0$ the former expression is equivalent to the latter one as follows: 

Let us consider Eq.~(\ref{eq:Grn}). 
Near the magnetic QCP, the thermal-expansion coefficient $\alpha$ is given in Eq.~(\ref{eq:a_SP}) 
and the specific heat $C_{V}$ is given in Eq.~(\ref{eq:CvT}). 
As shown in Sect.~\ref{sec:SCR}, for $t\ll 1$, the dominant contribution to $C_{V}$ comes from $C_{\rm a}$ as $C_{V}=C_{\rm a}-C_{\rm b}\approx C_{\rm a}$. 
Hence, $\Gamma$ is expressed for $t\ll 1$ as 
\begin{eqnarray}
\Gamma&\approx&\frac{\tilde{C}_{\rm b}}{C_{\rm a}}\frac{1}{t}\frac{1}{\kappa_{T}}\left(\frac{\partial{y}}{\partial{P}}\right)_{T}+\frac{1}{\kappa_T}\frac{1}{T_0}\left(\frac{\partial{T_0}}{\partial{P}}\right)_{T},
\nonumber
\\
&=&-\frac{\tilde{C}_{\rm b}}{C_{\rm a}}\frac{V}{t}\left(\frac{\partial{y}}{\partial{V}}\right)_{T}
-\frac{V}{T_0}\left(\frac{\partial{T_0}}{\partial{V}}\right)_{T}, 
\label{eq:Grn_lowt}
\end{eqnarray} 
where $\kappa_{T}$ defined in Eq.~(\ref{eq:comp}) has been used to derive the 2nd line. 
Since $T_0$ defined in Eq.~(\ref{eq:T0}) is the quantity at $T=0$ and does not depend on $T$, we have 
$\left(\frac{\partial{T_0}}{\partial{V}}\right)_{T}=\left(\frac{\partial{T_0}}{\partial{V}}\right)_{S}$. 
For sufficiently low temperatures, 
$\left(\frac{\partial{y}}{\partial{V}}\right)$ at a constant $T$ can be approximated as the one at a  
constant $S$, i.e., 
$\left(\frac{\partial{y}}{\partial{V}}\right)_{T}\approx\left(\frac{\partial{y}}{\partial{V}}\right)_{S}$. 
Then, it is confirmed explicitly for $t\ll 1$ that Eq.~(\ref{eq:Grn_lowt}) coincides with Eq.~(\ref{eq:Grn_QCP1}).  

Here, we remark the property of the isothermal compressibility $\kappa_{T}$ defined by Eq.~(\ref{eq:comp}). 
By differentiating the pressure $P$ in Eq.~(\ref{eq:dFdV}) with respect to the volume $V$ under a constant temperature $T$, $\left(\frac{\partial{P}}{\partial{V}}\right)_{T}$ can be calculated as discussed in Sect.~\ref{sec:a_PT_QCP}. 
It can be shown that the $t\to 0$ limit of $\left(\frac{\partial{P}}{\partial{V}}\right)_{T}$ is  finite but not zero at the QCP for each class specified by $z=3$ and $2$ in $d=3$ and $2$.   
Hence, the $t\to 0$ limit of the isothermal compressibility is finite at the QCP, i.e., $\lim_{t\to 0}\kappa_{T}={\rm const.}$

In the following subsections, the temperature dependence of $\Gamma$ at the QCP for each class will be analyzed on the basis of Eq.~(\ref{eq:Grn}). 
The specific heat $C_{V}$ in Eq.~(\ref{eq:CvT}) is calculated by the procedure in Sect.~\ref{sec:SCR} and the thermal-expansion coefficient 
$\alpha$ in Eq.~(\ref{eq:a_SP}) is calculated by the procedure in Sect.~\ref{sec:a}. 
With the use of Eq.~(\ref{eq:a_SP}), $\Gamma$ is expressed as 
\begin{eqnarray}
\Gamma=\Gamma_{\rm a}+\Gamma_{\rm b},
\label{eq:Grn_ab}
\end{eqnarray}
where $\Gamma_{\rm i}$ is defined by 
\begin{eqnarray}
\Gamma_{\rm i}\equiv\frac{\alpha_{\rm i}{V}}{C_{V}\kappa_{T}}
\label{eq:Grn_i}
\end{eqnarray}
for ${\rm i}={\rm a}, {\rm b}$. 
For $t\ll 1$, $\Gamma_{\rm a}$ and $\Gamma_{\rm b}$ lead to the 2nd and 1st terms in Eq.~(\ref{eq:Grn_lowt}), respectively. 
To plot the $t$ dependence of $\Gamma$, 
here we input $\kappa_{T}=0.1$ as a representative value (the reason for this parametrization is explained below),   
although given the first and second derivatives of $T_0$, $T_A$, $v_{4}$, and  $\eta_0/(Aq_{\rm B}^2)$
with respect to $V$, the temperature dependence of $\kappa_{T}$ can be computed explicitly.  
Other input parameters are the same as those set in Fig.~\ref{fig:Ct_t2dFM} and Fig.~\ref{fig:a_t_d3z3}. 

Here, we comment on the parametrization. The Gr\"{u}neisen parameter $\Gamma_{\rm a}(T=0)=-\frac{V}{T_0}\left(\frac{\partial{T_0}}{\partial{V}}\right)_{T=0}$  in heavy electron systems  is estimated to be in the same order of $-\frac{V}{T_{\rm K}}\left(\frac{\partial{T_{\rm K}}}{\partial{V}}\right)_{T=0}$, with $T_{\rm K}$ being the characteristic temperature called Kondo temperature, which typically has an enhanced value of $O(10)$, as will be shown in Sect.~\ref{sec:comp_exp} [see Eq.~(\ref{eq:Grn_HF})]. 
Since $\frac{1}{T_0}\left(\frac{\partial{T_0}}{\partial{P}}\right)_{T}=\Gamma_{\rm a}(T=0)\kappa_{T}=1.0$ was used in Sect.~\ref{sec:a} as a typical input value, here we input $\kappa_{T}=0.1$ as a typical value for the heavy electron system, giving rise to $\Gamma_{\rm a}(T=0)=10.0$. 

If one makes a comparison with the system with the normal metal where the Gr\"{u}neisen parameter is not enhanced in the Fermi-liquid region but has the value of $O(1)$ (e.g., d electron systems) , $\kappa_{T}$ of $O(1)$ is to be input for  $\frac{1}{T_0}\left(\frac{\partial{T_0}}{\partial{P}}\right)_{T}=1.0$, which gives $\Gamma_{\rm a}(T=0)$ of $O(1)$. Hence the vertical scales of the following Figs.~\ref {fig:G_t_d3z3}, \ref{fig:G_t_d3z2}, \ref{fig:G_t_d2z3}, and \ref{fig:G_t_d2z2} are an order of magnitude smaller in that case. 

To make more explicit comparison with experiments, the bulk modulus $\kappa_{T}^{-1}$ in the unit of GPa for each material is multiplied to $\frac{1}{T_0}\left(\frac{\partial{T_0}}{\partial{P}}\right)_{T}$ in the unit of GPa$^{-1}$ [see Eqs.~(\ref{eq:a_a_unit}) and (\ref{eq:a_b_unit})], giving rise to the dimensionless $\Gamma_{\rm a}$. 
These values can actually be determined from the measurements, which will be discussed in Sect.~\ref {sec:P_Ce}. 

\subsection{3d Ferromagnetic case}

\begin{figure*}
\includegraphics[width=15cm]{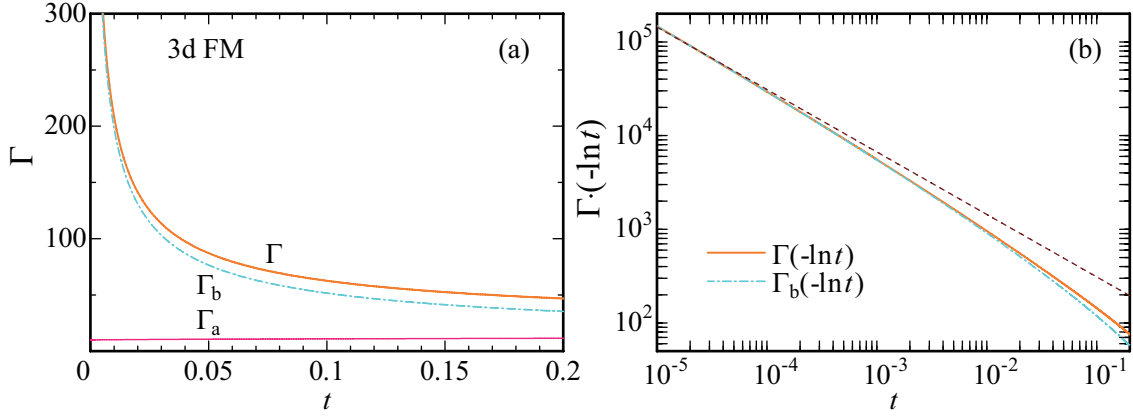}
\caption{(Color online) 
(a) Temperature dependence of the Gr\"{u}neisen parameter $\Gamma$ (thick solid line), 
$\Gamma_{\rm a}$ (thin solid line), and $\Gamma_{\rm b}$ (dash-dotted line) at the 3d 
FM
 QCP. 
(b) Temperature dependence of $\Gamma(-\ln{t})$ (thick solid line) and $\Gamma_{\rm b}(-\ln{t})$ (dash-dotted line). 
The dashed line represents the least-square fit of $\Gamma(-\ln{t})$ with $at^{-2/3}$ for $10^{-5}\le t\le 10^{-4}$. 
}
\label{fig:G_t_d3z3}
\end{figure*}

Figure~\ref{fig:G_t_d3z3}(a) shows the temperature dependence of the Gr\"{u}neisen parameter $\Gamma$ at the FM QCP $(z=3)$ in $d=3$. 
As $t$ decreases, $\Gamma$ increases and diverges at the ground state, which is mainly contributed from $\Gamma_{\rm b}$. 
For $t\ll 1$, the thermal-expansion coefficient is evaluated as $\alpha\approx\alpha_{\rm b}\sim t^{1/3}$ in Eq.~(\ref{eq:a_3dFM}) and the specific heat is evaluated as $C_{\rm V}\approx C_{\rm a}\sim-t\ln{t}$ in Eq.~(\ref{eq:Cv_d3z3}). Then, $\Gamma$ is evaluated as 
\begin{eqnarray}
\Gamma\approx\Gamma_{\rm b}\sim-\frac{t^{-\frac{2}{3}}}{\ln{t}} 
\label{eq:G_3dFM}
\end{eqnarray}
for $t\ll 1$. 
This is numerically confirmed in Fig.~\ref{fig:G_t_d3z3}(b), where $\Gamma(t)(-\ln{t})$ (thick solid line) behaves 
as $\sim t^{-2/3}$ (dashed line) for $t\ll 1$. 
The behavior of Eq.~(\ref{eq:G_3dFM}) is in accord with the RG theory~\cite{Zhu2003}, which appears at sufficiently low temperatures for $t\lsim 10^{-4}$, as shown in Fig.~\ref{fig:G_t_d3z3}(b). 

In the Curie-Weiss regime (see Fig.~\ref{fig:y_t}(a)), $\Gamma(t)$ shows a monotonic decrease as $t$ increases. The least-square fit of $1/\Gamma(t)$  in the form of $at^\gamma$ for $0.07\le t \le 0.20$ gives $\gamma=0.43$. Hence, $\Gamma(t)$ behaves as $\Gamma(t)\sim t^{-0.43}$ in the Curie-Weiss regime.

\subsection{3d Antiferromagnetic case}

\begin{figure*}
\includegraphics[width=15cm]{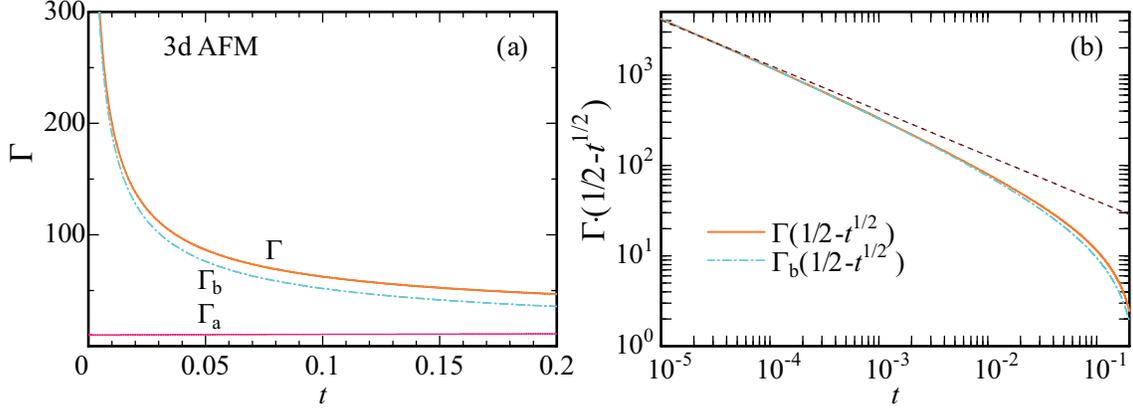}
\caption{(Color online) (a) Temperature dependence of the Gr\"{u}neisen parameter $\Gamma$ (thick solid line), 
$\Gamma_{\rm a}$ (thin solid line), and $\Gamma_{\rm b}$ (dash-dotted line) at the 3d AFM QCP.  (b) Temperature dependence of $\Gamma(\frac{1}{2}-t^{1/2})$ (thick solid line) and $\Gamma_{\rm b}(\frac{1}{2}-t^{1/2})$ (dash-dotted line). 
The dashed line represents the least-square fit of $\Gamma(\frac{1}{2}-t^{1/2})$ with $at^{-1/2}$ for $10^{-5}\le t\le 10^{-4}$. 
}
\label{fig:G_t_d3z2}
\end{figure*}

Figure~\ref{fig:G_t_d3z2}(a) shows the temperature dependence of the Gr\"{u}neisen parameter $\Gamma$ at the AFM QCP $(z=2)$ in $d=3$. 
As $t$ decreases, $\Gamma$ increases and diverges at the ground state, which is mainly contributed from $\Gamma_{\rm b}$. 
For $t\ll 1$, the thermal-expansion coefficient is evaluated as $\alpha\approx\alpha_{\rm b}\sim t^{1/2}$ in Eq.~(\ref{eq:a_3dAF}) and the specific heat is evaluated as $C_{\rm V}\approx C_{\rm a}\sim t ({\rm const.}-t^{1/2})$ in Eq.~(\ref{eq:Cv_d3z2}). Then, $\Gamma$ is evaluated as 
\begin{eqnarray}
\Gamma\approx\Gamma_{\rm b}\sim\frac{t^{-\frac{1}{2}}}{{\rm const.}-t^{1/2}} 
\label{eq:G_3dAF}
\end{eqnarray}
for $t\ll 1$. 
This is numerically confirmed in Fig.~\ref{fig:G_t_d3z2}(b), where $\Gamma(t)(\frac{1}{2}-t^{1/2})$ (thick solid line) behaves 
as $\sim t^{-1/2}$ (dashed line) for low $t$. 
The behavior of Eq.~(\ref{eq:G_3dAF}) is in accord with the RG theory~\cite{Zhu2003}, which appears at sufficiently low temperatures for $t\lsim 10^{-4}$, as shown in Fig.~\ref{fig:G_t_d3z2}(b). 

As for the Curie-Weiss regime (see Fig.~\ref{fig:y_t}(b)), the least-square fit of $1/\Gamma(t)$ in the form of $at^\gamma$ in the Curie-Weiss regime for $0.07\le t \le 0.20$ gives $\gamma=0.43$. 
Hence, $\Gamma(t)$ behaves as $\Gamma(t)\sim t^{-0.43}$ in the Curie-Weiss regime. 

\subsection{2d Ferromagnetic case}

\begin{figure*}
\includegraphics[width=15cm]{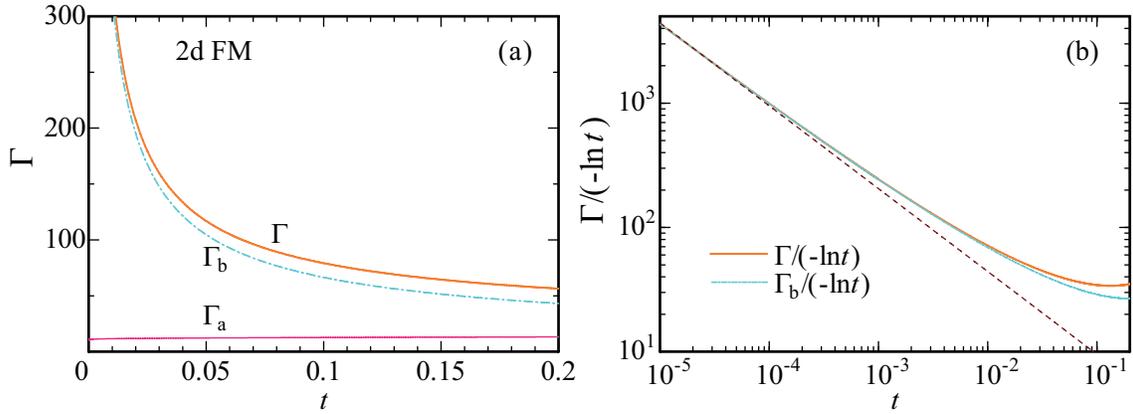}
\caption{(Color online) (a) Temperature dependence of the Gr\"{u}neisen parameter $\Gamma$ (thick solid line), 
$\Gamma_{\rm a}$ (thin solid line), and $\Gamma_{\rm b}$ (dash-dotted line) at the 2d FM QCP. (b) Temperature dependence of $\Gamma/(-\ln{t})$ (thick solid line) and $\Gamma_{\rm b}/(-\ln{t})$ (dash-dotted line). 
The dashed line represents the least-square fit of $\Gamma/(-\ln{t})$ with $at^{-2/3}$ for $10^{-5}\le t\le 10^{-4}$. 
}
\label{fig:G_t_d2z3}
\end{figure*}

Figure~\ref{fig:G_t_d2z3}(a) shows the temperature dependence of the Gr\"{u}neisen parameter $\Gamma$ at the FM QCP $(z=3)$ in $d=2$. 
As $t$ decreases, $\Gamma$ increases and diverges at the ground state, which is mainly contributed from $\Gamma_{\rm b}$.
For $t\ll 1$, the thermal-expansion coefficient is evaluated as $\alpha\approx\alpha_{\rm b}\sim -\ln{t}$ in Eq.~(\ref{eq:a_2dFM}) and the specific heat is evaluated as $C_{\rm V}\approx C_{\rm a}\sim t^{2/3}$ in Eq.~(\ref{eq:Cv_d2z3}). Then, $\Gamma$ is evaluated as 
\begin{eqnarray}
\Gamma\approx\Gamma_{\rm b}\sim-t^{-\frac{2}{3}}\ln{t} 
\label{eq:G_2dFM}
\end{eqnarray}
for $t\ll 1$. 
This is numerically confirmed in Fig.~\ref{fig:G_t_d2z3}(b), where $\Gamma(t)/(-\ln{t})$ (thick solid line) behaves 
as $\sim t^{-2/3}$ (dashed line) for low $t$. 
The behavior of Eq.~(\ref{eq:G_2dFM}) is in accord with the RG theory~\cite{Zhu2003}, which appears at sufficiently low temperatures for $t\lsim 10^{-4}$, as shown in Fig.~\ref{fig:G_t_d2z3}(b). 

As for the Curie-Weiss regime (see Fig.~\ref{fig:y_t}(c)), the least-square fit of $1/\Gamma(t)$ in the form of $at^\gamma$ in the Curie-Weiss regime for $0.07\le t \le 0.20$ gives $\gamma=0.50$. 
Hence, $\Gamma(t)$ behaves as $\Gamma(t)\sim t^{-0.50}$ in the Curie-Weiss regime. 

\subsection{2d Antiferromagnetic case}
\label{sec:G_d2z2}

\begin{figure*}
\includegraphics[width=15cm]{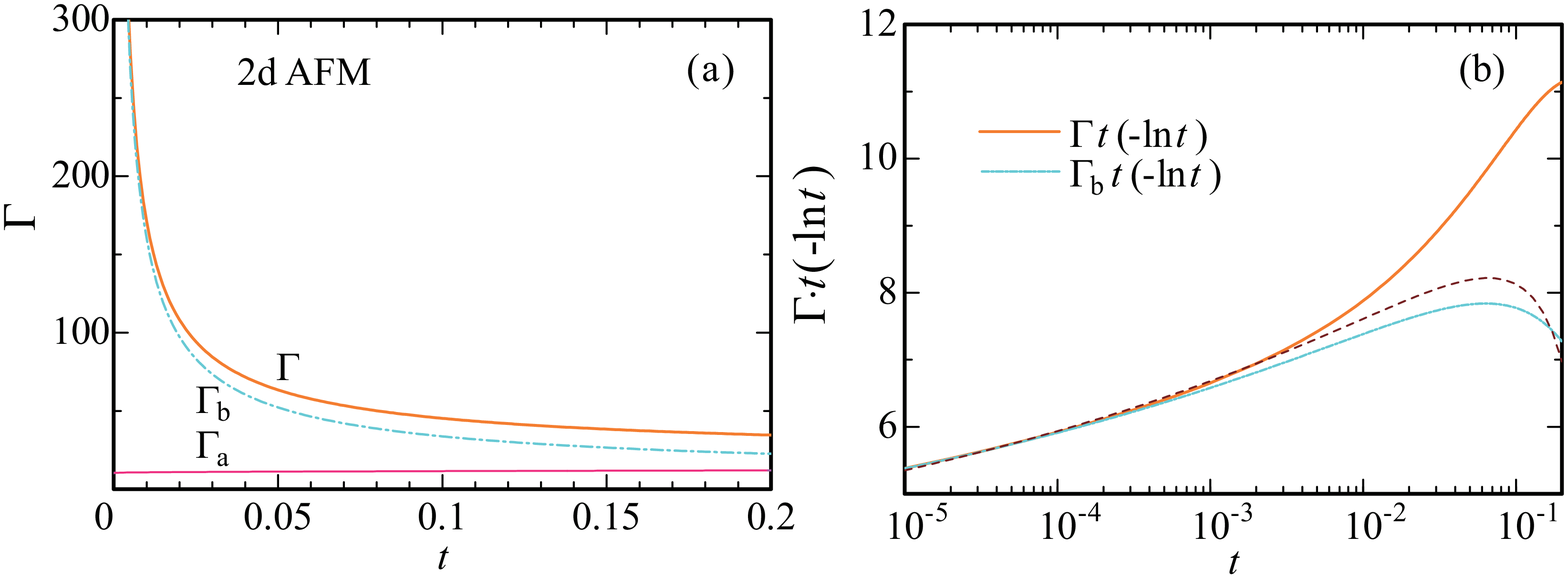}
\caption{(Color online) (a) Temperature dependence of the Gr\"{u}neisen parameter $\Gamma$ (thick solid line), 
$\Gamma_{\rm a}$ (thin solid line), and $\Gamma_{\rm b}$ (dash-dotted line) at the 2d 
AFM
 QCP.  (b) Temperature dependence of $\Gamma{t}(-\ln{t})$ (thick solid line) and $\Gamma_{\rm b}{t}(-\ln{t})$ (dash-dotted line). 
The dashed line represents the least-square fit of $\Gamma{t}(-\ln{t})$ with $
-
a\frac{\ln(-\ln{t})}{\ln\left(-\frac{t}{\ln{t}}\right)}$ for $10^{-5}\le t\le 10^{-4}$. 
}
\label{fig:G_t_d2z2}
\end{figure*}

Figure~\ref{fig:G_t_d2z2}(a) shows the temperature dependence of the Gr\"{u}neisen parameter $\Gamma$ at the AFM QCP $(z=2)$ in $d=2$. 
As $t$ decreases, $\Gamma$ increases and diverges at the ground state, which is mainly contributed from $\Gamma_{\rm b}$.
For $t\ll 1$, the thermal-expansion coefficient is evaluated in Eq.~(\ref{eq:a_t_d2z2}), 
whose precise form is 
$\alpha\approx
\alpha_{\rm b}\sim
-
\frac{\ln(-\ln{T})}{\ln\left(-\frac{T}{\ln{T}}\right)}$. 
The specific heat is evaluated as $C_{\rm V}\approx C_{\rm a}\sim-t\ln{t}$ in Eq.~(\ref{eq:Cv_d2z2}). Then, $\Gamma$ is evaluated as 
\begin{eqnarray}
\Gamma\approx\Gamma_{\rm b}\sim\frac{1}{t\ln{t}}\frac{\ln(-\ln{t})}{\ln\left(-\frac{t}{\ln{t}}\right)}
\label{eq:G_2dAF}
\end{eqnarray}
for $t\ll 1$. 
This is confirmed in Fig.~\ref{fig:G_t_d2z2}(b), where $\Gamma(t)t(-\ln{t})$ (thick solid line) behaves 
as $\sim
-
\frac{\ln(-\ln{t})}{\ln\left(-\frac{t}{\ln{t}}\right)}$ (dashed line) for low $t$. 
The behavior of Eq.~(\ref{eq:G_2dAF}) agrees with the critical part shown by the RG theory~\cite{Zhu2003} except for $\ln\left(-\frac{t}{\ln{t}}\right)$ in the denominator, which arises from the prefactor $(\partial{y}/\partial{P})_T$ in $\alpha_{\rm b}$ as  discussed in Sect.~\ref{sec:a_d2z2}
~\cite{GarstDrT}
. 

As for the Curie-Weiss regime (see Fig.~\ref{fig:y_t}(d)), the least-square fit of $1/\Gamma(t)$ in the form of $at^\gamma$ in the Curie-Weiss regime for $0.07\le t \le 0.20$ gives $\gamma=0.41$. 
Hence, $\Gamma(t)$ behaves as $\Gamma(t)\sim t^{-0.41}$ in the Curie-Weiss regime. 

The temperature dependence of $\Gamma$ for $t\ll 1$ at the QCP for each class 
is summarized in Table~\ref{tb:alpha_QCP}.

\section{Numerical results on Gr\"{u}neisen parameter $\Gamma$ near 
the magnetic QCP}
\label{sec:Grn_y0}

\begin{figure*}
\includegraphics[width=15cm]{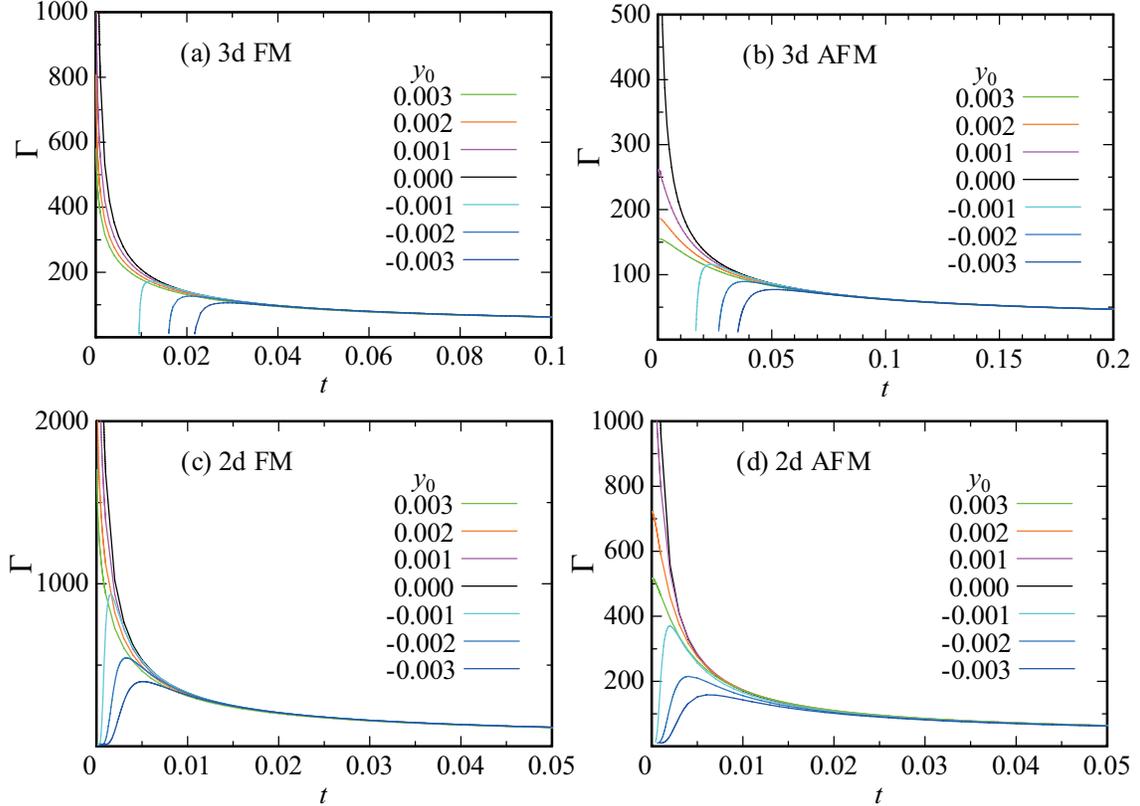}
\caption{
(Color online) Scaled temperature dependence of the Gr\"{u}neisen parameter for (a) 3d FM, (b) 3d AFM, (c) 2d FM, and (d) 2d AFM. 
}
\label{fig:Grn_t_y0}
\end{figure*}

In this section, we discuss the Gr\"{u}neisen parameter near the QCP for each class on the basis of the SCR theory. 
We calculate $\Gamma(t)$ in the paramagnetic phase by solving the SCR equation Eq.~(\ref{eq:SCReq2}) for $d+z>4$ and Eq.~(\ref{eq:SCReq3}) for $d+z=4$ in the paramagnetic region $(y_0>0)$ and also in the region where the magnetic order takes place $(y_0<0)$. The input parameters other than $y_0$ are the same as those in section~\ref{sec:Grn}. The results are shown in Figs.~\ref{fig:Grn_t_y0}(a)-(d) [see corresponding Figs.~\ref{fig:y_t}(a)-(d), respectively]. 

At the QCP specified by $y_0=0$, $\Gamma(t)$ shows the divergence for $t\to 0$ in each class. As getting away from the QCP in the paramagnetic region, $\Gamma(t)$ for $t\to 0$ becomes finite, whose value decreases as $y_0$ increases from 0, as shown in Figs.~\ref{fig:Grn_t_y0}(a)-(d). 

On the other hand, as $y_0$ decreases from 0, the magnetic order occurs at finite temperature  $t=t_{c}\equiv T_{\rm c}/T$ for 3d FM [Fig.~\ref{fig:Grn_t_y0}(a)] and 3d AFM [Fig.~\ref{fig:Grn_t_y0}(b)]. In the high-$t$ regime for $t\gsim 0.7$, $\Gamma(t)$ shows the Curie-Weiss behavior in each class as mentioned above. As $t$ decreases, $\Gamma(t)$ increases and turns to decrease, which finally converges into a finite value  $\Gamma(t)\to\Gamma_{\rm a}(t=t_{\rm c})=-\frac{V}{T_0}\left(\frac{\partial T_0}{\partial V}\right)_{T=T_{\rm c}}=10.0$ for $t\to t_{{\rm c}+}$. 
This is because $\tilde{C}_{\rm b}\to 0$ is realized for $y\to 0$ as $t$ approaches $t_{\rm c}$ from the high-$t$ side, making $\alpha_{\rm b}\propto\frac{\tilde{C}_{\rm b}(t_{\rm c})}{t_{\rm c}}\to 0$ in Eq.~(\ref{eq:a_b}).
Then, by Eq.~(\ref{eq:Grn_ab}), it turns out that the Gr\"{u}neisen parameter at $t_{\rm c}$ is expressed as $\Gamma(t)\to\Gamma_{\rm a}(t)$ for $t\to t_{{\rm c}+}$. 

For 2d FM and 2d AFM, the magnetic order occurs only at $t=t_{\rm c}=0$. The Gr\"{u}neisen parameters for $y_0<0$ in this case also show $\Gamma(t)\to\Gamma_{\rm a}(t)=-\frac{V}{T_0}\left(\frac{\partial T_0}{\partial V}\right)_{T=0}=10.0$ for $t\to 0_{+}$, as seen in Fig.~\ref{fig:Grn_t_y0}(c) and Fig.~\ref{fig:Grn_t_y0}(d), respectively.

\section{Discussion}

\subsection{Divergence of the Gr\"{u}neisen parameter and the characteristic energy scale at the QCP}

In the SCR theory, the characteristic temperature of spin fluctuation $T_0$ is not zero even at the QCP in general, as will be illustrated in Fig.~\ref{fig:T_P_Ce}. In Sect.~\ref{sec:Grn}, it was shown that the Gr\"{u}neisen parameter $\Gamma$ diverges at the magnetic QCP for each class $(z=3, 2$ in $d=3, 2)$. 
The origin of the divergence can be traced back to the entropy of the SCR theory. 
The entropy $S$ is expressed as the scaled form in Eq.~(\ref{eq:S}) with a variable $u$ defined by Eq.~(\ref{eq:u}). 
The volume dependence arises from $y$ in the numerator and  
$T_0$ in the denominator of Eq.~(\ref{eq:u}), which lead to the first and second terms in Eq.~(\ref{eq:Grn_QCP1}), respectively. 
The former gives rise to the divergence of $\Gamma$ for $t\to 0$ at the QCP and the latter gives the $V$ derivative of the characteristic temperature $T_0$ [see Eq.~(\ref{eq:Grn_QCP1}) or Eq.~(\ref{eq:Grn_lowt})]. 
The present study has clarified that 
the inverse susceptibility (renormalized by the mode-mode coupling of spin fluctuations) coupled to $V$ 
gives rise to the divergence of $\Gamma$ in addition to the ordinary contribution from the $V$ derivative of the characteristic temperature of the system. 

The temperature dependence of the dominant term of the thermal-expansion coefficient $\alpha$ and $\Gamma$ for $t\ll 1$ coincides with the critical part shown by the RG theory~\cite{Zhu2003} except for the temperature dependent $\left(\frac{\partial{y}}{\partial{P}}\right)_{T}$ in $\alpha$ and $\Gamma$ for $z=2$ in $d=2$ (see Sect.~\ref{sec:a_d2z2} and Sect.~\ref{sec:G_d2z2}). 
In Ref.~\cite{Zhu2003}, 
the Gr\"{u}neisen parameter defined by $\Gamma\equiv \alpha/C_{P}$ was analyzed and 
the relation $\Gamma\sim T^{-\frac{1}{\nu z}}$ was derived by assuming the hyperscaling relation, which is generally justified only below the upper critical dimension  $(d+z<4)$ within the $\Phi^4$ theory [see Eq.~(\ref{eq:Action})]. 
Here, $\nu$ is the exponent for the correlation length $\xi\sim |r|^{-\nu}$ with $r\equiv(P-P_{\rm c})/P_{\rm c}$. 
The results in Sect.~\ref{sec:Grn} are obtained above $(d+z>4)$ and just at the upper critical dimension $(d+z=4)$. 
For comparison, let us reexpress the specific heat $C_{V}$, $\alpha$, and $\Gamma$ in terms of $d$ and $z$ for each class in the following subsections. 

\subsubsection{3d Ferromagnetic case $(d=3, z=3)$}
For $t\ll 1$, the specific heat [Eq.~(\ref{eq:Cv_d3z3})] is expressed as   
$C_{V}\approx C_{\rm a}\sim -t^{\frac{d}{z}}\ln{t}$. 
The thermal expansion coefficient $\alpha$
[Eq.~(\ref{eq:a_3dFM})] is expressed as 
$\alpha\approx\alpha_{\rm b}\sim t^{\frac{d-2}{z}}$. 
Then, the Gr\"{u}neisen parameter [Eq.~(\ref{eq:G_3dFM})] is expressed as 
\begin{eqnarray}
\Gamma\approx\frac{\alpha}{C_{V}}\sim-\frac{t^{-\frac{2}{z}}}{\ln{t}}.
\label{eq:Grn_d3z3}
\end{eqnarray}
%

\subsubsection{3d Antiferromagnetic case $(d=3, z=2)$}
For $t\ll 1$, the specific heat [Eq.~(\ref{eq:Cv_d3z2})] is expressed as   
$C_{V}\approx C_{\rm a}\sim t^{\frac{d-1}{z}}\left({\rm const.}-t^{\frac{d}{z}-1}\right)$. 
The thermal expansion coefficient $\alpha$
[Eq.~(\ref{eq:a_3dAF})] is expressed as 
$\alpha\approx\alpha_{\rm b}\sim t^{\frac{d-2}{z}}$. 
Then, the Gr\"{u}neisen parameter [Eq.~(\ref{eq:G_3dAF})] is expressed as 
\begin{eqnarray}
\Gamma\approx\frac{\alpha}{C_{V}}\sim\frac{t^{\frac{d-2}{z}}}{t^{\frac{d-1}{z}}\left({\rm const.}-t^{\frac{d}{z}-1}\right)}.
\label{eq:Grn_d3z2}
\end{eqnarray}
%

\subsubsection{2d Ferromagnetic case $(d=2, z=3)$}
For $t\ll 1$, the specific heat [Eq.~(\ref{eq:Cv_d2z3})] is expressed as   
$C_{V}\approx C_{\rm a}\sim t^{\frac{d}{z}}$. 
The thermal expansion coefficient $\alpha$
[Eq.~(\ref{eq:a_2dFM})] is expressed as 
$\alpha\approx\alpha_{\rm b}\sim-\ln{t}$. 
Then, the Gr\"{u}neisen parameter [Eq.~(\ref{eq:G_2dFM})] is expressed as 
\begin{eqnarray}
\Gamma\approx\frac{\alpha}{C_{V}}\sim-\frac{t^{-\frac{d}{z}}}{\ln{t}}.
\label{eq:Grn_d2z3}
\end{eqnarray}
%

\subsubsection{2d Antiferromagnetic case $(d=2, z=2)$}
For $t\ll 1$, the specific heat [Eq.~(\ref{eq:Cv_d2z2})] is expressed as   
$C_{V}\approx C_{\rm a}\sim -t^{\frac{d}{z}}\ln{t}$. 
The thermal expansion coefficient $\alpha$
[Eq.~(\ref{eq:a_t_d2z2})] is expressed precisely as 
$\alpha\approx\alpha_{\rm b}\sim 
-
\ln(-\ln{t})/\ln\left(-\frac{t}{\ln{t}}\right)$. 
Then, the Gr\"{u}neisen parameter [Eq.~(\ref{eq:G_2dAF})] is expressed as 
\begin{eqnarray}
\Gamma\approx\frac{\alpha}{C_{V}}\sim\frac{t^{-\frac{d}{z}}}{\ln{t}}\frac{\ln(-\ln{t})}{\ln\left(-\frac{t}{\ln{t}}\right)}.
\label{eq:Grn_d2z2}
\end{eqnarray}

In Eqs.~(\ref{eq:Grn_d3z3}), (\ref{eq:Grn_d2z3}) and (\ref{eq:Grn_d2z2}), $\Gamma$ has the $t$ dependence as $t^{-2/z}$ with logarithmic corrections. In Eq.~(\ref{eq:Grn_d3z2}), if the first term of the denominator is neglected, $\Gamma$ has the $t$ dependence as $t^{\frac{z-d-1}{z}}$, which is also expressed as $t^{-2/z}$. 
Since the dynamical magnetic susceptibility with the quadratic momentum dependence in Eq.~(\ref{eq:chi}) yields $\nu=1/2$ in the SCR theory, as a result, all these $t$ dependence can be expressed as $t^{-\frac{1}{\nu z}}$ except for the logarithmic corrections.

\subsection{Comparison with the Moriya-Usami theory}

Moriya and Usami discussed the magneto-volume effect in nearly ferromagnetic metals~\cite{MU1980} on the basis of the volume strain $\omega_{\rm m}\equiv\frac{\delta{V}}{V}$ 
expressed as 
\begin{eqnarray}
\omega_{\rm m}(T)-\omega_{\rm m}(T_{\rm c})\propto y
\label{eq:wm}
\end{eqnarray}
for $T>T_{\rm c}$, where $T_{\rm c}$ is the ferromagnetic-transition temperature.   
The volume strain in nearly antiferromagnetic metals was also discussed on the basis of Eq.~(\ref{eq:wm}) where $T_{\rm c}$ is the N{\'e}el temperature~\cite{IM1998}. 
Since the thermal-expansion coefficient $\alpha$ is obtained by $\alpha=\frac{d\omega_{\rm m}}{dT}$ according to its definition in Eq.~(\ref{eq:a1}), Eq.~(\ref{eq:wm}) indicates that $\alpha$ is proportional to $\frac{dy}{dT}$, i.e., $\alpha\propto\frac{dy}{dT}$. 

It should be noted here that Eq.~(\ref{eq:wm}) was {\it not} shown to be derived from the free energy~\cite{MU1980,IM1998,Takahashi2006}. 
In this paper, we have derived the thermal-expansion coefficient $\alpha$ starting from the free energy (or equivalently from the entropy) in the SCR theory with the use of the stationary condition adequately, which results in Eq.~(\ref{eq:a_SP}). 
Then, we have obtained $\alpha\approx\alpha_{\rm b}\sim\frac{\tilde{C}_{\rm b}}{t}\left(\frac{\partial{y}}{\partial{P}}\right)_{T}$ for $t\ll 1$ at the QCP, as shown in Sect.~\ref{sec:a}. 
Hence, let us compare our result with the Moriya-Usami theory at $T_{\rm c}=0$. 

For $d+z>4$, the temperature dependence of $y$ and $\tilde{C}_{\rm b}$ is the same for $t\ll 1$, and  
$\left(\frac{\partial{y}}{\partial{P}}\right)_{T}\sim{\rm const.}$ for $t\to 0$, 
as shown in Sects.~\ref{sec:a_d3z3}, \ref{sec:a_d3z2}, and \ref{sec:a_d2z3}. 
Hence, it turns out that $\frac{\tilde{C}_{\rm b}}{t}\left(\frac{\partial{y}}{\partial{P}}\right)_{T}$ has the same temperature dependence as 
$\frac{dy}{dT}$ for $t\ll 1$. 

For $d+z=4$, the temperature dependence of $y\sim-\frac{t\ln(-\ln{t})}{\ln{t}}$ and $\tilde{C}_{\rm b}\sim t\ln(-\ln{t})$ for $t\ll 1$ are different. 
However, $\left(\frac{\partial{y}}{\partial{P}}\right)_{T}$ has the temperature dependence as 
$\left(\frac{\partial{y}}{\partial{P}}\right)_{T}\sim-\left[\ln\left(-\frac{t}{\ln{t}}\right)\right]^{-1}$ for $t\ll 1$, as shown in Sect.~\ref{sec:a_d2z2}. 
Hence, $\frac{\tilde{C}_{\rm b}}{t}\left(\frac{\partial{y}}{\partial{P}}\right)_T$ is expressed as $\sim-\frac{\ln(-\ln{t})}{\ln\left(-\frac{t}{\ln{t}}\right)}$. This still looks different from $\frac{dy}{dT}\sim-\frac{\ln(-\ln{t})}{\ln{t}}$. However, at low temperatures $\frac{\tilde{C}_{\rm b}}{t}\left(\frac{\partial{y}}{\partial{P}}\right)_T$ can be approximated as $\sim-\frac{\ln(-\ln{t})}{\ln{t}}$, which was confirmed numerically as shown in Fig.~\ref{fig:a_t_d2z2}(b). 
Thus, it can be regarded that in practice, $\frac{\tilde{C}_{\rm b}}{t}\left(\frac{\partial{y}}{\partial{P}}\right)_{T}$ 
has the same temperature dependence as $\frac{dy}{dT}\sim-\frac{\ln(-\ln{t})}{\ln{t}}$ as far as the leading term is concerned. 

Hence, as a consequence, our result in Eq.~(\ref{eq:a_SP}) and Moriya-Usami's $\alpha\propto\frac{dy}{dT}$ give the same 
(practically the same)
 $t$ dependence of $\alpha$ for $t\ll 1$ for $d+z>4$ 
$(d+z=4)$.  
This can be seen immediately by comparing $\frac{dy}{dT}\propto\frac{d\eta}{dT}$ in Table~\ref{tb:mag_QCP} with $\alpha(t)$ in Table~\ref{tb:alpha_QCP} for each class.  
Note that behavior in Table~\ref{tb:alpha_QCP} appears at sufficiently low temperatures as shown in Figs.~\ref{fig:a_t_d3z3}, \ref{fig:a_t_d3z2}, \ref{fig:a_t_d2z3}, and \ref{fig:a_t_d2z2}, and hence $\alpha\propto\frac{dy}{dT}$ does not hold at temperatures except for $t\ll 1$. 

\subsection{Comparison with experiments}
\label{sec:comp_exp}

The thermal-expansion coefficient is generally expressed as 
\begin{eqnarray}
\alpha=\alpha_{\rm e}+\alpha_{\rm ph}+\alpha_{\rm mag}, 
\label{eq:a_general}
\end{eqnarray}
where $\alpha_{\rm e}$ and $\alpha_{\rm ph}$ are contributions from itinerant electrons
and acoustic phonons, respectively. 
At low temperatures, 
$\alpha_{\rm e}$ behaves as $\alpha_{\rm e}=aT$, as shown by the free-electron model, 
and $\alpha_{\rm ph}$ is given by  $\alpha_{\rm ph}=\beta{T^3}$~\cite{Aschcroft}. 
In Eq.~(\ref{eq:a_general}), $\alpha_{\rm mag}$ arises from spin fluctuations, which become profound near the continuous magnetic-transition point, as discussed in Sect.~\ref{sec:alpha} (note that $\alpha_{\rm mag}$ was denoted as $\alpha$ in Sect.~\ref{sec:alpha}).  

The Gr\"{u}neisen parameter is generally expressed as
\begin{eqnarray}
\Gamma=\Gamma_{\rm e}+\Gamma_{\rm ph}+\Gamma_{\rm mag}, 
\label{eq:Grn_general}
\end{eqnarray}
where 
$\Gamma_{\rm i}$ is defined by Eq.~(\ref{eq:Grn_i}) with i$=$e, ph, and mag, 
corresponding to each term in Eq.~(\ref{eq:a_general}). 
For sufficiently lower temperatures than the Fermi temperature, 
$\Gamma_{\rm e}$ is given as a constant~\cite{Aschcroft}. 
In heavy electron systems, the characteristic temperature of the quasiparticles is the effective Fermi temperature, which is referred to as the Kondo temperature $T_{\rm K}$ in the lattice system. Then, the characteristic temperature $T^{*}$ in Eq.~(\ref{eq:Grn_TV}) is set to be $T_{\rm K}$, which leads to~\cite{UmeoPRB1996,Flouquet2005,Goltsev2005}   
\begin{eqnarray}
\Gamma_{\rm e}=-\frac{V}{T_{\rm K}}\left(\frac{\partial{T_{\rm K}}}{\partial{V}}\right)_{S}. 
\label{eq:Grn_TK}
\end{eqnarray}
To grasp the main property, let us take the view from the strong limit of onsite Coulomb repulsion of f electrons. 
By inputting $T_{\rm K}=De^{-\frac{1}{N_{\rm cF}J}}$ to Eq.~(\ref{eq:Grn_TK}), where $D$ and $N_{\rm cF}$ are the half band width and the density of states at the Fermi level of conduction electrons per ``spin", respectively, and $J$ is the effective Kondo exchange coupling $(J>0)$ in the lattice system~\cite{Ono1989,KLM1996}, we obtain for $N_{\rm cF}J\ll 1$ 
\begin{eqnarray}
\Gamma_{\rm e}\approx-\frac{1}{N_{\rm cF}J}\left\{\frac{V}{J}\left(\frac{\partial{J}}{\partial{V}}\right)_{S}+c\right\},  
\label{eq:Grn_HF}
\end{eqnarray}
where $c$ is a constant of $O(1)$ (e.g., $c=2/3$ for free conduction electrons in $d=3$). 
In heavy electron systems, $(JN_{\rm cF})^{-1}$ typically has a magnitude of $O(10)$. 
Thus we see that the factor $(N_{\rm cF}J)^{-1}$ gives rise to the enhancement of $|\Gamma_{\rm e}|$, which is often observed in the heavy electron metals with about 10-100 times larger values than those of ordinary metals~\cite{deVisser1989,UmeoPRB1996,Flouquet2005,Thompson1994}.  

When the system approaches the continuous magnetic-transition point by varying parameters, e.g., by applying pressure or magnetic field, or chemical doping, $\Gamma_{\rm mag}$ arising from spin fluctuations becomes predominant in Eq.~(\ref{eq:Grn_general}), as discussed in Sect.~\ref{sec:Grn} (note that $\Gamma_{\rm mag}$ was denoted as $\Gamma$ in Sect.~\ref{sec:Grn}). 
In the following subsections, 
let us discuss the pressure tuning to the magnetic QCP in the Ce- and Yb-based 
heavy electron
systems.  

\subsubsection{Pressure effects on Ce-based systems}
\label{sec:P_Ce}

In the Ce-based heavy electron systems, by applying pressure, the hybridization between f and conduction electrons $|V_{\rm fc}|$ increases and the f level $\varepsilon_{\rm f}$ increases in general. 
Hence, the Kondo coupling $J\sim\frac{V_{\rm fc}^2}{\varepsilon_{\rm F}-\varepsilon_{\rm f}}$ increases, giving rise to increase in the Kondo temperature $T_{\rm K}$. 
This yields $\left(\frac{\partial{J}}{\partial{V}}\right)_{S}<0$, which leads to $\Gamma_{\rm e}>0$ in Eq.~(\ref{eq:Grn_HF})~\cite{note_dJdV}. 

Namely, the system becomes more itinerant under pressure, which makes the characteristic temperature of spin fluctuation $T_0$ increase. 
Direct evaluation of $T_0$ in Eq.~(\ref{eq:T0}) gives $T_0=\tilde{A}v_{\rm F}\tilde{q}_{\rm B}/(\pi^2n^{2/3})$ in $d=3$, where $v_{\rm F}$ is the Fermi velocity and $n$ is the filling defined by $n\equiv\frac{N_{\rm e}}{2N}$. Here, $\tilde{q}_{\rm B}$ is given by $\tilde{q}_{\rm B}=q_{\rm B}$ for $z=3$ and $\tilde{q}_{\rm B}=Q$ for $z=2$, and $\tilde{A}$ is a dimensionless constant defined by the $q^2$ coefficient around the ordered vector $\bf Q$ in the irreducible susceptibility at $\omega=0$ (e.g., $\tilde{A}=\frac{1}{12}$ for the free electron model~\cite{Hertz}). 
Since $v_{\rm F}\sim T_{\rm K}$ holds, applying pressure induces increase in $v_{\rm F}$ 
reflecting the pressure-induced expansion of the effective band width of the quasiparticles. 
This effect contributes to increase in $T_0$ under pressure, i.e.,  
$\left(\frac{\partial{T_0}}{\partial{P}}\right)>0$. 

Indeed, 
in Ce$_7$Ni$_3$, it was observed that $T_0$ increases as pressure increases~\cite{Umeo1996}. 
Moreover, smooth variation of $T_0$ and $T_{\rm K}$ observed under pressure is also 
understandable, since $T_{0}(P)$ can get close to $T_{\rm K}(P)$ according to the parameters of $\tilde{A}$ and $n$.
The plot of $T_{0}(P)$ as well as $T_{\rm K}(P)$ determined from the measurements of the specific heat and resistivity in Ce$_7$Ni$_3$ under pressure~\cite{Umeo1996} enables us to estimate $\frac{1}{T_{\rm K}}\left(\frac{\partial{T_{\rm K}}}{\partial{P}}\right)=4.0$~GPa$^{-1}$, which is comparable to  $\frac{1}{T_0}\left(\frac{\partial{T_0}}{\partial{P}}\right)$~\cite{UmeoPRB1996}. The bulk modulus is observed as $\kappa_{T}^{-1}=24. 6$~GPa at room temperature. The Gr\"{u}neisen parameter is estimated to be $\Gamma\approx 100$~\cite{UmeoPRB1996}. 

\begin{figure}
\includegraphics[width=6.5cm]{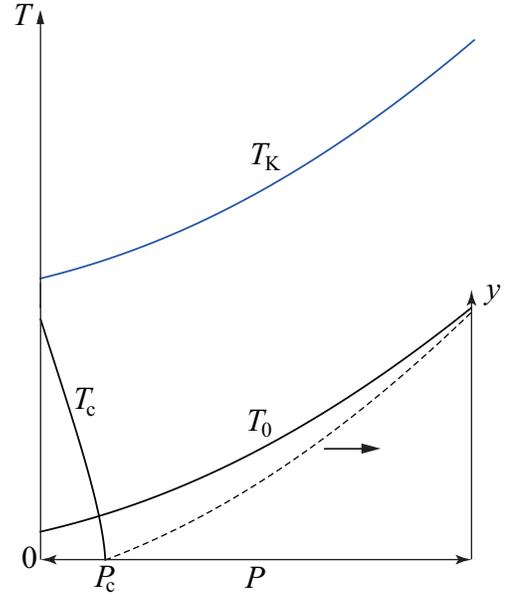}
\caption{(Color online) 
Schematic 
temperature-pressure phase diagram of Ce-based heavy-electron systems (right-pointing $P$ axis) and Yb-based 
heavy-electron 
systems (left-pointing $P$ axis) in $d=3$. 
The Kondo temperature $T_{\rm K}$ and the characteristic temperature of spin fluctuation $T_0$  are given by the solid lines (left axis). 
Note that vertical scales for $T_{\rm K}$ and $T_{0}$ can be different in general. 
Depending on the parameters in each material, it is possible that $T_{0}(P)$ gets close to $T_{\rm K}(P)$ (see text). 
The magnetic transition temperature $T_{\rm c}$ (solid line, left axis) is suppressed to 0 at the QCP denoted by $P_{\rm c}$.   
The dashed line gives $y=1/[2T_{A}\chi_{\bf Q}(0,0)]$ (right axis) for $P>P_{\rm c}$.  
Note that in $d=2$ for $z=2$ the dashed line starts to appear from $P_{\rm c}$ with a vanishing slope  $\left(\frac{\partial{y}}{\partial{P}}\right)_{T=0}=0$ (see Sect.~\ref{sec:a_d2z2}). 
Note that it is possible that a crossing of $T_{\rm K}(P)$ and $T_{\rm c}(P)$ occurs depending on the material parameters. 
 If the system has XY or Heisenberg symmetry, no magnetic transition occurs for finite $T$ in $d=2$. 
}
\label{fig:T_P_Ce}
\end{figure}

Figure~\ref{fig:T_P_Ce} with the right-pointing $P$ axis illustrates the $T$-$P$ phase diagram of the Ce-based 
heavy electron
 systems. 
As $P$ increases, the magnetic transition temperature $T_{\rm c}$ is suppressed to be absolute zero at the QCP denoted by $P_{\rm c}$. 
At $P=P_{\rm c}$, the magnetic susceptibility $\chi_{\bf Q}(0,0)$ diverges with $y=0$ [see Eq.~(\ref{eq:chi})]. 
Since $y$ increases as $P$ increases from $P_{\rm c}$ as shown by 
the
 dashed line, $\left(\frac{\partial{y}}{\partial{P}}\right)_{T=0}>0$ holds 
(For the AFM QCP $(z=2)$ in $d=2$, $y$ starts to appear with zero slope $\left(\frac{\partial{y}}{\partial{P}}\right)_{T=0}=0$ at $P_{\rm c}$, as discussed in Sect.~\ref {sec:a_d2z2}). 
Then, from Eq.~(\ref{eq:a_b}) and resultant Eq.~(\ref{eq:a_SP}), the positive thermal expansion coefficient appears $\alpha_{\rm mag}>0$ for $P>P_{\rm c}$ at low temperatures. 

This is understandable from the $P$ dependence of the entropy $S$. 
When the QCP is approached from the paramagnetic side for $P>P_{\rm c}$,  
$S/T\approx C_{V}/T$ (see Table~\ref{tb:mag_QCP}) increases toward $P_{\rm c}$ at the infinitesimal temperature. This gives $\left(\frac{\partial{S}}{\partial{P}}\right)_{T}<0$ for $P>P_{\rm c}$, leading to  $\alpha_{\rm mag}>0$ by Eq.~(\ref{eq:a_SP_def}) and hence $\Gamma_{\rm mag}>0$. 

On the other hand, when $P$ further decreases from $P_{\rm c}$, the continuous transition to the magnetically ordered phase makes $S/T$ decrease continuously for $P<P_{\rm c}$ at the infinitesimal $T$. This gives $\left(\frac{\partial{S}}{\partial{P}}\right)_{T}>0$ for $P<P_{\rm c}$,  leading to  $\alpha_{\rm mag}<0$ by Eq.~(\ref{eq:a_SP_def}) and hence $\Gamma_{\rm mag}<0$. 
Namely, the sign change of $\alpha_{\rm mag}$ and $\Gamma_{\rm mag}$ occurs at $P_{\rm c}$ as a consequence of the entropy accumulation near the QCP~\cite{Zhu2003,Garst2005}. 
It is noted that the sign change of $\alpha_{\rm mag}$ was shown in Ref.~\cite{Moriya}~\cite{noteMU1985} and the sign change of $\Gamma_{\rm mag}$ as well was discussed in Ref.~\cite{Garst2005}. 

\subsubsection{Pressure effects on Yb-based systems}

On the other hand, in the Yb-based 
heavy electron
systems, the electronic state with 4f$^{13}$ configuration for Yb$^{3+}$ is understood on the basis of the hole picture. 
Hence, the f-hole level $\varepsilon_{\rm f}$ decreases as pressure increases in the Yb-based systems. 
In case this effect outweighs increase in the f-c hybridization, 
the Kondo coupling $J\sim\frac{V_{\rm fc}^2}{\varepsilon_{\rm F}-\varepsilon_{\rm f}}$ decreases, giving rise to decrease in 
the Kondo temperature $T_{\rm K}$ under pressure. 
This yields $\left(\frac{\partial{J}}{\partial{V}}\right)_{S}>0$, which leads to $\Gamma_{\rm e}<0$ in Eq.~(\ref{eq:Grn_HF}) and hence the negative volume expansion $\alpha_{\rm e}<0$ in Eq.~(\ref{eq:a_general}).   

Namely, pressure induces the system where f electrons becomes more localized. 
When $T_{\rm K}$ decreases under pressure, decrease in the f-hole level is more effective than increase in the f-c hybridization, which makes $v_{\rm F}$ decrease. 
This effect contributes to the decrease in $T_0$ under pressure, i.e., $\left(\frac{\partial{T_0}}{\partial{P}}\right)<0$. 
This yields negative thermal-expansion coefficient and Gr\"{u}neisen parameter 
$\alpha_{\rm a}<0$ in Eq.~(\ref{eq:a_a}) and $\Gamma_{\rm a}<0$ in Eq.~(\ref{eq:Grn_i}). 

Figure~\ref{fig:T_P_Ce} with the left-pointing $P$ axis illustrates the $T$-$P$ phase diagram of the Yb-based 
heavy electron
 systems, where 
by applying pressure to the paramagnetic metal phase, the magnetic transition occurs at $T_{\rm c}$ starting from the QCP 
denoted
 by $P_{\rm c}$. 
As $P$ approaches $P_{\rm c}$,
the magnetic susceptibility $\chi_{\bf Q}(0,0)\propto y^{-1}$ increases and diverges at $P_{\rm c}$ for $T=0$. Hence, $\left(\frac{\partial{y}}{\partial{P}}\right)_{T}<0$ holds for $P<P_{\rm c}$, as shown by the dashed line. 
Thus, from Eq.~(\ref{eq:a_b}), $\alpha_{\rm b}<0$ appears, which results in the negative thermal expansion coefficient $\alpha_{\rm mag}<0$ and hence $\Gamma_{\rm mag}<0$ for $P<P_{\rm c}$ at low temperatures. 

On the other hand, when $P$ further increases from $P_{\rm c}$, the continuous transition to the magnetically ordered phase makes $S/T$ decrease continuously for $P>P_{\rm c}$ at the infinitesimal $T$. This gives $\left(\frac{\partial{S}}{\partial{P}}\right)_{T}<0$ for $P>P_{\rm c}$,  leading to  $\alpha_{\rm mag}>0$ by Eq.~(\ref{eq:a_SP_def}) and hence $\Gamma_{\rm mag}>0$. 
Namely, sign change of $\alpha_{\rm mag}$ and $\Gamma_{\rm mag}$ occurs at $P_{\rm c}$ due to the entropy accumulation near the QCP.

\subsection{Observation of $\alpha$ and $\Gamma$ near the magnetic QCP}

To detect 
the thermal expansion coefficient $\alpha$ and the Gr\"{u}neisen parameter $\Gamma$ near the QCP, experimental efforts have been devoted~\cite{Kambe1997,Kuchler2007,Steppke2013,Gegenwart2016}. 
So far, a few data have been reported to exhibit the quantum criticality shown in Table~\ref{tb:mag_QCP} and Table~\ref{tb:alpha_QCP} in the stoichiometric compounds. 
To access the QCP, 
chemical doping has often been performed for high accuracy measurement of $\alpha$ at ambient pressure.
However, the chemical doping 
 more or less brings about effects of disorder, which sometimes masks true critical behaviors expected in clean systems. 
In this subsection, keeping this aspect in mind, experimental data to be compared with the criticality in Table~\ref{tb:mag_QCP} and Table~\ref{tb:alpha_QCP} are discussed. 

In CeNi$_2$Ge$_2$, the specific heat $C_{4{\rm f}}/T\sim\gamma_0-a_{C}T^{1/2}$ and resistivity $\rho\sim T^{
n_{\rho}
}$ $(1.2\le 
n_{\rho}
\le 1.5)$ are observed in the low $T$ region at ambient pressure, suggesting close proximity to the 3d AFM QCP (see Table~\ref{tb:mag_QCP})~\cite{Kuchler2003}. 
The measured thermal expansion coefficient $\alpha=a_{\alpha}\sqrt{T}+b_{\alpha}T$ is in accordance with $\alpha_{\rm mag}\sim T^{1/2}$ in Eq.~(\ref{eq:a_3dAF}) and $\alpha_{\rm e}\sim T$ in Eq.~(\ref{eq:a_general}). 
The Gr\"{u}neisen parameter $\Gamma\approx 57$ at $T=5$~K is already enhanced reflecting the contribution from $\Gamma_{\rm a}$ in Eq.~(\ref{eq:Grn_ab}) and the heavy-electron background as noted around Eq.~(\ref{eq:Grn_HF}). Further enhancement of $\Gamma$ for lowering $T$ is observed as $\Gamma\approx 98\pm 10$ at $T\approx 0.1$~K, which suggests the contribution from $\Gamma_{\rm b}$  as shown in Eq.~(\ref{eq:G_3dAF}) (see Fig.~\ref{fig:G_t_d3z2}).  

As for the sign change of the thermal expansion coefficient, 
$\alpha_{\rm mag}<0$ in the AFM phase for $x<x_{\rm c}$ and $\alpha_{\rm mag}>0$ in the paramagnetic phase for $x>x_{\rm c}$ were observed in CeIn$_{3-x}$Sn$_x$ with $x_{\rm c}=0.67\pm0.03$~\cite{Kuchler2006} and in CeRhIn$_{5-x}$Sn$_x$ with $x_{\rm c}=0.48$~\cite{Donath2009}. 

Since the systematic study of $T_0(P)$ and $T_{\rm K}(P)$ has already been performed in Ce$_7$Ni$_3$~\cite{Umeo1996,UmeoPRB1996}, the measurements of $\alpha(T)$ and $\Gamma(T)$ at the QCP specified by $P_{\rm c}=0.39$~GPa and their analyses based on Eq.~(\ref{eq:a_SP}) and Eq.~(\ref{eq:Grn_ab}) are an interesting subject for future studies. 

Furthermore, 
experimental observation of the quantum criticality shown in Table~\ref{tb:mag_QCP} and Table~\ref{tb:alpha_QCP} for each class is also greatly desired. 
Near the FM QCP, $\left(\frac{\partial{y}}{\partial{P}}\right)_T$ can be directly observed by measuring the pressure dependence of the uniform susceptibility since $\chi_{\bf 0}(0,0)^{-1}\propto y$ holds. 
Near the AFM QCP, by measuring the pressure dependence of the NMR relaxation rate $(T_{1}T)^{-1}$ or resistivity $\rho(T)$ at low temperatures, $\left(\frac{\partial{y}}{\partial{P}}\right)_T$ can be extracted~\cite{Moriya,MU2003}.  
Observation of $T_0(P)$ as well as $T_{\rm K}(P)$ and evaluation of $\frac{1}{T_0}\left(\frac{\partial{T_0}}{\partial{P}}\right)_{T}$ as done in Ce$_7$Ni$_3$ and $\left(\frac{\partial{y}}{\partial{P}}\right)_{T}$ enables us to  make the complete analysis of $\alpha(T)$ and $\Gamma(T)$ at the QCP on the basis of Eq.~(\ref{eq:a_SP}) and Eq.~(\ref{eq:Grn_ab}). Such measurements are much to be desired.

\section{Summary
}

The properties of the thermal-expansion coefficient $\alpha$ and the Gr\"{u}neisen parameter $\Gamma$ near the magnetic QCP in itinerant electron systems have been discussed on the basis of the SCR theory in this paper. 

By taking into account the zero-point as well as thermal spin fluctuation, we have calculated the specific heat $C_{V}$ at the magnetic QCP by considering the stationary condition of the SCR theory correctly.  
For each class of the FM QCP $(z=3)$ and AFM QCP $(z=2)$  in $d=3$ and $2$,  
$C_{V}$ was shown to be expressed as $C_{V}=C_{\rm a}-C_{\rm b}$, where $C_{\rm a}$ is dominant for $t\ll 1$. 
The criticality of $C_{\rm a}$ reproduces the results obtained by the past SCR theory, which was endorsed by the RG theory. 

Then, we have derived the thermal-expansion coefficient $\alpha$ starting from the expression of the entropy in the SCR theory, which has been proven to be equivalent to $\alpha$ derived from the expression of the free energy in the SCR theory. 
The result shows that $\alpha$ is expressed as $\alpha=\alpha_{\rm a}+\alpha_{\rm b}$ with $\alpha_{\rm a}=\frac{1}{V}\frac{C_{\rm a}}{T_0}\left(\frac{\partial{T_0}}{\partial{P}}\right)_T$ and 
$\alpha_{\rm b}=\frac{1}{V}\frac{\tilde{C}_{\rm b}}{t}\left(\frac{\partial{y}}{\partial{P}}\right)_T$  
where $\alpha_{\rm b}$ is dominant for $t\ll 1$. 
An important result is that $\alpha_{\rm b}$ contains the temperature-dependent $\left(\frac{\partial{y}}{\partial{P}}\right)_T$, which contributes to the crossover from the quantum-critical to Curie-Weiss regimes for each universality class and even affects the critical behavior for $t\ll 1$ in the case of upper critical dimension, i.e., $z=2$ in $d=2$. 

On the basis of these correctly calculated $C_{V}$ and $\alpha$, we have derived the Gr\"{u}neisen parameter $\Gamma$. 
The results show that $\Gamma$ is expressed as $\Gamma=\Gamma_{\rm a}+\Gamma_{\rm b}$, where $\Gamma_{\rm a}$ and $\Gamma_{\rm b}$ contain 
$\alpha_{\rm a}$ and $\alpha_{\rm b}$, respectively. 
For $t\ll 1$, $\Gamma_{\rm a}$ is given by $\Gamma_{\rm a}=-\frac{V}{T_0}\left(\frac{\partial{T_0}}{\partial{V}}\right)_T$,  
which has an enhanced value of typically $O(10)$ in the heavy electron systems. 
A remarkable result is that for $t\ll1$, $\Gamma_{\rm b}$ is expressed as   
$\Gamma_{\rm b}=-\frac{\tilde{C}_{\rm b}}{C_{\rm a}}\frac{V}{t}\left(\frac{\partial{y}}{\partial{V}}\right)_{T}$, which diverges at the QCP for each universality class. 
This result shows that 
the inverse susceptibility (renormalized by the mode-mode coupling of spin fluctuations) coupled to $V$ 
gives rise to the divergence of the Gr\"{u}neisen parameter even though the characteristic energy scale $T_0$ remains finite at the QCP. 

The obtained results give the complete expressions of $\alpha$ and $\Gamma$, which consist of not only the critical part but also non-critical part with their coefficients as well as the temperature dependences. 
The temperature dependences of $\alpha_{\rm b}$ and $\Gamma_{\rm b}$ for $t\ll 1$ coincide with the critical parts shown by the RG theory for each universality class except for the case $z=2$ in $d=2$, where the temperature dependent $\left(\frac{\partial{y}}{\partial{P}}\right)_T$ affects the criticality. 
The complete expressions of $\alpha$ and $\Gamma$ clarify their whole temperature dependences from the quantum-critical regime to the Curie-Weiss regime, and are useful for comparison with experiments. 
The temperature dependence of $\alpha$ coincides with the Moriya-Usami theory for $t\ll 1$ where $\alpha\propto \frac{dy}{dt}$ holds 
for $d+z>4$ and approximately holds for $d+z=4$. 

Our study has made it possible to evaluate the temperature dependence of the Gr\"{u}neisen parameter in the Curie-Weiss regime. The results are $\Gamma(T)\sim T^{-0.43}$ for the $d=3$ FM and AFM QCPs, $\Gamma(T)\sim T^{-0.50}$ for the $d=2$ FM QCP, and $\Gamma(T)\sim T^{-0.41}$ for the $d=2$ AFM QCP. These results are also useful for comparison with experiments.

In the heavy electron systems, 
the Gr\"{u}neisen parameter in the Fermi-liquid regime is shown to be enhanced by a factor of $(JN_{\rm cF})^{-1}\gg 1$, where $J$ is the Kondo coupling and $N_{\rm cF}$ is the density of states of conduction electrons at the Fermi level. 
When the QCP is approached, further enhancement caused by spin fluctuation is added to the heavy-electron background, and $\Gamma$ eventually diverges at the QCP. 

The characteristic temperature of spin fluctuation is shown to be proportional to the Kondo temperature in the lattice system, $T_0\propto T_{\rm K}$. 
At sufficiently low temperatures, $\alpha>0$ and $\Gamma>0$ appear in the paramagnetic phase for 
$P>P_{\rm c}$, while $\alpha<0$ and $\Gamma<0$ appear in the magnetically-ordered phase for $P<P_{\rm c}$ in the Ce-based heavy electron systems. 
On the other hand, $\alpha<0$ and $\Gamma<0$ appear in the paramagnetic phase for 
$P<P_{\rm c}$, while $\alpha>0$ and $\Gamma>0$ appear in the magnetically-ordered phase for $P>P_{\rm c}$ in the Yb-based heavy electron  systems. 

 
\begin{acknowledgments}
We thank K. Umeo for informative discussions on the experimental data of Ce$_7$Ni$_3$.
This work was supported by JSPS KAKENHI Grant Numbers JP24540378, JP25400369, 
JP15K05177, JP16H01077, and JP17K05555. 
\end{acknowledgments}


\appendix

\section{Gr\"{u}neisen parameter in Fermi liquid at low temperatures}
\label{sec:lowT_free}

In this appendix, it is shown that 
the Gr\"{u}neisen parameter for free electrons in the isotropic three dimensional system is easily derived from the specific heat at low temperatures. 

At low temperatures, the specific heat at a constant volume is given by
\begin{eqnarray}
C_{V}=Nk_{\rm B}\frac{\pi^2}{2}\frac{T}{T_{\rm F}}
=Nk_{\rm B}\frac{T}{T^*}, 
\label{eq:Cv_free}
\end{eqnarray}
where $T^{*}$ is defined as   
$T^*\equiv\frac{2}{\pi^2}T_{\rm F}$. 
Then, the entropy $S$ is given by 
\begin{eqnarray}
S=\int_{0}^{T}\frac{C_V}{T}dT=Nk_{\rm B}\frac{T}{T^*}.
\label{eq:S_free_lowT}
\end{eqnarray}
Note that the Fermi temperature $T_{\rm F}$ is expressed as
$
T_{\rm F}=\frac{\varepsilon_{\rm F}}{k_{\rm B}}
$
with the Fermi energy 
$\varepsilon_{\rm F}\equiv\frac{\hbar^2k_{\rm F}^2}{2m}$, 
where $m$ and $k_{\rm F}$ are 
mass of an electron and the Fermi wave number 
$k_{\rm F}=\left(3\pi^2\frac{N_{\rm e}}{V}\right)^{1/3}$, respectively. 
Then, by differentiating Eq.~(\ref{eq:S_free_lowT}) with respect to the volume $V$ under a constant entropy $S$, we obtain
\begin{eqnarray}
\left(
\frac{\partial{T^*}}{\partial{V}}
\right)_{S}
=-\frac{2}{3}\frac{T^*}{V}.
\label{eq:dTdV_free}
\end{eqnarray}
Hence, by Eq.~(\ref{eq:Grn_TV2}), the Gr\"{u}neisen parameter $\Gamma$ is obtained as
\begin{eqnarray}
\Gamma=\frac{2}{3}, 
\label{eq:Grn_free_1b}
\end{eqnarray}
which reproduces the result of 
the free-electron model~\cite{Aschcroft}. 

\section{Quantum criticality in $d=3$}
\label{sec:Ld3}

In this appendix, it is explained that the quantum criticality at the QCP in $d=3$ 
is given by Eq.~(\ref{eq:y_t}) 
by analyzing the solution of the SCR equation [Eq.~(\ref{eq:SCReq2})] at low temperatures~\cite{Moriya_text}. 
The $x$ integral in the r.h.s. of Eq.~(\ref{eq:SCReq2}) is defined by Eq.~(\ref{eq:L_def}) as 
\begin{eqnarray}
L\equiv\int_{0}^{x_{\rm c}}dxx^{d+z-3}
\left\{
{\ln}u-\frac{1}{2u}-\psi(u)
\right\}, 
\label{eq:L}
\end{eqnarray}
where $d$ is the spatial dimension and $z$ is the dynamical exponent.
By changing the integral variable as $x'=x/t^{\frac{1}{z}}$, Eq.~(\ref{eq:L}) is expressed as 
\begin{eqnarray}
L=t^{\frac{d+z-2}{z}}
\int_{0}^{\frac{x_{\rm c}}{t^{\frac{1}{z}}}}dx'(x')^{d+z-3}
\left\{
{\ln}u-\frac{1}{2u}-\psi(u)
\right\}, 
\label{eq:L2}
\end{eqnarray}
where $u$ is given by
\begin{eqnarray}
u=(x')^{z-2}\left\{\frac{y}{t^{\frac{2}{z}}}+(x')^2\right\}.
\end{eqnarray}
We see that at low temperatures $t\ll 1$ for  
\begin{eqnarray}
\frac{y}{t^{\frac{2}{z}}}\to 0, 
\label{eq:yt2z}
\end{eqnarray}
the $x'$ integral in $d=3$ converges in Eq.~(\ref{eq:L2}) where the upper bound of the integral is set to be $\infty$. 
Hence, the $t$ dependence of $L$ is evaluated as
\begin{eqnarray}
L\propto t^{\frac{z+1}{z}}.  
\end{eqnarray}
Then, from the SCR equation [Eq.~(\ref{eq:SCReq2})], the following solution 
\begin{eqnarray}
y\propto t^{\frac{z+1}{z}}
\end{eqnarray}
is immediately obtained 
at the QCP where $y_0=0.0$ is set in Eq.~(\ref{eq:SCReq2}). 
It is confirmed that 
this result satisfies the condition of Eq.~(\ref{eq:yt2z}), i.e., 
$y/t^{\frac{2}{z}}\ll 1$, 
for $t\ll 1$. 

\section{Solution of SCR equation for $z=3$ in $d=2$}
\label{sec:d2z3}

The derivation of the solution of the SCR equation [Eq.~(\ref{eq:SCReq2})] for $z=3$ in $d=2$ is shown in this appendix~\cite{HM1995}. 

By using the approximation formula 
\begin{eqnarray}
{\ln}u-\frac{1}{2u}-\psi(u)\approx
\frac{1}{2u(1+6u)}, 
\label{eq:approx}
\end{eqnarray}
in Eq.~(\ref{eq:L}), the $x$ integration can be performed and 
the leading terms are evaluated as 
\begin{eqnarray}
L\approx
\frac{t}{4}
\ln
\left[
\frac{1}{y}\left(\frac{t}{6}\right)^{\frac{2}{3}}
\right]   
\end{eqnarray}
for $y \ll t^{\frac{2}{3}} \ll 1$. 
Then, the solution of the SCR equation [Eq.~(\ref{eq:SCReq2})] $y=y_0+y_{1}L$ at the QCP with $y_0=0.0$ is obtained as follows. 
\begin{eqnarray}
y=-\frac{y_1}{12}t\ln{t}. 
\end{eqnarray}
Since the correlation length $\xi$ is given by $y\propto\xi^{-2}$,  
this coincides with the result of the RG theory for $z=3$ 
in $d=2$ in Ref.~\cite{Millis}.

\section{Equivalence of SCR solution and renormalization group for $z=2$ in $d=2$}
\label{sec:d2z2}

In this appendix, it is shown that the solution of the SCR equation [Eq.~(\ref{eq:SCReq3})] for $z=2$ in $d=2$ coincides with the result of the RG theory by Millis~\cite{Millis}. 

By using the approximation formula Eq.~(\ref{eq:approx})
%
%
in the SCR equation [Eq.~(\ref{eq:SCReq3})], the $x$ integration can be performed as 
\begin{eqnarray}
y=y_0+\frac{y_1}{2}\left(
y{\ln}y+\frac{t}{2}
\left\{
{\ln}\frac{x_{\rm c}^2+y}{y}
-
{\ln}\frac{x_{\rm c}^2+y+\frac{t}{6}}{y+\frac{t}{6}}
\right\}
\right). 
\nonumber
\\
\end{eqnarray}
At the QCP with $y_0=0.0$, the leading terms are evaluated as  
\begin{eqnarray}
-{\ln}2y\approx\frac{t}{2y}{\ln}\frac{t}{2y},
\label{eq:d2z2A} 
\end{eqnarray}
for $y\ll t\ll 1$. Then, one finds that 
\begin{eqnarray}
y=-t\frac{{\ln}(-{\ln}t)}{2{\ln}t}
\end{eqnarray}
is the solution of Eq.~(\ref{eq:d2z2A}) up to the order of $\ln{t}$, which coincides with 
the result of the RG theory for $z=2$ in $d=2$ in Ref.~\cite{Millis}. 

\section{Derivation of Eq.~(\ref{eq:dFdV})}
\label{sec:Pressure}

In this appendix, the derivation of Eq.~(\ref{eq:dFdV}) is explained.

By differentiating the free energy in the SCR theory [Eq.~(\ref{eq:freeE})] with respect to $V$ under a constant temperature, we obtain 
\begin{widetext}
\begin{eqnarray}
\left(\frac{\partial\tilde{F}}{\partial V}\right)_{T}
&=& \frac{1}{\pi}
\sum_{q}\int_{0}^{\omega_{\rm c}}d\omega
\left[
\frac{\partial}{\partial\Gamma_{q}}\left(\frac{\Gamma_{q}}{\omega^2+\Gamma_{q}^2}\right)
\right]_{T}
\left(\frac{\partial\Gamma_{q}}{\partial V}\right)_{T}
\nonumber
\\
& &\times\left[\frac{\omega}{2}+T{\ln}\left(1-{\rm e}^{-\frac{\omega}{T}}\right)\right]
-\frac{1}{2N_{\rm F}}\left(\frac{\partial\eta}{\partial V}\right)_{T}\langle\varphi^2\rangle_{\rm eff}
\nonumber
\\
&+&\left[
\frac{\eta_0-\eta}{2N_{\rm F}}+\frac{6v_4}{N}\langle\varphi^2\rangle_{\rm eff}
\right]\left(\frac{\partial\langle\varphi^2\rangle_{\rm eff}}{\partial V}\right)_{T}
\nonumber
\\
&+&\left\{
\left(\frac{\partial\eta_0}{\partial V}\right)_{T}\frac{1}{2N_{\rm F}}
+(\eta_0-\eta)\left[\frac{\partial}{\partial V}\left(\frac{1}{2N_{\rm F}}\right)\right]_{T}
\right\}
\langle\varphi^2\rangle_{\rm eff}
\nonumber
\\
&+&\frac{3}{N}\left(\frac{\partial v_4}{\partial V}\right)_{T}\langle\varphi^2\rangle_{\rm eff}^2. 
\label{eq:dFdV_deriv}
\end{eqnarray}
\end{widetext}
Since the second term of Eq.~(\ref {eq:dGdV}) is expressed as $(\partial\Gamma_{q}/\partial y)_{T}(\partial y/\partial V)_{T}$, by substituting  
\begin{eqnarray}
\left(\frac{\partial\eta}{\partial{V}}\right)_{T}=Aq_{\rm B}^2\left(\frac{\partial{y}}{\partial{V}}\right)_{T}+y\left[\frac{\partial(Aq_{\rm B}^2)}{\partial{V}}\right]_{T} 
\label{eq:deta_dV}
\end{eqnarray}
into the second term of Eq.~(\ref{eq:dFdV_deriv}), the first and second terms of Eq.~ (\ref{eq:dFdV_deriv}) are expressed as 
\begin{widetext}
\begin{eqnarray}
& &\frac{1}{T_0}\left(\frac{\partial T_0}{\partial V}\right)_{T}I
\nonumber
\\
& &
+
\left\{
\frac{1}{\pi}
\sum_{q}\int_{0}^{\omega_{\rm c}}d\omega
\left[
\frac{\partial}{\partial\Gamma_{q}}\left(\frac{\Gamma_{q}}{\omega^2+\Gamma_{q}^2}\right)
\right]_{T}
\left(\frac{\partial\Gamma_{q}}{\partial y}\right)_{T}
\left[\frac{\omega}{2}+T{\ln}\left(1-{\rm e}^{-\frac{\omega}{T}}\right)\right]
-T_{A} \langle\varphi^2\rangle_{\rm eff}
\right\}
\left(\frac{\partial y}{\partial V}\right)_{T}
\nonumber
\\
& &
-y\left[\frac{\partial(Aq_{\rm B}^2)}{\partial{V}}\right]_{T}\frac{1}{2N_{\rm F}}\langle\varphi^2\rangle_{\rm eff},   
\label{eq:F1and2terms}
\end{eqnarray}
\end{widetext}
where $I$ is defined by Eq.~(\ref{eq:I_integral}) and the definition of $T_A$ [Eq.~(\ref{eq:TA})] has been used. 
Here, we note that the $\{\cdots\}$ part vanishes because of the stationary condition $(\partial\tilde{F}/\partial y)_{T}=0$ applied to  
\begin{eqnarray}
\left(\frac{\partial\tilde{F}}{\partial V}\right)_{T}=\left(\frac{\partial y}{\partial V}\right)_{T}\left(\frac{\partial\tilde{F}}{\partial y}\right)_{T}+\cdots. 
\end{eqnarray}
This implies that the coefficient multiplied to $(\partial y/\partial V)_{T}$ in the calculation of $(\partial\tilde{F}/\partial V)_{T}$ vanishes, which is nothing but the $\{\cdots\}$ part in Eq.~(\ref{eq:F1and2terms}). 
This can also be directly confirmed by noting the fact that the term in the second line of Eq.~ (\ref{eq:F1and2terms}) equals to $(\partial F_{\rm eff}/\partial y)_{T}$, which is expressed as
\begin{eqnarray}
\left(\frac{\partial F_{\rm eff}}{\partial y}\right)_{T}&=&Aq_{\rm B}^2\frac{T}{2}\sum_{q}\sum_{l}\frac{1}{\eta+Aq^2+C_{q}|\omega_{l}|}
\nonumber
\\
&=&T_{A}\langle\varphi^2\rangle_{\rm eff}. 
\label{eq:dFeff_dy}
\end{eqnarray}
Here, $F_{\rm eff}$ was defined by Eq.~(\ref{eq:F_eff}) and the definition of $\langle\varphi^2\rangle_{\rm eff}$ [Eq.~(\ref{eq:phi2_def})] has been used to derive the last line.
Then, the last term inside of $\{\cdots\}$ in Eq.~(\ref{eq:F1and2terms}) is subtracted from Eq.~(\ref{eq:dFeff_dy}), which results in zero.  

On the third line of Eq.~(\ref {eq:dFdV_deriv}), the $[ \cdots ]$ part multiplied to $(\partial\langle\varphi^2\rangle_{\rm eff}/\partial V)_{T}$ vanishes because of the SCR equation [Eq.~(\ref{eq:SCReq})]. 

On the fourth line of Eq.~(\ref {eq:dFdV_deriv}), the $\{\cdots\}$ part is expressed as 
\begin{widetext}
\begin{eqnarray}
\left(\frac{\partial\eta_0}{\partial V}\right)_{T}\frac{1}{2N_{\rm F}}
+(\eta_0-\eta)\left[\frac{\partial}{\partial V}\left(\frac{1}{2N_{\rm F}}\right)\right]_{T}
\nonumber
\\
=
\left[\frac{\partial}{\partial{V}}\left(\frac{\eta_0}{Aq_{\rm B}^2}\right)\right]_{T}
T_{A}
+\left(\frac{\eta_0}{Aq_{\rm B}^2}-y\right)\left(\frac{\partial{T_A}}{\partial{V}}\right)_{T}
+y\left[\frac{\partial(Aq_{\rm B}^2)}{\partial{V}}\right]_{T}\frac{1}{2N_{\rm F}}.  
\label{eq:terms3} 
\end{eqnarray}
\end{widetext}
Then, it turns out that the contribution from the last term is cancelled by the last term of Eq.~(\ref {eq:F1and2terms}). 

Eventually, the remaining terms are the first term of Eq.~ (\ref {eq:F1and2terms}), the contributions from the first and second terms in the r.h.s. of Eq.~(\ref{eq:terms3}), and the last term of Eq.~(\ref{eq:dFdV_deriv}), which result in Eq.~ (\ref{eq:dFdV}). 

\section{Derivation of the last three terms in Eq.~(\ref{eq:ak2})}
\label{sec:last3}

In this appendix, the last three terms in Eq.~(\ref{eq:ak2}) are derived from the last three terms in Eq.~(\ref{eq:ak}). 

The last three terms in Eq.~(\ref{eq:ak}) are calculated by using the SCR equation [Eq.~(\ref{eq:SCReq})], as follows:
\begin{widetext}
\begin{eqnarray}
-\left.
\frac{\partial}{\partial{T}}
\left\{
\left[\frac{\partial}{\partial{V}}\left(\frac{\eta_0}{Aq_{\rm B}^2}\right)\right]_{T}
T_{A}\langle\varphi^2\rangle_{\rm eff}
\right\}
\right|_{V}
&=&
-\left(\frac{\partial{y}}{\partial{t}}\right)_{V}
\frac{N}{6v_4}\frac{T_{A}^2}{T_0}
\left[\frac{\partial}{\partial{V}}\left(\frac{\eta_0}{Aq_{\rm B}^2}\right)\right]_{T},
\label{eq:dV_eta0}
\\
-\left.\frac{\partial}{\partial{T}}
\left[
\left(\frac{\eta_{0}}{Aq_{\rm B}^2}-y\right)
\left(\frac{\partial{T_A}}{\partial{V}}\right)_{T}
\langle\varphi^2\rangle_{\rm eff}
\right]
\right|_{V}
&=&\left(\frac{\partial{y}}{\partial{t}}\right)_{V}
\frac{2}{T_0}
\langle\varphi^2\rangle_{\rm eff}
\left(\frac{\partial{T_A}}{\partial{V}}\right)_{T},
\label{eq:dV_NF}
\\
-\left.
\frac{\partial}{\partial{T}}
\left[
\frac{3}{N}\left(\frac{\partial{v_4}}{\partial{V}}\right)_{T}\langle\varphi^2\rangle_{\rm eff}^2
\right]
\right|_{V}
&=&
-
\left(\frac{\partial{y}}{\partial{t}}\right)_{V}
\frac{T_A}{T_0}\langle\varphi^2\rangle_{\rm eff}
\frac{1}{v_4}\left(\frac{\partial{v_4}}{\partial{V}}\right)_{T}. 
\label{eq:dV_v4}
\end{eqnarray}
\end{widetext}
These are the last three terms in Eq.~(\ref{eq:ak2}), respectively.   

\section{Derivation of Eq.~(\ref{eq:Izero_1_res})}
\label{sec:Izero_log}

The derivation of Eq.~(\ref{eq:Izero_1_res}) is shown in this appendix.

The pressure dependence of $C_1$ and $C_2$ appears via the characteristic temperature $T_0$, 
as seen in Eqs.~(\ref{eq:C1}) and (\ref{eq:C2}), respectively. 
Hence, differentiation of $C_1$ and $C_2$ with respect to the pressure $P$ under a constant temperature gives 
\begin{eqnarray}
\left(\frac{\partial{C_1}}{\partial{P}}\right)_{T}
&=&-\frac{2}{T_0}\left(\frac{\partial{T_0}}{\partial{P}}\right)_{T}
\int_{0}^{x_{\rm c}}dxx^{d+z-3}\frac{\omega_{{\rm c}T_0}^2}
{\omega_{{\rm c}T_0}^2+x^{2z}}, 
\nonumber
\\
& &
\label{eq:C1P}
\\
\left(\frac{\partial{C_2}}{\partial{P}}\right)_{T}
&=&-\frac{4}{T_0}\left(\frac{\partial{T_0}}{\partial{P}}\right)_{T}
\int_{0}^{x_{\rm c}}dxx^{d+z-5}\frac{\omega_{{\rm c}T_0}^2x^{2z}}
{\left(\omega_{{\rm c}T_0}^2+x^{2z}\right)^2}, 
\nonumber
\\
& &
\label{eq:C2P}
\end{eqnarray}
respectively. 

Near the QCP, the second term in $\left\{\cdots\right\}$ of Eq.~(\ref{eq:dIdT_zero}) 
can be expanded around $y=0$ as
\begin{eqnarray}
\int_{0}^{x_{\rm c}}dxx^{d+z-3}\frac{\omega_{{\rm c}T}^2}{\omega_{{\rm c}T}^2+u^2}
=
\int_{0}^{x_{\rm c}}dxx^{d+z-3}\frac{\omega_{{\rm c}T_0}^2}
{\omega_{{\rm c}T_0}^2+x^{2z}}
& &
\nonumber
\\
-2\int_{0}^{x_{\rm c}}dxx^{d+z-5}\frac{\omega_{{\rm c}T_0}^2x^{2z}}
{\left(\omega_{{\rm c}T_0}^2+x^{2z}\right)^2}
y+\cdots
& &
\nonumber
\\
& &
\label{eq:Izero_1}
\end{eqnarray}
By substituting Eqs.~(\ref{eq:C1P}) and (\ref{eq:C2P}) into the first and second terms 
in the r.h.s. of Eq.~(\ref{eq:Izero_1}), respectively, we obtain
\begin{eqnarray}
-\frac{1}{T_0}\left(\frac{\partial{T_0}}{\partial{P}}\right)_{T}
\int_{0}^{x_{\rm c}}dxx^{d+z-3}\frac{\omega_{{\rm c}T}^2}{\omega_{{\rm c}T}^2+u^2}
\nonumber
\\
=\frac{1}{2}
\left\{
\left(\frac{\partial{C_1}}{\partial{P}}\right)_{T}
-\left(\frac{\partial{C_2}}{\partial{P}}\right)_{T}y
\right\}+\cdots,
\label{eq:Izero_1_res_P}
\end{eqnarray}
for small $y$, which holds near the QCP.
This gives Eq.~(\ref {eq:Izero_1_res}).

\section{Derivation of Eq.~(\ref{eq:a_P2})}
\label{sec:deriv_a}

The derivation of Eq.~(\ref{eq:a_P2}) is explained in this appendix. 

By substituting Eq.~(\ref {eq:Izero_log_res}) and Eq.~(\ref {eq:Izero_1_res}) into the third line and the fourth line of Eq.~(\ref{eq:a_P}), respectively, the terms in the second to fourth lines of Eq.~(\ref{eq:a_P}) are expressed as 
\begin{widetext}
\begin{eqnarray}
\frac{1}{V}\left(\frac{\partial{y}}{\partial{t}}\right)_V
\left[
-\frac{1}{T_0}\left(\frac{\partial{T_0}}{\partial{P}}\right)_{T}\tilde{C}_{\rm b}
\right.
\nonumber
\\
\left.
+\frac{Nd}{2}
\left\{
\frac{1}{T_0}\left(\frac{\partial{T_0}}{\partial{P}}\right)_T
\left(
C_1-C_{2}y_{0}
\right)
+\left(\frac{\partial{C_1}}{\partial{P}}\right)_{T}
-\left(\frac{\partial{C_2}}{\partial{P}}\right)_{T}y_{0}
\right.
\right.
\nonumber
\\
\left.
\left.
+
\left[
\frac{1}{T_0}\left(\frac{\partial{T_0}}{\partial{P}}\right)_T
\left(
2-C_{2}y_{1}\frac{d}{2}
\right)
-\left(\frac{\partial{C_2}}{\partial{P}}\right)_{T}
y_{1}\frac{d}{2}
\right]
L
\right\}
\right], 
\label{eq:a_i}
\end{eqnarray}
\end{widetext}
where the SCR equation [eq.~(\ref{eq:SCReq2})], which is written as $y=y_{0}+(d/2)y_{1}L$ using the definition of $L$ [Eq.~(\ref{eq:L_def})], is substituted into $y$ in the r.h.s. of Eq.~(\ref {eq:Izero_log_res}) and Eq.~(\ref {eq:Izero_1_res}). 

Then, one realizes that 
the second line inside the outermost $[\cdots]$ in Eq.~(\ref {eq:a_i}) can be expressed in the form as  
\begin{eqnarray}
N\frac{2}{y_1}\left(\frac{\partial{y_0}}{\partial{P}}\right)_{T}, 
\label{eq:dy0dPterm}
\end{eqnarray}
{\it as far as} the terms with the pressure derivative of $T_0$, $C_1$, and $C_2$ in Eq.~(\ref {eq:dy0dP}) are concerned. 
Similarly, one realizes that 
the third line inside the outermost $[\cdots]$ in Eq.~(\ref {eq:a_i}) can be expressed in the form as  
\begin{eqnarray}
N\frac{d}{y_1}\left(\frac{\partial{y_1}}{\partial{P}}\right)_{T}L,
\label{eq:dy1dPterm}
\end{eqnarray}
{\it as far as} the terms with the pressure derivative of $T_0$, $C_1$, and $C_2$ in Eq.~(\ref {eq:dy1dP}) are concerned. 

As for the last two lines in Eq.~(\ref{eq:a_P}), it turns out that the terms other than noted above in Eq.~(\ref{eq:dy0dP}) and Eq.~(\ref{eq:dy1dP}) complement the remaining terms in $(\partial{y_0}/\partial{P})_T$ in Eq.~(\ref{eq:dy0dPterm}) and $(\partial{y_1}/\partial{P})_T$ in Eq.~(\ref{eq:dy1dPterm}), respectively, as follows: 

The term with 
$[\partial(\eta_0/Aq_{\rm B}^2)/\partial{P}]_T$
 in Eq.~(\ref{eq:a_P}) can be expressed in the form as Eq.~(\ref{eq:dy0dPterm}), {\it as far as} the term with the pressure derivative of 
$\eta_0/(Aq_{\rm B}^2)$
 in Eq.~(\ref{eq:dy0dP}) is concerned. 

On the term with 
$(\partial{T_A}/\partial{P})_T$
 in Eq.~(\ref{eq:a_P}), by substituting 
Eqs.~(\ref{eq:psi2_zero}) and (\ref{eq:psi2_th}) into the expression of 
$\langle\varphi^2\rangle_{\rm eff}$ 
[Eq.~(\ref{eq:psi2})] and using the SCR equation [Eq.~(\ref{eq:SCReq2})], 
we obtain
\begin{widetext}
\begin{eqnarray}
-\frac{2}{T_0}\langle\varphi^2\rangle_{\rm eff}\left(\frac{\partial{T_A}}{\partial{P}}\right)_{T}
\nonumber
\\
=
-Nd\frac{1}{T_A}\left(\frac{\partial{T_A}}{\partial{P}}\right)_T
\left[
C_1-C_{2}y_0+\left(2-C_{2}y_{1}\frac{d}{2}\right)L 
\right].
\end{eqnarray}
\end{widetext}
On the last term in Eq.~(\ref{eq:a_P}), we similarly obtain
\begin{widetext}
\begin{eqnarray}
\frac{T_A}{T_0}\langle\varphi^2\rangle_{\rm eff}\frac{1}{v_4}\left(\frac{\partial{v_4}}{\partial{P}}\right)_T
\nonumber
\\
=\frac{Nd}{2}\frac{1}{v_4}\left(\frac{\partial{v_4}}{\partial{P}}\right)_T
\left[
C_1-C_{2}y_0+\left(2-C_{2}y_{1}\frac{d}{2}\right)L 
\right].
\end{eqnarray}
\end{widetext}
Hence, the last two terms in Eq.~(\ref{eq:a_P}) can be expressed in the form as the summation of Eq.~(\ref{eq:dy0dPterm}) and Eq.~(\ref{eq:dy1dPterm}), 
{\it as far as} 
the terms with the pressure derivative of 
$T_{A}$
 and $v_4$ in Eqs.~(\ref{eq:dy0dP}) and (\ref{eq:dy1dP}), respectively, are concerned. 

Thus, the summation of all these terms noted above lead to the summation of Eq.~(\ref{eq:dy0dPterm}) and Eq.~(\ref{eq:dy1dPterm}), that is  nothing but the second line of Eq.~(\ref{eq:a_P2}).
Since the first term with $C_{\rm a}$ and the third term with $\tilde{C}_{\rm b}$ in Eq.~(\ref{eq:a_P}) directly appear in the first line of Eq.~(\ref{eq:a_P2}) as the first and second terms, respectively, by combining the result of the second line of Eq.~(\ref{eq:a_P2}) derived above, we obtain Eq.~(\ref{eq:a_P2}).   



\begin{thebibliography}{99}
\bibitem{note_Heisenberg} It is straightforward to extend the present results to the case of magnetic fluctuations of XY or Heisenberg type, where the factor of 2 or 3 is multiplied to the free energy in Eq.~(\ref{eq:freeE}), respectively 
(Note that the solution of the SCR equation correctly satisfies the Mermin-Wagner theorem for the Heisenberg symmetry as shown in Fig.~\ref{fig:y_t}). 
Hence, the factor is also to be multiplied to the entropy in Eq.~(\ref{eq:S}) and the specific heat in Eqs.~(\ref{eq:Ca}) and (\ref{eq:tildaCb}). In real materials, there exists more or less anisotropy of the spin space, leading to anisotropic spin fluctuation. Therefore, we plot $\alpha$ as well as $C_V$ for a single component of the spin fluctuation with the factor of 1 in a series of Figures for convenience of comparison with experiments. 
As for the Gr\"{u}neisen parameter $\Gamma$, the results shown in Figs.~\ref{fig:G_t_d3z3}, \ref{fig:G_t_d3z2}, \ref{fig:G_t_d2z3}, and \ref{fig:G_t_d2z2} are not affected by this factor 
except for $\kappa_{T}$, since the factors in $\alpha$ and $C_{V}$ are cancelled each other in Eq.~(\ref{eq:Grn_i}).
\bibitem{note_d2_z2} Although the criticality in $d=2$ for $z=2$, $y\sim-\frac{t}{{\ln}t}$was reported in Ref.~\cite{IM1998}, more precisely it can be shown that $y\sim -t\frac{{\ln}(-{\ln}t)}{{\ln}t}$ is the solution of the SCR equation, which coincides with the result of the renormalization group discussed in Ref.~\cite{Millis} (see \ref{sec:d2z2}).
\bibitem{note_dJdV} It is generally expected that the first term in Eq.~(\ref{eq:Grn_HF}) dominates over the second term since the volume dependence of the Kondo exchange coupling is much stronger than that of the density of states of conduction electrons. 
If we adopt the tight-binding picture with the distance dependence of the hybridization between the f and conduction electrons $V_{\rm fc}\sim 1/r^{\ell+\ell'+1}$ with the azimthal quantum numbers $\ell=3$, $\ell'$, respectively~\cite{Andersen1,Andersen2,Harrison}, we estimate   $\frac{V}{J}\left(\frac{\partial{J}}{\partial{V}}\right)\sim-\frac{1}{d}(2\ell'+6+O(\frac{\varepsilon_{\rm f}}{\varepsilon_{\rm F}-\varepsilon_{\rm f}}))$ in the $d$-dimensional system. 
The absolute value is larger than $c\equiv\frac{V}{N_{\rm F}}\left(\frac{\partial{N_{\rm F}}}{\partial{V}}\right)\sim\frac{2}{d}$ estimated from the nearly-free-electron picture, since $\varepsilon_{\rm f}$ is not so close to $\varepsilon_{\rm F}$ in typical heavy-electron systems.  
This reflects the fact that the locality of the 4f wavefunction reacts severely to the volume change of the system.
\end{thebibliography}

\end{document}